\documentclass[10pt,aps,prx,twocolumn,superscriptaddress,showpacs,
floatfix]{revtex4-2}
\usepackage{amsmath,amssymb,amsthm,amsfonts,graphics,epsfig,epstopdf,
color,verbatim,tabularx,bm,multirow,appendix,url}
\let\oldproofname=\proofname
\renewcommand{\proofname}{\rm\bf{\oldproofname}}
\usepackage[utf8]{inputenc}
\usepackage[T1]{fontenc}
\usepackage{xcolor}
\usepackage{dsfont}
\usepackage{textcomp}
\usepackage{yfonts}

\usepackage{bm}
\usepackage{subfigure}
\usepackage{mathrsfs}
\usepackage{graphicx}
\usepackage{verbatim}
\usepackage[colorlinks=true, citecolor=black, linkcolor=black, urlcolor=black]{hyperref}
\usepackage{multirow}
\usepackage{enumitem,kantlipsum}

\usepackage{braket}
\usepackage[normalem]{ulem}
\usepackage{tikz}
\usetikzlibrary{calc}
\usetikzlibrary{shapes.multipart}
\usepackage[compat=1.1.0]{tikz-feynman}

\usepackage{pifont}
\newcommand{\cmark}{\ding{51}}
\newcommand{\xmark}{\ding{55}}

\DeclareMathOperator{\tr}{tr}

\newcommand{\di}{\mathrm{d}}

\newcommand{\ei}{\mathrm{e}} 

\newcommand{\be}{\begin{equation}}
\newcommand{\ee}{\end{equation}}
\newcommand{\bea}{\begin{eqnarray}}
\newcommand{\eea}{\end{eqnarray}}
\newcommand{\ii}{\mathrm{i}}

\renewcommand{\vec}[1]{{\mathbf #1}}

\newcommand{\comments}[1]{}

\newcommand{\refeq}[1]{\textmd{Eq.\ }(\ref{#1})}
\newcommand{\reffg}[1]{\textmd{Fig.\ }\ref{#1}}

\renewcommand{\qed}{QED$_3$}


\begin{document}

\title{Emergent Gauge Flux in Mixed \texorpdfstring{QED$_3$}{QED3} with Flavor Chemical Potential:
Implications for Magnetized U(1) Dirac Spin Liquids}

\author{Chuang Chen}
\affiliation{Department of Physics and HK Institute of Quantum Science and Technology, The University of Hong Kong, Pokfulam Road,  Hong Kong SAR, China}
\affiliation{State Key Laboratory of Optical Quantum Materials, The University
of Hong Kong, Pokfulam Road,  Hong Kong SAR, China}

\author{Urban F. P. Seifert}
\affiliation{Institute for Theoretical Physics, University of Cologne,
Z\"ulpicher Strasse 77a, 50937 Cologne, Germany}

\author{Kexin Feng}
\affiliation{Department of Physics and HK Institute of Quantum Science and Technology, The University of Hong Kong, Pokfulam Road,  Hong Kong SAR, China}
\affiliation{State Key Laboratory of Optical Quantum Materials, The University
of Hong Kong, Pokfulam Road,  Hong Kong SAR, China}

\author{Oleg A. Starykh}
\affiliation{Department of Physics and Astronomy, University of Utah, Salt Lake City, Utah 84112, USA}

\author{Leon Balents}
\affiliation{Kavli Institute for Theoretical Physics, University of
California, Santa Barbara, California 93106, USA}
\affiliation{French American Center for Theoretical Science, CNRS, KITP, Santa
Barbara, California 93106-4030, USA}
\affiliation{Canadian Institute for Advanced Research, Toronto, Ontario,
Canada}

\author{Zi Yang Meng}
\email{zymeng@hku.hk}
\affiliation{Department of Physics and HK Institute of Quantum Science and Technology, The University of Hong Kong, Pokfulam Road,  Hong Kong SAR, China}
\affiliation{State Key Laboratory of Optical Quantum Materials, The University
of Hong Kong, Pokfulam Road,  Hong Kong SAR, China}

\begin{abstract}
We design a lattice model of a ``mixed'' U(1) gauge field coupled to fermions
with a flavor chemical potential and solve it with large-scale determinant
quantum Monte Carlo simulations. In this mixed construction, dynamical
magnetic monopole events are suppressed in the temporal-continuum limit. For
zero flavor chemical potential, the model realizes three-dimensional quantum
electrodynamics (QED$_3$) which has been argued to describe the ground state
and low-energy excitations of the Dirac spin liquid phase of quantum
antiferromagnets.
At finite flavor chemical potential, which corresponds to a Zeeman field
perturbing the Dirac spin liquid, we find a ``chiral flux'' phase which is
characterized by the generation of a finite mean emergent gauge flux and,
accordingly, the formation of relativistic Landau levels for the Dirac
fermions. We use field-theory arguments to argue that in this state a
U(1)$_m$ magnetic symmetry is spontaneously broken, leading to a gapless
photon mode. Because of spin-flux attachment this mode would contribute to the
low-energy longitudinal spin response. We numerically compute longitudinal and
transverse spin structure factors which match our continuum and lattice
mean-field theory predictions. In a different region of the phase diagram,
strong fluctuations of the emergent gauge field give rise to an
antiferromagnetically ordered state with gapped Dirac fermions coexisting with
a deconfined gauge field. We also find an interesting intermediate phase where
the chiral flux phase and the antiferromagnetic phase coexist. We argue that
our results pave the way to testable predictions for magnetized Dirac spin
liquids in frustrated quantum antiferromagnets.
\end{abstract}

\date{October 31, 2026}
\maketitle

\section{Introduction}
\label{sec:I}
\subsection{Motivation}
The emergence of nontrivial quantum field theories in condensed matter
systems is a remarkable phenomenon.  Among the most interesting are strongly
interacting gapless field theories with enhanced (generalized) symmetries.
Three-dimensional (two space plus one time) quantum electrodynamics, or \qed,
is a striking example of such a field theory.
Composed of $N$ two-component Dirac spinors interacting with a U(1) gauge field,
\qed\ is known to realize a conformal field theory (CFT) for sufficiently
large
$N$ (it is commonly believed that $N \geq 4$ is sufficient while we caution
that distinguishing genuine conformal behavior and pseudocriticality in
numerical simulations is challenging)
\cite{karthikNoEvidence2016,giombi16,di2017scaling,he22}.
This CFT contains a rich collection of scaling operators including fermion
bilinears (SU($N$) flavor currents and mass terms) and monopole operators,
which generate U(1) magnetic fluxes, and which carry nontrivial
representations of flavor.  As a relatively simple strongly interacting gauge
and conformal field theory, \qed\ is a test case for nonperturbative methods
in the high-energy community. In condensed matter, \qed\ has been argued to
arise as a ground state of certain quantum antiferromagnets, and in this
context is known as  a \emph{Dirac spin liquid}
(DSL)~\cite{Hermele2004,Hermele2005,ranProjected2007}.
In the DSL, the U(1) gauge field and fermions are \emph{emergent}: the
microscopic formulation begins at a lattice level with only a spin-1/2 direct
product Hilbert space.  The appearance of \qed\ in such a situation is
striking and fascinating, worthy of detailed study and verification.

In our opinion, the existence of \qed\ as a stable phase of matter
(i.e.~robust to symmetry-allowed perturbations) is well established in some
circumstances {(this presupposes that a UV symmetry-allowed SO(6) rank-2
symmetric tensor operator is irrelevant; recent conformal-bootstrap studies
have analyzed $N=4$ QED$_3$ under spectral assumptions concerning this
operator \cite{he22,Albayrak2022}, but do not by themselves establish its
irrelevance)}.
From the field-theory perspective, basic properties of \qed\ are understood,
e.g. the set of primary fields, rough determinations of their scaling
dimensions \cite{di2017scaling,chester2016towards,he22,Albayrak2022}, and how
microscopic (UV) symmetries are implemented \cite{song19,WietekQED3}, allowing
for a symmetry-based analysis of perturbations
\cite{LuoTwistedDSL22,NambiarMonopole23,SeifertSpinPeierls24}.
However, more detailed properties such as the computation of multi-point
correlation functions, and renormalization group flows under various
perturbations, are as yet unknown.

From the condensed matter side, the DSLs have been proposed and investigated
in spin-1/2 Heisenberg models on kagom\'e
\cite{Hastings2003,ranProjected2007,Iqbal2011,He2017,
luUnraveling2026} and triangular
\cite{Imada2014,White2015,Hu2015,Iqbal2016,Gong2017,
Saadamand2017,He2019,WietekQED3,FerrariStability2024} lattices by a variety of
analytical and numerical techniques. Yet, the understanding of their physical
response functions remains limited \cite{Ferrari2019,Sherman2023,Drescher2023,
WillsherDynamics2025}. 

In this paper, we extend the understanding of \qed\ and the DSL by exploring
the effect of a particularly important physical perturbation: an external
magnetic field in the condensed matter realization, which couples to the spins
via Zeeman interaction.  From the \qed\ perspective, this field appears as a
flavor chemical potential.  In experiments on quantum magnets, the study of
the phase diagram enriched by the field axis is a routine and powerful way to
probe the physics.  In \qed, we will see that the flavor chemical potential
enjoys an intriguing interplay with the emergent flux, and may, for the DSL,
make the emergent structure of the system more apparent.
In order to decouple the complicated physics of magnetic monopoles from that
of the emergent orbital flux and relativistic Landau levels of fractionalized
spinons, which is the focus of our study, we propose to investigate a
``mixed'' lattice gauge theory that suppresses magnetic monopole events when
one takes the continuum limit along the temporal direction. We stress that
within this mixed theory, a magnetic flux through spatial plaquette remains
well defined. All our simulations reported below adhere to this central idea.
Our simulations therefore isolate the field-induced flux physics in a
controlled setting, establishing the chiral flux (CF) phase and its characteristic
signatures in the dynamical spin response within the mixed theory while
disentangling them from the additional effects of dynamical monopoles. How
these monopole effects reshape the resulting phase in a fully compact gauge
theory, as is relevant to microscopic spin models, remains an important
question for future work.

\begin{figure*}[htp!]
\includegraphics[width=\textwidth]{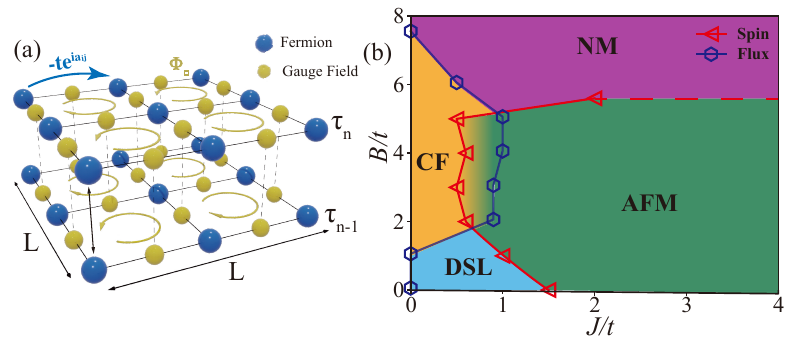}
\caption{\textbf{Lattice model and DQMC phase diagram.} (a) Fermions are on
the sites of cubic
          lattice represented by filled blue balls while the gauge fields are
          on the bonds of the lattice represented by filled yellow balls. The
          gauge field on the temporal direction is fixed to $0$ (denoted by
the dashed lines).
          Fermions hop between nearest neighbor sites with a phase $e^{\pm i
a_{ij}}$. Each plaquette is attached to a flux $\Phi_\Box$,
          computed from gauge field on the four bonds. Black arrows indicate
spatial
          and temporal directions, with lattice constant $1$ for spatial and
          $0.1$ for temporal directions. $\tau_n$ and $\tau_{n-1}$ are
adjacent temporal layers.  (b) Phase
          diagram obtained from DQMC simulation. The red line is
          determined from spin correlation ratio $r_{\text{AFM}}$ while the
blue line is from flux
          Binder cumulant $U_{\text{flux}}$. Right of red line is the AFM
phase which
          breaks the U(1)$_f$ symmetry of spin rotation when $B\neq 0$.
        Left of blue line is chiral flux state (CF), in which the
$\mathcal{T}$
        symmetry of flux of the gauge field is broken. There is a
nonmagnetic
        (NM) phase at the top, Dirac spin liquid (DSL) phase at the bottom
        corner and a possible coexisting CF-AFM phase, represented by a color
        gradient between CF and AFM phases. 
        Note that the finite extent of the DSL phase for $B>0$ is due to the
nonzero temperature accessible in our simulations. We expect the DSL to give
way to the CF phase for infinitesimal Zeeman fields $B>0$ in the
zero-temperature thermodynamic limit.}
        \label{fig:fig1}
\end{figure*}

Our investigation is inspired by work of Ran et al.~\cite{ranSpontaneous2009}
who considered the response to a Zeeman field in a DSL on a kagom\'e lattice,
using a parton mean-field approach and a variational wave function.  It is also
related to recent work discussing spontaneous symmetry breaking in \qed\ in
zero field for $N=2$, for which there is believed not to be a stable CFT
\cite{DumitrescuSymmetryQED325}.
Yet at the same time, recent lattice model quantum Monte Carlo simulations of
\qed\ have shown signatures of DSL and its phase transitions to confined
phases at the spatial scale of $20\times
20$~\cite{xu2019monte,wangDynamics2019,lukasConfinement2020}.

\subsection{Model}

Here we carry out a numerically exact determinant quantum Monte Carlo (DQMC)
study of a lattice model for \qed\ derived from the one in
Refs.~\cite{xu2019monte,wangDynamics2019} on a cubic space-time lattice
(square spatial lattice). The model is defined by the action
\begin{align}
S & = \sum_{i,n} \left[\psi^\dagger_{i}(\tau_n) (\psi_{i}(\tau_n) -
\psi_{i}(\tau_{n-1})) \right. \nonumber \\
& \left. - \frac12  B \psi^\dagger_{i}(\tau_n) \sigma^z
\psi_{i}(\tau_n) \right]  \nonumber \\
& - t \sum_{\langle ij\rangle,n} \left[ e^{i a_{ij}(\tau_n)}
\psi^\dagger_{i}(\tau_n) \psi_{j}(\tau_n) + \textrm{h.c.} \right] \nonumber \\
& + \frac{1}{J } \sum_{\langle ij\rangle,n} \left[
a_{ij}(\tau_{n})-a_{ij}(\tau_{n-1})\right]^2
\label{eq:action}
\end{align}
where $\psi_i = \begin{pmatrix} \psi_{i,\uparrow} \\
  \psi_{i,\downarrow} \end{pmatrix}$ is a fermionic spinor, $a_{ij}$ is a
bosonic scalar and it acts as a U(1) gauge field, $i,j \in [1,L^2]$ labels
sites of the square
lattice, and $\langle ij \rangle$ is the nearest neighbor bond, with the
simulated spatial system size $L=6,8,10,\cdots,20$. The temporal system size
is $N_{\tau}=\frac{1}{T\Delta\tau}=20L$, where $\tau_n \in
[0,\beta=\frac{1}{T}=2L]$ labels the discrete imaginary time with step
$\Delta\tau=0.1$ in our setting and $n\in[0,N_\tau]$ is its integer index, as
shown Fig.~\ref{fig:fig1} (a). The partition function $Z = {\rm Tr} \,
\{e^{-S}\}$ is obtained by tracing out all fermion and boson degrees of
freedom, see Eq.~\eqref{eq: partition}.

The model in Eq.~\eqref{eq:action} is a somewhat unconventional variant of the
gauge theory.  In particular, while the fermion terms are fully periodic under
a
$2\pi$ shift of $a_{ij}(\tau_n)$ on any single bond at any single imaginary
time
$\tau_n$, the ``Maxwell'' term in the last line of Eq.~\eqref{eq:action} lacks
this periodicity.  Thus the action lacks such periodicity and instead has a
\emph{subsystem} periodicity: it is invariant under the shift $a_{ij}(\tau_n)
\rightarrow a_{ij}(\tau_n) + 2\pi m_{ij}$ with any set of integers $m_{ij}$
which are time independent.  This choice was made in order to be as close as
possible to the noncompact theory (which lacks any periodicity) while still
allowing the model to dynamically choose a nontrivial flux through spatial
slices (horizontal surfaces in space-time defined by
$\tau_n=\mathrm{const}.$).
We refer to our model as the ``mixed'' compact-noncompact theory, or
    simply the ``mixed'' theory.

It is well known that the compact (fully periodic) theory includes ``magnetic
monopoles'': pointlike defects in spacetime (instantons) associated with
nonzero quantized values of magnetic three-flux integrated over a closed
space-time surface. These defects are dynamical degrees of freedom and to be
summed over in a partition function for a fully compact U(1) gauge theory.
The critical phases of \qed\ can suffer instabilities driven by monopole
proliferation.  When exactly such instability occurs depends upon the phase in
question and subtle lattice physics due to Berry phase effects~\cite{song19}.
Our simulation has been tailored to suppress monopoles in order to evade such
instabilities.

The behavior of monopoles in the mixed theory is somewhat subtle.
As we show explicitly in an analytically tractable model in Supplemental Material (SM)
\cite{suppl}\nocite{gradshteyn15}, the theory Eq.~\eqref{eq:action} still
admits  monopole excitations owing to the periodic (in $a_{ij}$) fermion-gauge
coupling, albeit with a highly constrained mobility (i.e.~quantized magnetic
flux can only penetrate surfaces perpendicular to the temporal direction).
As
a result, taking the continuum limit along the temporal direction suppresses
these monopole excitations, so that the mixed theory behaves as a noncompact
gauge theory, except that the net flux through a full spatial slice is free and
energetically chosen by the model.
Our numerical simulations are carried out well into this limit, which
corresponds to a highly anisotropic space-time lattice (with $N_\tau \gg N$).
Therefore in our simulation we expect stable phases with both free and
strongly interacting photons.  We remark that, even when monopoles are
suppressed,  one may consider monopole operators as nondynamical
``insertions'' defined by modifying the partition function appropriately.

Our analysis of the DQMC results rests on an understanding of the symmetries
of the problem.  In this paper, we follow modern conventions and only refer to
symmetries of those operations which act nontrivially on physical states,
i.e., on gauge-invariant observables.
In particular, gauge invariance does not qualify as a symmetry, but rather
reflects a constraint on the Hilbert space.
We list important \emph{physical} symmetries of the lattice model
\eqref{eq:action} which are relevant to our analysis:
\begin{enumerate}
    \item {Considering the model with $n$ flavors of fermions (with flavor
index $\alpha=1\ldots n$), in zero applied Zeeman field $B = 0$, there is an
SU($n$) flavor symmetry, which we denote SU($n$)$_f$. Here we focus on $n=2$.
    This symmetry is lowered upon application of a finite Zeeman field $B\neq
0$ (explicitly breaking SU(2)$_f$ symmetry) to U(1)$_f$, associated with
rotations around the field axis in flavor space, which acts on Dirac fermions
as $\mathrm{U(1)}_f : \psi_i(\tau_n) \to e^{-i \alpha \sigma^z}
\psi_i(\tau_n)$ with $\alpha \in [0,2\pi)$ and leaves the U(1) gauge field
invariant.}
    \item {In the time-continuum limit, as discussed above, there is a
conserved magnetic flux through the spatial slices of the lattice.   This may
be written $\Phi(\tau_n) = \sum_{(r_x,r_y)} \textrm{mod}((\nabla\times
a)_{(r_x,r_y,\tau_n)},2\pi)$ and the conservation means that $\Phi(\tau_n)$ is
independent of $\tau_n$ for finite-action configurations in the Monte Carlo
procedure in the continuum time limit.  This conserved flux defines a U(1)$_m$
``magnetic'' symmetry on quantum states, i.e. the conserved flux is the
symmetry generator for the U(1)$_m$ symmetry. The definition of the symmetry
generator {\it{is}} the way a symmetry is defined in quantum mechanics. In a
phase in which fermions are gapped, this symmetry can be made manifest through
an explicit duality transformation of the mixed theory, which is presented in
Supplemental Material \cite{suppl}.  This reveals that the spontaneous breaking
of this symmetry manifests in a linearly dispersing ``dual'' Goldstone mode,
i.e. a photon.}
    \item {There is a discrete time-reversal invariance $\mathcal{T}$
($\psi_i(\tau) \rightarrow \psi^\dagger_i(-\tau)$, $\psi^\dagger_i(\tau)
\rightarrow - \psi_i(-\tau)$, $a_{ij}(\tau) \rightarrow - a_{ij}(-\tau)$)
which is a symmetry of Eq.~\eqref{eq:action} also at \emph{finite Zeeman
fields} $B \neq 0$. Hence, $\mathcal{T}$ is distinct from the usual
time-reversal symmetry for spin-1/2 particles, the latter being broken
\emph{explicitly} by finite Zeeman fields $B\neq 0$.}
    \item {The model further enjoys spatial translation symmetries (along with
point group symmetries, which are not of immediate relevance for the remainder
of our work). Translations along $\hat{x}$ act as $T_x: \psi_i(\tau) \to
\psi_{T_x(i)}(\tau)$ and $a_{ij}(\tau_n) \to a_{T_x(i),T_x(j)}(\tau_n)$ with
$T_x(i) \equiv i + \hat{x}$, and analogously for $T_y$.}
\end{enumerate}

We summarize the key aspects of the “mixed” gauge theory, and in particular
the consequence of the non-periodicity in $a_{ij}$ of the time-derivative
Maxwell term. For nonzero Trotterization, monopoles are allowed, but they are
very strongly suppressed with diminishing time discretization, and become
fully absent in the time-continuum limit.    In this limit, a perfectly
conserved magnetic flux $\Phi$ defines a $U(1)_m$ symmetry.

We emphasize that the lattice model of Eq.~\eqref{eq:action} is a form of
\qed\ in that it describes fermions interacting with a bosonic U(1) gauge
field, but it is not a continuum model.  Standard continuum \qed\ contains $N$
species of two-component Dirac fermions coupled to such a gauge field.  In a
certain regime such a theory arises as a continuum limit of
Eq.~\eqref{eq:action} (see below).  Because of fermion doubling, the low energy
continuum theory has $N=2n$ Dirac fermions and the SU($n$)$_f$ lattice
symmetry is enlarged to an emergent SU($N$)$_f$ = SU($2n$) = SU(4)$_f$
symmetry.
However, we should keep in mind that residual effects originating from lattice
corrections to the continuum limit preserve only the SU(2)$_f$ subgroup.
Accordingly, we focus on the robust exact symmetries of the lattice model to
characterize the system.

\subsection{Summary of results}

Aided by these symmetries, our DQMC results, obtained on finite systems of
dimensions $\beta \times L \times L$, as explained above, are summarized in
the low-temperature phase diagram of Fig.~\ref{fig:fig1} (b), spanned by the
axes of gauge fluctuation $J/t$ and magnetic field $B/t$. To characterize the
symmetries and their breaking, we introduce order parameters.  There are two
manifest order parameters: $N^+ = (-1)^i\langle\psi^\dagger_{i} \sigma^+
\psi_i^{\vphantom\dagger}\rangle$, which describes spontaneous breaking of the
U(1)$_f$ flavor symmetry to form antiferromagnetic (AFM) spin order in the XY plane,
and $\chi = \langle\sin \Phi_\Box\rangle$, a chirality, which describes broken
$\mathcal{T}$ symmetry.  Here $\Phi_\Box = \nabla\times a$ is the flux through
a spatial plaquette, given by the lattice curl of the gauge field.   A third
order parameter, which is hidden, characterizes the U(1)$_m$ magnetic
symmetry: this is a monopole operator $\mathcal{M}^+$, which creates a $2\pi$
flux in the gauge field.  We do not currently measure $\mathcal{M}^+$ in our
simulations, but its presence can be deduced from an understanding of the
corresponding Goldstone mode: the Goldstone mode of the U(1)$_m$ symmetry is
the \emph{photon} of the gauge field.

We observe three manifestly symmetry-broken phases as visible in the phase
diagram Fig.~\ref{fig:fig1}: an antiferromagnetic phase, in which $N^+$
is nonzero but $\chi=0$, a chiral flux phase in which $\chi \neq 0$ but
$N^+=0$, and an overlapping chiral AFM phase in which both $N^+$ and $\chi$
are nonzero.
We also find two symmetric phases: a Dirac spin liquid phase and a
polarized phase at strong field in which the spins are nearly fully aligned
with the field (full polarization is not possible at finite temperatures where
our simulations are done).
In all the phases except the DSL, the U(1)$_m$ symmetry is spontaneously
broken, i.e. $\mathcal{M}^+ \neq 0$, witnessed by the presence of a linearly
dispersing free-photon mode. We provide an overview of various phases of the
model and the symmetries that they preserve and/or spontaneously break in
Table~\ref{tab:overview}.

\begin{table}[htb]
\begin{tabular}{l||c|c|c|c|c}
& $\mathrm{U(1)}_f$ & $\mathrm{U(1)}_m$ & $\mathrm{U(1)}_m + \mathrm{U(1)}_f$
& $\mathcal{T}$ & $T_x,T_y$ \\ \hline\hline
DSL & \cmark & \cmark & \cmark & \cmark & \cmark \\
CF & \xmark & \xmark & \cmark & \xmark & \cmark \\
AFM & \xmark & \xmark & \xmark & \cmark & \xmark \\
AFM + CF & \xmark & \xmark & \xmark & \xmark & \xmark \\
NM & \cmark & \xmark & \xmark & \cmark & \cmark \\
\end{tabular}
\caption{Phases and their (spontaneously broken) symmetries: Here, ``\cmark''
indicates that a symmetry is preserved in a phase while ``\xmark'' indicates
the spontaneous breaking of a symmetry. We use ``$\mathrm{U(1)}_m +
\mathrm{U(1)}_f$'' to denote a diagonal subgroup of $\mathrm{U(1)}_m \times
\mathrm{U(1)}_f$, see also the discussion in Sec.~\ref{sec:CF}.
\label{tab:overview}}
\end{table}

In addition to the phase diagram, we explore the magnetic spectra of the gauge-invariant fermion bilinears -- the spin operators in the condensed matter
language -- of \qed.  Universal features of these spectra provide signatures
of fractionalization and emergent gauge fields in quantum magnets proximate to
a DSL phase.  We are particularly interested in uncovering structures that are
unexpected without spin liquid physics, and which might be striking aspects to
seek experimentally.  In that regard, we focus particularly on the chiral
phase in which the emergence of Landau-Hofstadter bands that result from the
fractional flux in the gauge field may be reflected in the spin correlations.
We present DQMC results for the longitudinal and transverse dynamical spin
susceptibilities, and demonstrate a strong correlation between these
correlations and a mean-field result which precisely describes fermions in
Hofstadter bands.  We observe a clear splitting of spectral weight into
multiple features that can be understood from the effective enlargement of the
magnetic unit cell of fermions experiencing a fractional flux per plaquette,
despite the lack of translational symmetry breaking in the system.
In some cases features reminiscent of the Landau-level nature of spinon bands
are visible in the numerically obtained magnetic spectra.
Our DQMC results for the longitudinal spin structure factor further show
signatures of a gapless mode as the possible manifestation of the gapless
photon, originating from gauge field fluctuations and therefore a manifestly
beyond-mean-field result.

\subsection{Guide to the paper}
The paper is organized as follows: 
Sec.~\ref{sec:II} lays out the lattice implementation of the
mixed \qed\ model and its DQMC numerical algorithm (Sec.~\ref{sec:IIA}), and
we further explain the Monte Carlo update scheme of the gauge field to
overcome its slow dynamics and anisotropic space-time gauge choice
(Sec.~\ref{sec:IIB}). Sec.~\ref{sec:rpa}, on the other hand, provides the
theoretical discussion of the different phases (DSL, CF, AFM) in the phase diagram at the continuum limit with random phase approximation (RPA).
Section~\ref{sec:III} presents the phase diagram.
Section~\ref{sec:IIIA} provides the lattice mean-field analysis at $J=0$,
while Sec.~\ref{sec:IIIB} determines the phase boundaries using DQMC
simulations at finite $J$ and $B$.
Sec.~\ref{sec:IV} focuses on the data and analysis of the magnetic spectra in
the CF and AFM states, starting from the field-theoretical calculation
(Sec.~\ref{sec:IVA}), the lattice mean-field calculation (Sec.~\ref{sec:IVB})
and more importantly, to the DQMC simulation results and the discussion of
similarity and difference between the numerically unbiased solution and the
mean-field analysis (Sec.~\ref{sec:IVC}). Sec.~\ref{sec:conclusion} provides
the discussions of the results and makes connection with potential DSL and CF
phases in frustrated magnets and their experimental detections.

Sections I and II of the Supplemental Material (SM)~\cite{suppl}
describe the magnetic spectra calculated using lattice mean-field and
low-energy field-theory approaches, respectively. Section III reviews
Larmor's theorem and its application to the magnetic spectra at the
$\Gamma$ point. Section IV presents flux-correlation results for the pure
gauge theory and for the DSL and AFM phases. Section V explains the
$z$-direction flux insertion and global update scheme used in the DQMC
simulations. Section VI discusses the U(1)$_m$ symmetry and the duality map.
Section VII compares the mean-field and DQMC magnetic spectra at $B/t=2$.
Section VIII presents a finite-size analysis of the order parameters.

\section{Methods and Theoretical Expectations}
\label{sec:II}
\subsection{\texorpdfstring{$\text{QED}_3$}{QED3} with lattice DQMC}
\label{sec:IIA}

As mentioned in Sec.~\ref{sec:I}, we consider the QED$_3$ theory on the cubic
lattice, with the action in Eq.~\eqref{eq:action}. The mixed nature of the
gauge coupling penalizes monopoles more than in a compact gauge theory, and
indeed they are parametrically suppressed in the temporal continuum limit.  In
that regime we expect our mixed gauge theory to act like a noncompact gauge theory.
We discard the $K$ term used in prior work (Ref.~\cite{xu2019monte}), because
it adds an undesirable additional pinning field favoring $\pi$-flux in the
gauge field, masking the emergence of non-$\pi$-flux states with nonzero
Zeeman field.

In the DQMC study of the action above, the partition function takes the form 
\begin{align}
Z=\int D(a, \bar{\psi}, \psi) \ e^{-\left(S_a+S_f\right)}=\int Da \ e^{-S_a}
\operatorname{Tr}_\psi\left[e^{-S_f}\right],
\label{eq: partition}
\end{align}
where $S_a$ is the pure gauge field action (the last line in
Eq.~\eqref{eq:action}) and $S_f$ contains all the remaining contributions
which are quadratic in fermions. The gauge field $a_{ij}$ is an unconstrained
continuous variable living on the nearest spatial bonds of cubic lattice as
shown in
Fig.~\ref{fig:fig1} (a).   The quadratic fermion $\psi_i$ can be traced out to
give
\begin{align}
  \operatorname{Tr}_\psi\left[e^{-S_f}\right]=&\prod_\alpha
\left[\operatorname{det}\left(\mathbf{I}+\prod_{n=1}^{N_\tau}
\mathbf{B}_{\tau_n,\alpha}\right)\right] \nonumber\\
=&\prod_\alpha\operatorname{det}\mathbf{M}_\alpha.
\label{eq:eq3}
\end{align}
where $\mathbf{B}_{\tau,\alpha}=e^{-\mathbf{V}_{\tau,\alpha}}$ is the
exponential of
fermion-gauge coupling matrix $\mathbf{V}_{\tau,\alpha}$, whose elements are
determined by gauge field $-t \ e^{ia_{ij}(\tau)}$ and Zeeman term.
$\alpha=\uparrow,\downarrow$ in our case, and the matrix $$\mathbf{M}_\alpha=
\begin{pmatrix}
\mathbf{1} & 0 & 0 & \cdots & 0 & \mathbf{B}_{\tau_{N_\tau},\alpha} \\
-\mathbf{B}_{\tau_1,\alpha} & \mathbf{1} & 0 & \cdots & 0 & 0 \\
0 & -\mathbf{B}_{\tau_2,\alpha} & \mathbf{1} &\cdots & 0 & 0 \\
\vdots & \vdots & \vdots & \ddots & \vdots & \vdots \\
0& 0 & 0 & \cdots & \mathbf{1} & 0 \\
0 & 0 & 0 & \cdots &-\mathbf{B}_{\tau_{N_\tau-1},\alpha} & \mathbf{1}
\end{pmatrix}$$ is of the size of space-time volume $NN_{\tau} \times N
N_{\tau}$. We simulate the system size of $N=L\times L=6\times6,
8\times8,10\times10, \cdots, 16\times 16, 20\times 20$, with $N_{\tau}= 20L$
($\beta=\frac{1}{T}=2L$ and $\Delta\tau=0.1$) to ensure the low temperature.
For the case without Zeeman field, it can be shown that
$\operatorname{det}\mathbf{M}_\alpha\in \mathbb{R}$ and
$\operatorname{det}\mathbf{M}_\uparrow=\operatorname{det}
\mathbf{M}_\downarrow$, thus sign-problem-free for Monte Carlo simulation \cite{xu2019monte,panSign2024}. With nonzero
Zeeman field,
$\operatorname{det}\mathbf{M}_\alpha$ is complex. However, from particle-hole
symmetry between
the two spin flavors, $\operatorname{det}\mathbf{M}_\uparrow =
\operatorname{det}\mathbf{M}_\downarrow^\dagger$ is guaranteed so that the
full weight remains sign-problem-free as well. Thus the partition function Eq.~\eqref{eq: partition} is
amenable to DQMC method~\cite{blankenbecler1981monte,
assaad2008world,
xuRevealing2019}.

\subsection{Update scheme in DQMC simulations}
\label{sec:IIB}
In a nonzero Zeeman field $B>0$, the preferred configurations of the gauge
field have a flux $\Phi_\Box$ different from $\pi$ per plaquette at the mean
field level~\cite{ranProjected2007}.     Beyond mean field, the flux on each
plaquette remains a gauge-invariant observable and the energy of the system
depends on the
flux configuration.  We expect in the DQMC simulation, the probability
distribution of the flux will thus generally deviate from one centered at
$\pi$. Therefore, a Monte Carlo update method that changes the flux
directly would be highly desirable from the perspective of efficiency. To this
end, we combine a local update of the gauge field on each bond $a_{ij}(\tau_n)
\rightarrow a_{ij}(\tau_n) + \delta a$ with a global update that can change
the flux directly in the entire lattice.

To perform the space-time global update, based on the $z$-direction flux
insertion technique~\cite{alf2.0,jiangMonte2022}, one can update the gauge
field
$a_{ij}(\tau_n)$ for all spatial bonds and imaginary time layers
simultaneously with the
following scheme,
\begin{equation}
  a_{ij}(\tau_n) \rightarrow a_{ij}(\tau_n) + a_{o,ij}
\end{equation}
with $a_{o,ij}$ the introduced extra orbital field. In the Landau gauge,
$a_{o,ij}$ for bonds in the
$x$ direction is $-\frac{2\pi y N_\Phi
}{L^2}$. While for bonds in the $y$ direction, depending on the location,
$a_{o,ij} = 0$ for bonds away from boundary and
$a_{o,ij} = \frac{2\pi x N_\Phi}{L}$ for those cross boundary in $y$
direction. $x,y$ in the expression is the position of the bond in unit of
lattice constant and $N_\Phi$ is an integer. For detailed derivation of the
Landau gauge on the lattice with $z$-direction flux insertion, please refer to
Sec.~V of the SM~\cite{suppl}.

At each global update, $N_\Phi$ is randomly chosen and we apply the same
$a_{o,ij}$ (according to the random $N_\Phi$ and the Landau gauge) for all
imaginary time layers. If the update is accepted, a uniform flux $\frac{2\pi
N_\Phi}{L^2}$ is inserted into the model. In DQMC simulations, such an update requires the
computation of determinant in Eq.~\eqref{eq:eq3} from scratch and it is
expensive. We thus choose to perform a combination of a sweep of Metropolis
local
updates plus $4$ times of such global updates, which is defined as one
complete
sweep in DQMC simulation. In this way, the number of global updates does not
scale with system size. As mentioned above, global update helps to quickly
evolve to
the desired flux sector and local update will explore the whole phase space
ergodically.

\subsection{Continuum limit and random phase approximation}
\label{sec:rpa}

To provide a framework to understand the numerical results, we discuss the
continuum limit of Eq.~\eqref{eq:action} and analyze it in the random phase
approximation (RPA), allowing for the possibility of spontaneous AFM order.
To obtain the continuum limit, we assume first small Zeeman field $B\ll 1$, so
that the flux $\Phi_\Box \approx \pi$.  We then take $a_{i,i+\mu}(\tau_n) =
\bar{a}_{i,i+\mu} + A_\mu(x_i,y_i,\tau_n)$, where $\mu=x,y$ and
$\bar{a}_{i,i+\mu}$ is a c-number background gauge field representing the
$\pi$ flux, $\nabla \times \bar{a} = \pi$, and $A_\mu(x,y,\tau)$ is a slowly
varying continuum field describing small deviations from $\pi$ flux.  When $J
\lesssim 1$, we may assume the fluctuations of $A_\mu$ are weak, and that the
fermionic action may be approximated by the low energy form near zero
frequency and the zero energy Dirac points of the $\pi$ flux problem.

For concreteness, we take the gauge $\bar{a}_{i,i+x}=0$ and $\bar{a}_{i,i+y}=
\pi x_i$, which makes the hopping on the $y$-oriented bonds with odd $x_i$ to
be negative relative to other hoppings. This choice doubles the unit cell
along the $x$-direction.
Then the lattice fermion $\psi_{i,\sigma}$ (with $\sigma$ a spin index) can be
decomposed into slow-varying continuum fields $\Psi_{s,v,\sigma}(x,y,\tau)$
with $s=\textrm{mod}(x_i,2) = 0,1$ a sublattice index, $v=0,1$ a valley index,
as follows:
\begin{align}
  \psi_{i,\sigma}(\tau_n) \sim & \sum_{v=0,1}(i (-1)^v )^{x_i+y_i}
\Psi_{\textrm{mod}(x_i,2),v,\sigma}(x_i,y_i,\tau_n), \nonumber \\
    \psi^\dagger_{i,\sigma}(\tau_n) \sim & \sum_{v=0,1}  (-i (-1)^v)^{x_i+y_i}
\Psi^\dagger_{\textrm{mod}(x_i,2),v,\sigma}(x_i,y_i,\tau_n).
\label{eq:cont1}
\end{align}
We insert this into the action and gradient expand assuming slow variations of
$A_\mu$ and $\Psi$, so that the spatial part of the fermion action [the hopping term in Eq.~\eqref{eq:action}] turns into the sum of two valley contributions
$\propto -i t (-1)^v \int d\tau d^2 x \, \Psi^\dagger_{s,v,\sigma}
(\tau^x_{s,s'} \partial_x + \tau^z_{s,s'} \partial_y) \Psi_{s',v,\sigma}$.
This is followed by the transformations $\Psi_{s,1,\sigma} \rightarrow
\tau^y_{ss'} \Psi_{s',1,\sigma}$, $\Psi^\dagger_{s,0,\sigma} \rightarrow
\bar{\Psi}_{s',0,\sigma}\tau^y_{s's}$, which are not unitary but allowed since
$\bar{\Psi}$ and $\Psi$ are independent in the path integral, and the
``renaming'' of the Grassman integration field in the valley $v=1$ as
$\Psi_{s,1,\sigma}^\dagger \to \bar{\Psi}_{s,1,\sigma}$.  The result is the
continuum action $S = \int \! d\tau d^2x\, \mathcal{L}$, with
\begin{align}
  \label{eq:cont2}
  \mathcal{L} =\sum_{\mu=0}^2& \bar{\Psi} \gamma^\mu v_\mu (\partial_\mu + i
A_\mu)\Psi  - \frac{b}{2} \bar{\Psi}\gamma^0 \sigma^z\Psi \nonumber \\
& + \frac{1}{g} \sum_{\mu=x,y}(\partial_\tau A_\mu)^2,
\end{align}
where $v_0=1$, $v_1=v_2\equiv v=2t/\Delta \tau$, $b=B/\Delta\tau$, and
$g = J/\Delta \tau$. We can let $\gamma^0 = \tau^y$,
$\gamma^1=\tau^z$, $\gamma^2=-\tau^x$, where the Pauli matrices
$\bm{\tau}$ are defined to act in the sublattice space, $\bm{\sigma}$
acts in the spin space, and we suppressed all the sublattice, valley, and spin
indices.

Eq.~\eqref{eq:cont2} describes a continuum Dirac theory with $2\times 2=4$
flavors arising from the spin and valley, coupled to the continuum U(1) gauge
field $A_\mu$, which is precisely \qed\ with the gauge choice $A_\tau=A_0=0$.

Using the same transformations, we can also obtain the N\'eel order parameter
\begin{align}
    N^+ = \psi_i^\dagger \sigma^+ \psi^{\vphantom{\dagger}}_i (-1)^{x_i+y_i}
\sim \bar{\Psi} \sigma^+ \mu^x \Psi,
\end{align}
where $\sigma^+ = (\sigma^x+i\sigma^y)/2$ and we introduced the $\bm{\mu}$
Pauli matrices acting in the valley space.

Such continuum theory is capable of describing the quantum phases close to the
zero field DSL, which includes the DSL itself, the CF phase, and the AFM.  We
discuss each in turn, with an eye to the predictions for various correlation
functions and spectral properties to be tested in the DQMC simulations in
Secs.~\ref{sec:III} and \ref{sec:IV}.

\subsubsection{DSL}
\label{sec:IIC1}
The DSL occurs for $b=0$, in which case this is precisely noncompact \qed\
without any applied potential.  This has been analyzed extensively and is
believed to describe a scale-invariant conformal field theory.
Consequently, power law behavior is expected for all gauge-invariant
observables.
Because of fermion doubling in DSL phase, the CFT is expected to have emergent
SU(4) ($\sim \mathrm{SO}(6)$ up to a sign) symmetry, which enlarges the
microscopic SU(2) spin symmetry and some discrete operations in the space
group.  In the CFT of SU(4) \qed, the set of primary fields with low scaling
dimensions are as follows:
\begin{itemize}
\item the set of SU(4) conserved currents $J_\mu^a = \bar{\Psi} \gamma^\mu
\mathsf{T}^a\Psi$, where $\mu$ is a space-time index and $a$ ranges over the
15 generators $\mathsf{T}^a$ of SU(4).  Like all conserved currents in a
$(2+1)$-dimensional CFT, the currents have the exact scaling dimension $\Delta_J =
2$.
\item fermion bilinears or mass terms: a singlet $M_s = \bar{\Psi} \Psi$ and a
set of adjoints $M_a = \bar{\Psi} \mathsf{T}_a\Psi$, where $a=1\ldots 15$
range over the SU(4) generators.  \emph{A priori}, these two sets have
independent scaling dimensions, $\Delta_s$ and $\Delta_{\rm adj}$,
respectively.  An estimate from Ref.~\cite{di2017scaling} is $\Delta_s \approx
2.3$, $\Delta_{\rm adj} \in (1.4,1.7)$.

\item a set of 6 monopole operators $\mathcal{M}_q$ (and their conjugates
$\mathcal{M}^\dagger_q$), which form an antisymmetric tensor representation
of SU(4), or a real vector representation of SO(6), hence $q=1,\ldots, 6$. The
estimated scaling dimension from the large $N_f$
expansion~\cite{chester2016towards} is $\Delta_M \approx 1.0$.
\end{itemize}
These scaling operators and their composites (products) and descendants
(derivatives) appear in the long-distance correlations of lattice quantities.
For example, we expect that the DSL should exhibit flux-flux correlations
dictated by $\Delta_J=2$, which implies that, for example, the long-time
correlations of the local flux should behave as $\langle \Phi_{\Box}(\tau)
\Phi_{\Box}(0) \rangle \sim \tau^{-4}$.  Note that this is different from the
result in nonconformal phases described below in which the flux correlations
arise from the Goldstone mode of spontaneously broken U(1)$_m$ symmetry, and
correspond to that of a free-photon theory with $\langle \Phi_{\Box}(\tau)
\Phi_{\Box}(0) \rangle \sim \tau^{-3}$ (our numerical results for these
temporal
flux correlation functions are discussed in Sec.~IV of the SM~\cite{suppl}).

The spin correlations, i.e., the dynamic structure factor, in the DSL probe
CFT operators, which can appear in the continuum limit of individual spin
operators.  In this \emph{noncompact} theory, the spin operators (like all
operators we consider) conserve flux, and hence monopoles cannot appear in
their expansion.  Rather, the fermion bilinears $M_{a>0}$ are expected to
dominate.  One of these operators (see below in the discussion of the AFM
phase) corresponds to the N\'eel field, so that the staggered spin
correlations should show power-law behavior with a decay exponent
$2\Delta_{adj} \in (2.8,3.5)$.
Our DQMC results indeed reveal the consistent power-law behavior in spin
correlation with $2\Delta = 3.2(1)$, as shown in
Fig.~\ref{fig:fig_u1d_spectra} (b) below.
In fact, previous QMC simulation for finite sizes of $20\times 20$ have found
similar power-law decays of the spin and dimer operators, see Fig. 4 in
Ref.~\cite{xu2019monte}.

It is possible in principle to measure monopole correlations by explicitly
including a monopole-antimonopole insertion in the partition function, but we
have not done this in the current simulation as it requires a significant
technical development.  In the DSL phase, this correlation is expected to
decay with a power law of $2\Delta_M \approx 2$.

\subsubsection{CF phase}
\label{sec:CF}

In the CF phase, a spontaneous flux $\langle \nabla \times A\rangle \neq 0$
develops.  Any nonzero flux causes the Dirac cones to split into Landau
levels with energies $\epsilon_n = \omega_c \sqrt{|n|} \operatorname{sgn}(n)$, with
$n \in \mathbb{Z}$, and $\omega_c = \sqrt{2|\phi|}v$ the cyclotron energy when
the average flux is $\phi$.  The latter is determined by the condition that
the 0$^{\rm th}$ Landau levels for both valleys are full of up spin fermions
and empty of down.  This fixes the flux to density relation, $\langle \nabla
\times A\rangle = \phi = \pm \pi \langle n_\uparrow-n_\downarrow\rangle$.
Note that a finite flux breaks time-reversal symmetry $\mathcal{T}$, with a
spontaneously chosen sign.
The Zeeman field accordingly acts as an opposite Fermi level for the two spin
polarizations, lying between the 0$^{\rm th}$ and the $n=\pm 1$ Landau levels
for the up and down spins, respectively.  Hence $\epsilon_{F\alpha} = \alpha b/2 =
0.62\alpha\omega_c$ for $\alpha=\pm 1$ corresponding to up and down spins, respectively.
This relation between the Zeeman field $b$ and the spontaneous orbital flux
$\phi$, and hence $\omega_c$, follows from the mean-field analysis in
\cite{ranSpontaneous2009}.

We proceed to analyze the effect of this mean-field flux by analyzing the
fluctuations of the gauge field and the system's response.  Accordingly, we
write $A_\mu = \bar{A}_\mu + a^c_\mu$, where $ \nabla \times \bar{A} = \phi$,
and $a^c_\mu$ describes the fluctuations of the internal ``charge'' gauge
field. To probe the response, we furthermore add a \emph{probe} spin gauge
field $a^s_\mu$ which couples to the conserved U(1) \emph{spin} three-current of
the fermions. This field is fictitious but will be used as an infinitesimal
source to generate correlation functions and to characterize the spin
response.  Note that an infinitesimal change of Zeeman field $b \rightarrow b
+ \delta b$ is equivalent to including a small time component of the spin
gauge field, $a^s_0 = \ii \,\delta b$.
Observe that $a^s_0$ couples to the density of up or down spin fermions, $\alpha
= \pm 1$, with opposite signs.

The full Lagrangian including these fluctuations can therefore be written
\begin{align}
\label{eq:SLL}
    \mathcal{L} = & \frac{1}{g} (\partial_\tau a_j^c)^2 + \sum_{v=0}^1
\sum_{\alpha=\pm 1} \mathcal{L}_v^{\textrm{DLL}}\left(a^c_\mu+\frac{\alpha}{2}
a^s_\mu,\epsilon_F = \alpha \frac{b}{2}\right),
\end{align}
where $\mathcal{L}_v^{\textrm{DLL}}(a_\mu,\epsilon_F)$ is the Lagrangian for
valley $v$ of Dirac fermion Landau levels coupled to a total gauge field
$a_\mu$ and with Fermi level $\epsilon_F$.

Now we proceed to carry out an RPA treatment, integrating out the fermions to
quadratic order in the gauge fields.  Since each spin and valley of fermion is
decoupled, their contributions can be added.  Each contributes a standard
effective action for a system of fully filled and empty Landau levels,
consisting of a \emph{leading} Chern-Simons term whose coefficient is
$1/(4\pi)$ times the Hall conductivity of those fermions, and a
\emph{subleading} Maxwell term, representing the polarizability of the
fermions \cite{Fradkin2013}.
The Hall conductivity of each valley of up or down fermion is $\pm 1/2$,
capturing the change of Hall conductivity by one unit for occupying or emptying a
Landau level, and accounting for particle-hole symmetry.

Consequently, we obtain
\begin{align}
    \mathcal{L}_{\rm eff}  = & \sum_{\alpha=\pm 1} i \,\frac{\alpha}{4\pi}
\epsilon^{\mu\nu\lambda} \left(a_\mu^c+ \frac{\alpha}{2} a_\mu^s \right)
\partial_\nu \left(a_\lambda^c + \frac{\alpha}{2} a_\lambda^s \right)\nonumber
\\
    & + \frac{1}{g} (\partial_\tau a_j^c)^2 + \frac{1}{2\tilde{g}}
\sum_{\alpha = \pm 1} \left(f_{\mu\nu}^c+ \frac{\alpha}{2}
f_{\mu\nu}^s\right)^2 ,
\end{align}
where $f_{\mu\nu}^\sigma = \partial_\mu a_\nu^\sigma - \partial_\nu
a_\mu^\sigma$, and the \emph{quantized} Chern-Simons terms occupy the first
line (note that these terms appear with a factor of $i$ in our Euclidean field
theory), and the cutoff dependent and unquantized Maxwell terms the second.

Expanding and regrouping the terms, we find
\begin{align}
    \mathcal{L}_{\rm eff}  = & \,i \,\frac{1}{2\pi} \epsilon^{\mu\nu\lambda}
a_\mu^s \partial_\nu a_\lambda^c + \frac{1}{g'} (f_{\mu\nu}^c)^2 +
\frac{1}{4\tilde{g}} (f_{\mu\nu}^s)^2,
    \label{eq:mCS}
\end{align}
consisting of a \emph{mixed} Chern-Simons term for spin and charge, and
Maxwell contributions, describing renormalization of gauge field charges and
velocities that depend on the regularization scheme \cite{Zee2010}.

Eq.~\eqref{eq:mCS} is short but encodes several important conclusions, which
we now discuss.

\paragraph{Gapless mode.}

First set the probe field $a^s=0$.  Then Eq.~\eqref{eq:mCS} becomes just a
Maxwell term for the fluctuating gauge field.  This describes a massless
photon, i.e. a single branch of linearly dispersing mode $\omega = c |k|$ with
$c$ the speed of ``light''.  This can be regarded as simply arising from the
original gauge field in the lattice model, and is protected by gauge symmetry.
We discuss the alternative view as a Goldstone mode of spontaneously broken
U(1)$_m$ symmetry below.

From this effective action, one can calculate the flux-flux correlations,
$\langle \Phi_{\Box}(\tau) \Phi_{\Box}(0) \rangle \sim \tau^{-3}$.  This
behavior, different from the DSL, is characteristic of the Goldstone mode
phase.
An explicit demonstration of this conclusion on the lattice is given in
Sec.~IV of the SM~\cite{suppl}.

\paragraph{Low energy spectral weight.}

The gapless photon also appears in the correlation of $S^z$ operators.  To see
this, note that the time component of the probe field, $a_0^s$ couples to the
conserved spin density $S^z = \frac12 \bar{\Psi} \gamma^0 \sigma^z \Psi$.
Hence the derivative $\delta \mathcal{L}_{\rm eff}/\delta (\delta b)$
gives the representation of the spin operator in the effective free photon
theory.  This gives $S^z \sim \nabla \times a^c$, i.e. the low-energy spin
correlations are identical to those of the emergent magnetic flux.
Calculating the latter, one obtains
\begin{equation}
\label{eq:Sz=flux}
    \langle S^z S^z\rangle_{\omega,{\bf q}} \sim \frac12 \chi c q
\delta(\omega - c q),
\end{equation}
for small $\omega,q$, where $\chi = \frac{1}{N}\sum_i\partial   \langle
S_i^z\rangle/\partial B >0$ is the susceptibility.  Here we used the
\emph{compressible} nature of the CF phase and the fluctuation-dissipation
theorem to determine the prefactor, since $\chi$ is given by the limit $\omega
\rightarrow 0$ followed by $q\rightarrow 0$ of the longitudinal dynamical
susceptibility.
We present our DQMC results for the longitudinal susceptibility, and analyze
them in this context, in Sec.~\ref{sec:IVC}.

\paragraph{Monopole condensate.}

In the language of generalized symmetry, the gapless photon should be viewed
as a Goldstone mode of a spontaneously broken U(1)$_m$ symmetry.  However,
there are \emph{two} microscopic U(1) symmetries in the model, and only a
single Goldstone mode.  This means there must remain an unbroken U(1) subgroup
of the original U(1)$_m \times$U(1)$_f$ symmetry.  This statement is also
seemingly evident from Eq.~\eqref{eq:SLL}, which is diagonal in spin $\sigma$
so still invariant under U(1)$_f$ rotations.

To precisely identify the broken symmetry, we would like to write down the
corresponding order parameter.  Because it should transform under the U(1)$_m$
symmetry, it involves an insertion of $2\pi$ flux, which we associate with a
monopole operator $\mathcal{M}$.  The meaning of $\mathcal{M}_p(\tau)$ is that
it creates a flux $2\pi$ through plaquette $p$ at time $\tau$.

However, the order parameter is \emph{not} just the ``bare'' monopole operator
$\mathcal{M}$.  This can be seen from the mixed Chern-Simons term in
Eq.~\eqref{eq:mCS}, which implies that the charge flux $\nabla \times a^c$ is
tied to the spin density $S^z$ (conjugate to $a^s_0$).  Hence, to have a
nonzero expectation value, the order parameter must in addition to creating
the flux $2\pi$ also create the associated change in spin $\Delta S^z = 1$.
Consequently, the order parameter for the broken U(1) symmetry is
\begin{equation} \label{eq:m_ord}
    \mathcal{M}_{\mathrm{ord}} = \mathcal{M} S^+,
\end{equation}
which combines the insertion of flux with a spin flip.  From the above
arguments, we expect that
\begin{equation} \label{eq:m_ord_exp}
    \langle \mathcal{M}_{\mathrm{ord}}\rangle \neq 0.
\end{equation}
This operator, as required, breaks U(1)$_m$ symmetry but \emph{also} U(1)$_f$
symmetry.
This means that the CF state has a \emph{hidden} breaking of the U(1)$_f$
spin-rotation symmetry.
Why is it hidden?  It is because the order parameter (and hence the state
itself) preserves the combination $\mathcal{M} \rightarrow
\mathcal{M}e^{i\chi}$, $S^+ \rightarrow S^+ e^{-i\chi}$.  The latter
corresponds to the residual U(1) subgroup in the CF phase.  This ensures that
the expectation value of any pure spin operator that changes $S^z$ such as
$\langle S_i^\pm \rangle=0$ vanishes, and there is no antiferromagnetic order.
Only composite observables which involve a change of the flux can detect the
breaking of the spin rotation symmetry.  Consistent with this observation, the
gapless photon does not contribute as an intermediate state in the transverse
spin structure factor, which therefore shows a full gap in the CF phase (see
also the discussion of gapped transverse susceptibility obtained in our DQMC
results as presented in Sec.~\ref{sec:IVC}).

\subsubsection{AFM}
\label{sec:AFM}
In the AFM phase, the average flux is zero, but spontaneous AFM order has
developed.  While the AFM order arises from Eq.~\eqref{eq:cont2} by the effect
of gauge fluctuations, we can model it phenomenologically by introducing an
AFM order parameter $N^+(x,y,\tau)$, which weakly fluctuates and couples to
the Dirac fermions.  Since the system is ordered, it is sufficient to assume
$N^+ = |N| e^{i\theta(x,y,\tau)}$.  The effective Lagrangian in
Eq.~\eqref{eq:cont2} is then replaced by
\begin{align}
    \mathcal{L} =&\sum_{\mu=0}^2\bar{\Psi}  \gamma^\mu v_\mu (\partial_\mu + i
A_\mu)\Psi  - \frac{b}{2} \bar{\Psi}\gamma^0 \sigma^z\Psi  \nonumber \\
    & - |N| \left(e^{i\theta} \bar{\Psi}\sigma^-\mu^x\Psi+
\textrm{h.c.}\right)  + \frac{1}{g} \sum_{\mu=x,y}(\partial_\tau A_\mu)^2.
    \label{eq:cont3}
\end{align}
Now we make the unitary transformation $\Psi \rightarrow e^{i\theta
\sigma^z/2}\Psi$, $\bar{\Psi} \rightarrow \bar{\Psi} e^{-i\theta \sigma^z/2}$,
which is chosen to remove the dependence of the action on a constant phase
$\theta$.  When this phase depends upon space and time, the result is
\begin{align}
    \mathcal{L} & \rightarrow \sum_{\mu=0}^2\bar{\Psi}  \gamma^\mu v_\mu
(\partial_\mu + i A_\mu+\frac{i}{2}\partial_\mu\theta \sigma^z)\Psi  -
\frac{b}{2} \bar{\Psi}\gamma^0 \sigma^z\Psi  \nonumber \\
    & - |N|  \bar{\Psi}\sigma^x\mu^x\Psi  + \frac{1}{g}
\sum_{\mu=x,y}(\partial_\tau A_\mu)^2.
    \label{eq:cont4}
\end{align}
Now we can again perform the RPA, integrating out the fermions to quadratic
order in $A_\mu$ and $\partial_\mu \theta$.  An explicit calculation is
involved due to the need to regularize the Dirac theory, which must be done
with care to maintain charge conservation and gauge invariance \cite{Zee2010}.
Fortunately, the result can be understood on the grounds of symmetries.  The
general form of the RPA effective action is
\begin{align}
    S_{\rm eff} & = \frac12 \int\! d^3x d^3x' \, \nonumber \\
& \quad \times \left(A_\mu(x') \;
\partial_\mu \theta(x')\right) \mathsf{\Pi}_{\mu\nu}(x-x') \begin{pmatrix}
A_\nu(x) \\ \partial_\nu\theta(x) \end{pmatrix},
\end{align}
where here $x,x'$ are space-time coordinates, and $\mathsf{\Pi}_{\mu\nu}$ is a
generalized polarization tensor.

Owing to the presence of the gap in the fermion spectrum, the Fourier
transform of the polarization must be analytic in frequency and momentum at
scales below the gap, allowing a Taylor expansion.  Furthermore, charge
conservation (which enforces the continuity equation $\partial_\mu j_\mu=0$ at
the operator level) requires that $\partial_\mu \mathsf{\Pi}^{AA}_{\mu\nu}
=\partial_\nu \mathsf{\Pi}^{AA}_{\mu\nu} = 0$, and $\partial_\mu
\mathsf{\Pi}^{A\theta}_{\mu\nu} =\partial_\nu \mathsf{\Pi}^{\theta A}_{\mu\nu}
= 0$, where the superscripts $A,\theta$ indicate the blocks within the
polarization tensor.  Assuming, as before, the temporal gauge $A_0=0$, writing
the most general allowed form of the polarization tensor at lowest order in
three-momenta, and Fourier transforming back to space-time, we obtain
\begin{align}
\label{eq:Atheta}
    \mathcal{L}_{\rm eff} & = \frac14 c_1 F_{0j}^2 + \frac14 c_2 F_{ij}^2 +
\frac12 c_3 (\partial_0 \theta)^2 + \frac12 c_4 (\partial_j \theta)^2
\nonumber\\
    & + i c_5 (\partial_i A_i) (\partial_0 \theta),
\end{align}
where $F_{\mu\nu} = \partial_\mu A_\nu-\partial_\nu A_\mu$ is the
field-strength tensor, $i,j$ indicate spatial indices and $0$ is a time index.
The constants $c_{1-4}>0$ for stability. The factor of $i$ in the last term is
required by Hermiticity. The Lagrangian \eqref{eq:Atheta} is invariant under
the residual gauge transformations $A_j \to A_j + \partial_j f$ with
$\tau$-independent $f$. Finite frequency excitations of \eqref{eq:Atheta} are
represented by the standard transverse mode with speed $v_T=\sqrt{2c_2/c_1}$ and a
longitudinal sound mode which mixes $\theta$ and the longitudinal part of $A$,
with speed $v_L=\sqrt{(c_4+2c_5^2/c_1)/c_3}$.

To illustrate the logic leading to the Lagrangian in \eqref{eq:Atheta}, we
comment on the last, cross term. Analyticity of
$\mathsf{\Pi}^{A\theta}_{\mu\nu}$ and charge conservation require that in the
momentum space $\mathsf{\Pi}^{A\theta}_{\mu\nu}(q) \to
\mathsf{\Pi}^{(1)}_{\mu\nu\lambda} q_\lambda$, where the coefficient of $q$ is
a tensor of constants antisymmetric in indices $\mu,\lambda$:
$\mathsf{\Pi}^{(1)}_{\mu\nu\lambda} = \epsilon_{\mu\lambda\kappa} c_{\nu
\kappa}$ and $c_{\nu \kappa}$ are some constants.
In space-time, this leads to $c_{\nu\kappa} \epsilon_{\mu\lambda\kappa}
\partial_\lambda A_\mu \partial_\nu \theta$. For $c_{\nu\kappa} =
\delta_{\nu\kappa}$, this term vanishes.
Na\"ively, spatial rotation symmetry and time-reversal symmetry require
$c_{01}=c_{02}=0$, $c_{12}=-c_{21}$ and $c_{11}=c_{22}$.  With this form, the
$c_{11}=c_{22}$ contribution is a pure boundary term.
Then, there remains $c_{00}\neq 0$ term, and the above general form reduces to
$F_{12}\partial_0 \theta$ and $F_{i0} \partial_i \theta$ and terms equivalent
to these under integration by parts.
We next observe that Eq.~\eqref{eq:cont4} is symmetric under paritylike
transformation $x\to -x, A_x \to -A_x, \Psi \to \gamma^x \mu^z \Psi$ and
$\bar{\Psi} \to - \bar{\Psi} \gamma^x \mu^z$, which, however, changes the sign
of the flux $F_{12} \to - F_{12}$ and rules out the first option but preserves
the second one, $\propto \partial_i A_i \partial_0 \theta$.

The Lagrangian in Eq.~\eqref{eq:Atheta} is a coupled
Maxwell theory for $A_\mu$ and a free massless scalar field theory for
$\theta$, which describes two massless linearly dispersing modes.  The
presence of two such modes indicates that both the U(1)$_m$ \emph{and}
U(1)$_f$ symmetries are spontaneously broken in the AFM phase.  The gapless
$\theta$ mode will appear as a spin-wave-like mode in the \emph{transverse}
spin structure factor, with high intensity as it approaches the Bragg peak
associated with the AFM order (we show our DQMC results for the transverse
magnetic spectra in Sec.~\ref{sec:IVC}).

Because of the presence of the gapless photon, we also expect power-law
correlations
of the flux in the AFM phase, of the same form as in the CF phase (but
distinct
from those in the DSL). We refer the reader to our DQMC results for temporal
flux correlation functions in Sec.~IV of the SM~\cite{suppl}.  These power-law flux correlations, together with the gapless spin-wave branch in the structure factor, are signatures of the two broken symmetries of the AFM state.

\section{Phase diagram}
\label{sec:III}
Our numerically obtained phase diagram is presented in Fig.~\ref{fig:fig1}
(b). At zero magnetic field, there exists a stable Dirac
spin liquid phase, which is a nontrivial critical phase in which a
configuration with $\pi$-flux of the gauge field on each
plaquette is dynamically selected.  Such a small-$J$ DSL phase has also been
reported in the compact case
in Ref.~\cite{xu2019monte}.  There are theoretical arguments which indicate
the DSL is unstable in the compact case (see Sec.~\ref{sec:conclusion}). We
conclude there are strong finite-size effects which affect the compact model,
since for a $20\times 20$ lattice, the expected power-law correlations of the
fermion bilinears in spin operators with decay exponent of $2\Delta \sim 3$
(see Sec.~\ref{sec:IIC1}) inside DSL phase have been observed.
Similarly, the finite extent of the DSL phase for $B>0$ in Fig.~\ref{fig:fig1}
(b) is due to the nonzero temperature accessible in our simulations. We
expect the DSL to give way to the CF phase for infinitesimal Zeeman fields in
the zero-temperature thermodynamic limit.

At finite $B$,
the flux per plaquette starts to deviate from $\pi$ and the system enters a
chiral flux phase.   In this phase the N\'eel order parameter $N^+$
remains zero, but the flux per
plaquette deviates from $\pi$ (and $0$).  Two possible distinct states -- with
flux in $(0,\pi)$ or in $(\pi,2\pi)$ -- are possible, one of which is
spontaneously chosen, resulting in spontaneous breaking of the time-reversal
symmetry $\mathcal{T}$.
This is characterized by the chiral order parameter $\langle \sin
\Phi_\Box\rangle$. We stress that $\mathcal{T}$ is broken due to the nontrivial
gauge flux in the system: in contrast, a finite Zeeman field or spin polarization
at vanishing flux preserves $\mathcal{T}$.
As discussed in Sec.~\ref{sec:CF}, there is also a more subtle breaking of
U(1)$_m \times$U(1)$_f$ symmetry, which is reflected in a gapless photon mode
and the nonzero susceptibility of the CF phase.

As will be further discussed
in Sec.~\ref{sec:IIIB}, one can define a scale-invariant correlation ratio
$U_{\text{flux}}$ of the chiral order and at finite $B$ and small $J$, one
sees that $U_{\text{flux}}$ is large while the correlation ratio of the
antiferromagnetic phase $r_{\text{AFM}}$ is small (meaning no antiferromagnetic long-range order).
While at finite $B$ and large $J$, the trend is opposite, suggesting the
vanishing of the CF order and the establishment of the U(1)$_f$ symmetry
breaking in the form of an antiferromagnetic long-range order in the $x-y$
components (AFM).
To clearly discuss these results, we first provide the mean-field analysis of
the phase diagram at $J=0$ and then discuss the DQMC results at finite $J$.

\begin{figure}[tbp]
\includegraphics[width=\columnwidth]{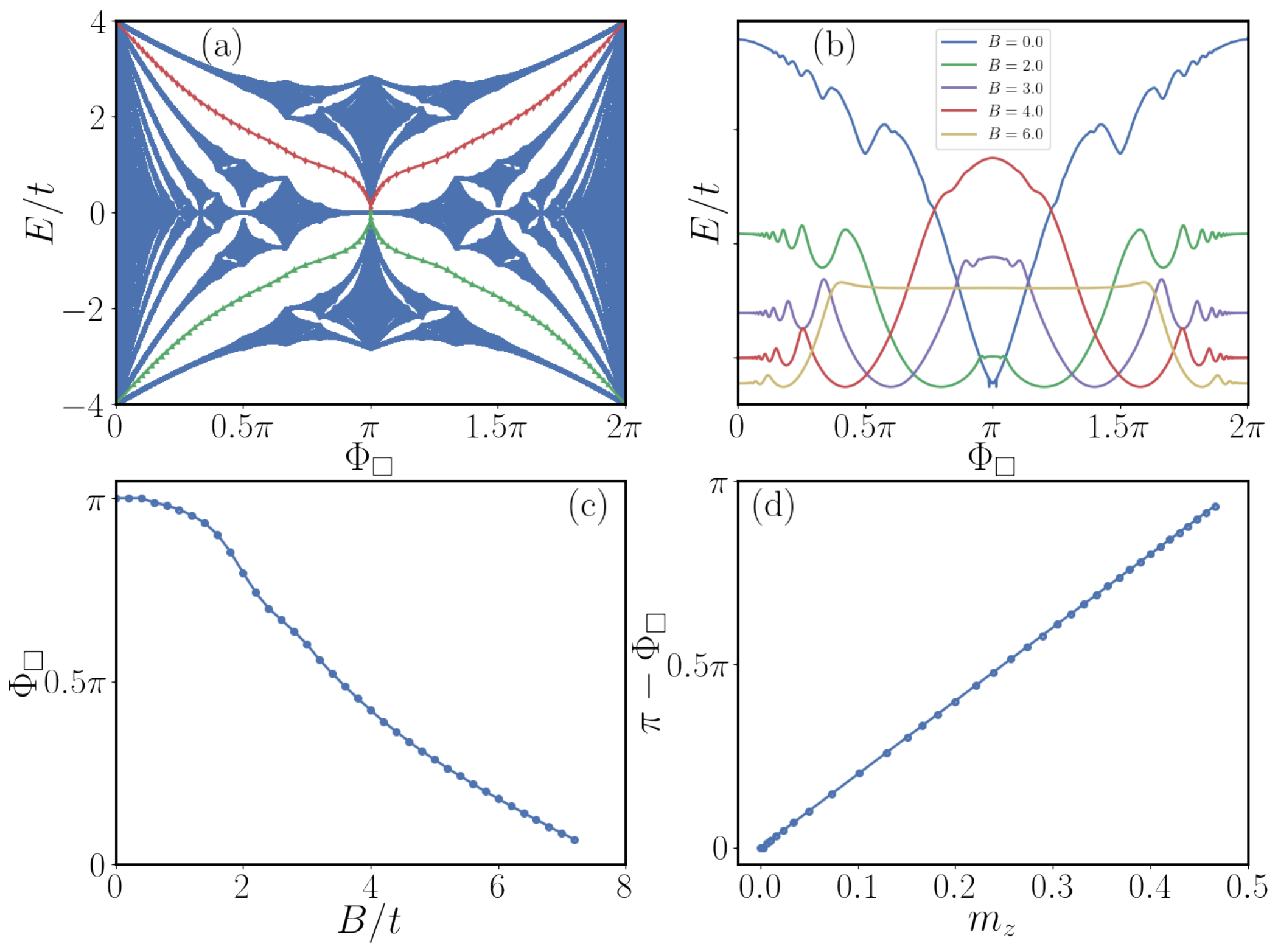}
\caption{\textbf{Lattice mean-field calculations at $L=32$.} (a) Eigenvalues
of
  the fermion hopping matrix at different flux sector from $0$ to $2\pi$,
known
  as Hofstadter's butterfly~\cite{hofstadter1976}. The red(green) line shows
the location of Fermi level for up(down) fermion. (b) Total energy of system
versus flux $\Phi_\Box$ for different values of Zeeman field $B$. With finite
Zeeman field, the flux sectors with the minimal energy deviate from $\pi$ and
are symmetric about $\pi$. (c) The flux sector with the minimal total energy
versus Zeeman field. The data points are obtained from (b). (d)
$\pi-\Phi_\Box$ versus magnetization $m_z$, shows linear relation and
indicates the induced orbital magnetic field is proportional to
magnetization.}
\label{fig:fig2}
\end{figure}

\subsection{Mean-field analysis at \texorpdfstring{$J=0$}{J=0}}
\label{sec:IIIA}
The deviation from $\pi$-flux at nonzero Zeeman field can be demonstrated by
a
mean field analysis of the lattice model, which forms a basis for
understanding the DQMC results. The mean-field approximation consists of
neglecting the dynamics of the gauge field, i.e. assuming $a_{ij}(\tau_n)$ is
independent of $\tau_n$.  Inspecting the final term in Eq.~\eqref{eq:action},
one sees that this approximation becomes exact in the limit $J \rightarrow 0$,
as configurations of the gauge field with any time dependence have infinite
action in this limit.
The fermionic path integral for a particular time-independent flux
configuration gives, then, simply the free fermion partition function with
this flux.  At zero temperature ($\beta \rightarrow\infty$), this is
$\exp(-\beta E_0(\Phi_\Box))$, where $E_0(\Phi_\Box)$ is the fermionic ground
state energy with the given flux, and hence the flux becomes (in the
thermodynamic and zero temperature limits) constrained to take the value which
minimizes the energy. To find the latter, we manually set the flux of each
plaquette in Landau gauge
and calculate the energies as a sum of the fermion kinetic energy and the
Zeeman energy. The
input parameters of the mean-field calculation are the Zeeman field $B$, which
acts
like an opposite chemical potential for fermion with opposite spin, and the
flux sector
$\Phi_\Box$, which in turn determines the gauge field within Landau gauge,
which we choose.   We can then construct and
diagonalize the fermion hopping matrix, whose eigenvalues, the single-particle
energies, are shown in
Fig.~\ref{fig:fig2}(a) for different flux sectors.   The Zeeman field $B$
controls the filling of the fermions, from which we obtain the magnetization
$m_z=\frac{1}{N}\sum_i^{N}\frac{1}{2}(n_{i,\uparrow}-n_{i,\downarrow})$. We
then can obtain the total energy of the system as a function of the specific
flux sector
$\Phi_\Box$ and Zeeman field $B$, from which we extract the flux sector with
minimal energy.  We observe a significant deviation from $\pi$-flux for
nonzero Zeeman
field. The conclusion is consistent with the mean-field analysis in
Ref.~\cite{ranSpontaneous2009} for a Kagom\'e lattice model.

The flux sector corresponding to the minimum energy versus Zeeman field is
shown in
Fig.~\ref{fig:fig2}(c). At zero Zeeman field, the system favors the
$\pi$-flux gauge field arrangement. However, for nonzero Zeeman field, the
favorable flux sectors deviate from $\pi$-flux and are symmetric about
$\pi$-flux from Fig.~\ref{fig:fig2}(b) data, indicating broken
$\mathcal{T}$
symmetry. By interpreting $\pm B/2$ as an opposite chemical
potential for up and down fermions, the $\Phi_\Box$ vs $B/t$ relation gives
the spin-dependent Fermi levels, shown in
Fig.~\ref{fig:fig2}(a) by red and green lines.  In this plot, spin up and down
fermions occupy states below the red and green levels, respectively. In
Fig.~\ref{fig:fig2}(d), we show $\pi-\Phi_\Box$ vs
magnetization $m_z$, which demonstrates a linear relation.  The constant slope
indicates that the bands between the two Fermi levels have a fixed total
Chern number of 1 for the full range of fields.

\subsection{DQMC simulations with \texorpdfstring{$J\neq 0$}{J not equal to 0}}
\label{sec:IIIB}
\begin{figure}[tbp]
\includegraphics[width=\columnwidth]{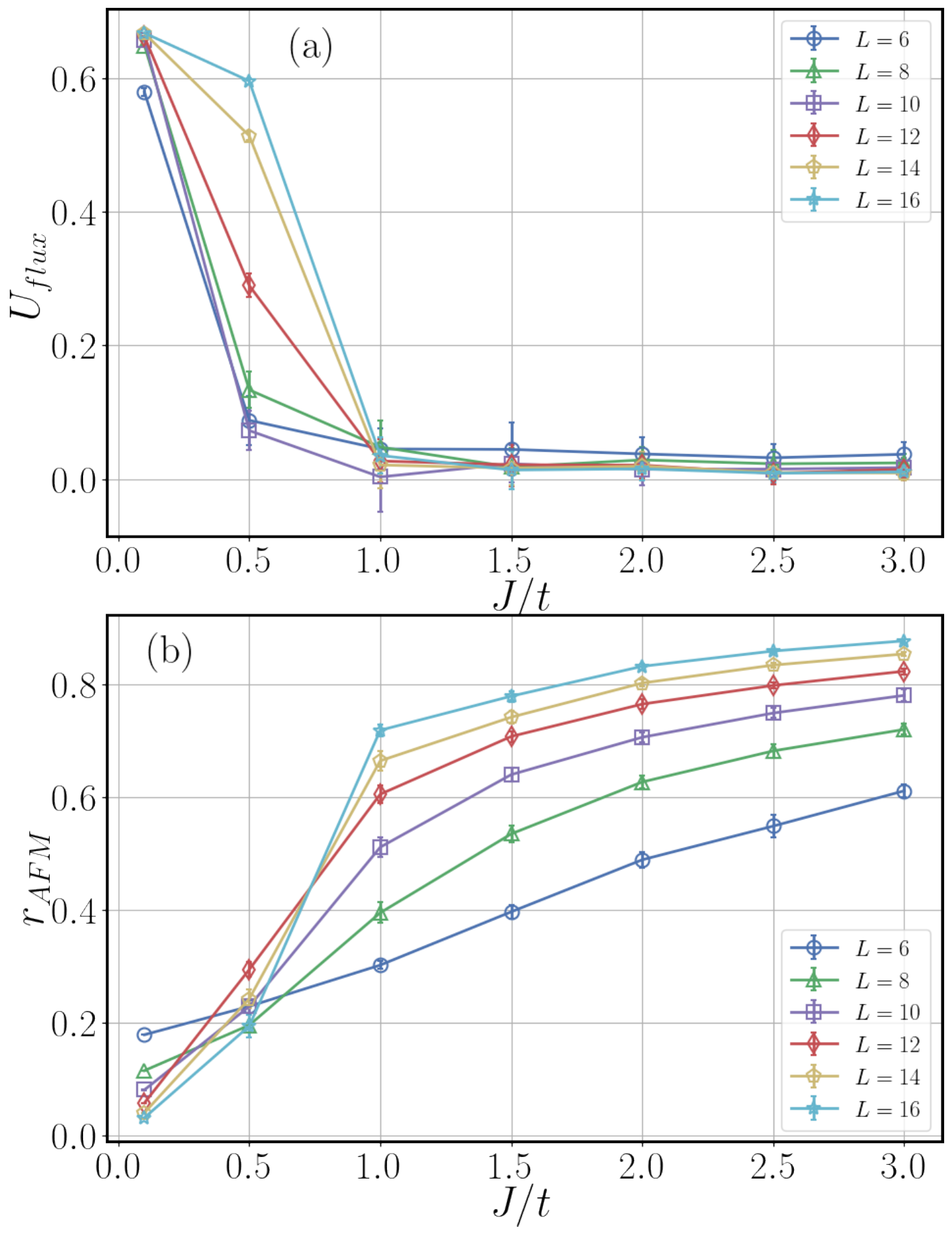}
\caption{\textbf{Determination of the phase boundary in Fig.~\ref{fig:fig1}
(b) at finite $B$.} Data are at Zeeman field $B/t=2$ with system sizes
$L=6,8,\cdots,16$.  (a) Critical $J/t \sim 0.9$ determined from flux Binder
cumulant $U_{\text{flux}}$. (b) Critical $J/t \sim 0.6$ determined from
transverse spin correlation ratio $r_{\text{AFM}}$. The phase boundaries of
the CF and AFM phases in Fig.~\ref{fig:fig1} (b), at $B/t=1,2,3, \cdots, 5$,
are determined in this way.}
        \label{fig:fig3}
\end{figure}

\begin{figure}[tbp]
\includegraphics[width=\columnwidth]{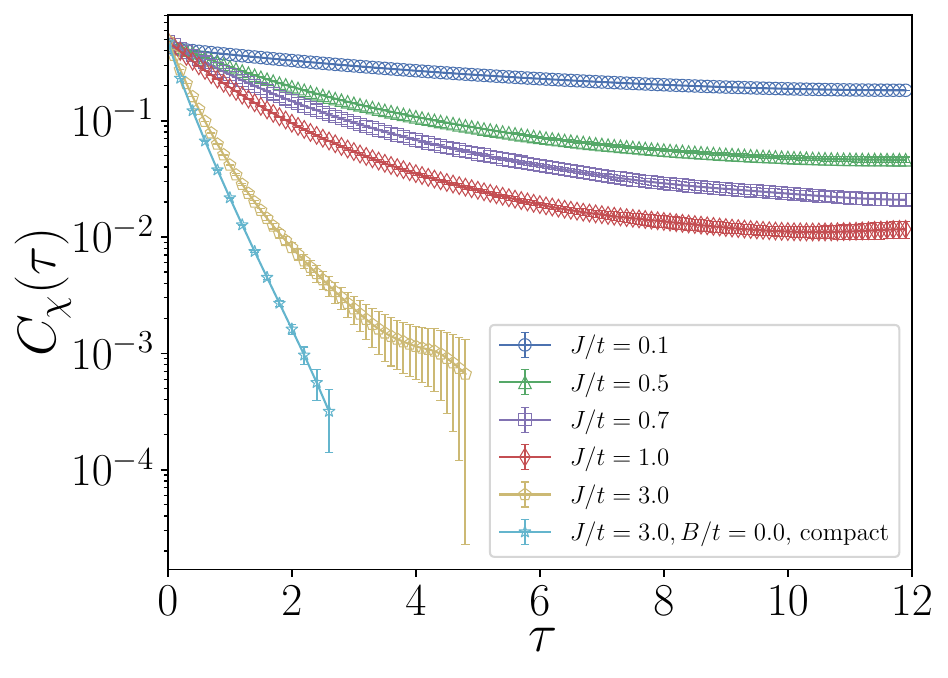}
\caption{\textbf{Dynamical flux correlation function $C_{\chi}(\tau)$.}
For system size $L=12$ and inverse temperature $\beta=24$, all values of $J/t$
at
Zeeman field $B/t=3$ exhibit clear deviations from exponential decay. This
behavior stands in contrast to the compact case at $J/t=3, B/t=0$, where the
correlation function shows well-defined exponential decay, characteristic of a
confined phase~\cite{xu2019monte}. In the mixed case with large $J/t=3$ and
$B/t=3$, the correlators display persistent curvature in the semilogarithmic
plot, indicating a breakdown of exponential scaling and suggesting deconfined
gauge field even inside AFM phase.}
\label{fig:fig4}
\end{figure}

With $J\neq 0$, the fluctuations of the gauge field render the problem in
Eq.~\eqref{eq:action} strongly correlated and one has to rely on the DQMC
results. As mentioned in Sec.~\ref{sec:IIA}, in our DQMC lattice simulations,
we denote the system size of the
cubic lattice in Fig.~\ref{fig:fig1}(a) by $L$ and $\beta$. We scan the phase
diagram in the unit of $B/t$ and $J/t$ and scale the
inverse temperature $\beta=\frac{1}{T}=2L$ so that we approach zero
temperature in the large system limit (in our setting $\Delta\tau=0.1$ and the
integer index in the temporal direction $N_{\tau}=\beta/\Delta\tau=20L$).

To analyze the symmetry breaking patterns, we consider both chiral flux and
antiferromagnetic orders.
We first verify that the chiral flux (CF) order is uniform and that the
antiferromagnetism is the standard two-sublattice staggered type by
considering their corresponding equal time structure factors.  For the chiral
order, this is
\begin{equation}
S_{\chi}(\mathbf{q})=\frac{1}{N_\Box}\sum_{i,j}\langle
\sin(\Phi_{\Box,i})\sin(\Phi_{\Box,j})\rangle e^{-i\mathbf{q}\cdot
\mathbf{r}_{ij}},
\end{equation}
where $N_\Box=L^2$ is the number of plaquettes. $i,j$ are the position of
plaquettes labeled by their lower left-hand site. The maximum of $S_{\chi}$
fixes the ordering wave vector, which is at $\mathbf{q}=\Gamma$.  For the AFM
order, the corresponding (transverse) equal time spin structure factor is
\begin{equation}
S^{\pm}(\mathbf{q})=\frac{1}{N}\sum_{ij} \langle S^{+}_iS^{-}_j+h.c. \rangle \
e^{-i\mathbf{q}\cdot \mathbf{r}_{ij}}
\label{eq:eq6}
\end{equation}
where $S^{+}_i=c^{\dagger}_{i,\uparrow}c_{i,\downarrow}$,
$S^{-}_i=c^{\dagger}_{i,\downarrow}c_{i,\uparrow}$ and $N=L^2$.  The maximum
structure factor again fixes the ordering wave vector, which is at the
$\bm{M}=(\pi,\pi)$ point for the AFM and the AFM order parameter discussed in
Sec.~\ref{sec:AFM}, $|N^+| = \lim_{N\to\infty}\sqrt{S^{\pm}(\bm{M})/(2N)}$.

With the ordering wave vectors determined, we proceed to find the phase
boundaries in the $J$-$B$ plane by choosing appropriate measures of the two
order parameters.   An example of data and analysis for a cut at fixed $B/t=2$
is shown in Fig.~\ref{fig:fig3}.  For the chiral order (panel (a)), we measure
the Binder cumulant of the sine of the flux on a plaquette.  We define
$f_{\chi}=\frac{1}{N_\Box}\sum_{\Box}\sin(\Phi_{\Box})$ and then the Binder
cumulant $U_{\text{flux}}$ is
\begin{equation}
  U_{\text{flux}}=1-\frac{\langle f_{\chi}^{4} \rangle}{3\langle f_{\chi}^{2}
\rangle^{2}}.
\end{equation}
Inside the CF phase, $U_{\text{flux}} \to \frac{2}{3}$ and once the
$\mathcal{T}$ symmetry is recovered, for example, by increasing $J$ with fixed
$B$, $U_{\text{flux}} \to 0$.  Hence a crossing is expected in the
thermodynamic limit within a plot of $U_{\text{flux}}$ along a line crossing
the boundary between a phase with chiral order (broken $\mathcal{T}$) and one
without.

For the antiferromagnetic order, we use dimensionless
correlation ratio
\begin{equation}
  r_{\text{AFM}} = 1-\frac{S^{\pm}(\mathbf{q}+\delta
\mathbf{q})}{S^{\pm}(\mathbf{q})}
\end{equation}
in which $\mathbf{q}=M$ is the ordering wave vector and $\mathbf{q}+\delta
\mathbf{q}$ is the closest adjacent wave vector in momentum space. In the
disordered phase, $r_{\text{AFM}} \to 0$ as the structure is flat in the
momentum space, and in the ordered phase $r_{\text{AFM}} \to 1$, once the
Bragg peak at $\mathbf{q}$ is fully developed in the thermodynamic limit.
Like for the Binder ratio of CF order, we thus expect a crossing of
$r_{\text{AFM}}$ on cuts crossing the phase boundary between AFM ordered and
disordered regions, for large system sizes, as shown in Fig.~\ref{fig:fig3}
(b) for $B/t=2$ case.

The results of this analysis using multiple cuts are summarized in the phase
diagram in
Fig.~\ref{fig:fig1}(b).  We observe substantial regions with CF order, AFM
order, and a small overlapping region in which both CF and AFM order coexist.
We will present and discuss the magnetic spectra in CF and AFM phases in
Sec.~\ref{sec:IV}.

Before doing so, we check the DQMC simulations for evidence of the gauge field
dynamics.  To this end, we consider the dynamical flux correlation function
\begin{equation}
    C_{\chi}(\tau) = \frac{1}{N_\Box}\sum_\Box\langle \sin(\Phi_{\Box}(\tau))
\sin(\Phi_{\Box}(0))  \rangle.
\end{equation}
Fig.~\ref{fig:fig4} shows the imaginary time decay of the $C_{\chi}(\tau)$ at
$B/t=3$ as a function of $J$.  For small $J/t<1$ (inside CF phase) the decay
saturates to a constant at ``large'' $\tau$, indicating the presence of
spontaneous chiral order.  At large $J/t>1$ (inside AFM phase), it decays to a
negligible value.  However, even well in the AFM phase, e.g. $J/t=3$, the form
of the decay is clearly subexponential.  This indicates power-law
correlations consistent with the expected gapless photon mode (see
Sec.~\ref{sec:AFM}).
In fact, as shown in Sec.~IV of the SM~\cite{suppl}, our DQMC simulations find the flux
correlation functions inside the AFM phase as $ C_{\chi}(\tau) \sim \tau^{-3}$
and for the DSL phase as $C_{\chi}(\tau) \sim \tau^{-4}$, consistent with
theoretical expectations.
For comparison, we also plot the same flux correlator for the \emph{compact}
model~\cite{xu2019monte} for $J/t=3,B/t=0$ (inside the confined AFM phase).
One can clearly see exponential decay in that case, which is due to
confinement physics that is absent in the mixed model. While it is difficult
to separate the power-law decay from the saturation within the CF phase, a
power-law approach to the large-time saturation is also expected in these
cases as well.
We conclude that through the phase diagram, the gauge field is ``deconfined''
in the mixed model.

\section{Magnetic spectra}
\label{sec:IV}

A key implication of the generation of a finite flux of the emergent gauge
field is that the Dirac fermions experience an orbital magnetic field and form
Landau levels~\cite{ranProjected2007}. While the single-spinon spectrum is not
gauge-invariant, one may inspect how the formation of such Landau levels
impacts the dynamic spin structure factor, where we consider both the
transverse component $S^\pm(\omega,\bm{q}) = \frac{1}{N}\sum_{ij} \int \di t
\langle S^+_i(t) S^-_j(0) + S^-_i(t) S^+_j(0) \rangle \ei^{\ii (\omega t -
\bm{q} \cdot \bm{r}_{ij})}$ as well as the longitudinal component
$S^{zz}(\omega, \bm{q})$.
Formally, in the zero-temperature limit, the dynamic structure factor can be obtained via the
fluctuation-dissipation theorem, $S^a(\omega,\bm{q}) = -2 \Theta(\omega) \operatorname{Im}
\chi^a(\ii \omega \to \omega+\ii 0^+,\bm{q})$ from the dynamical susceptibility
(i.e.~response) $\chi^a(\ii \omega,\bm{q})$ to an external probe field which
couples to the spin density at momentum $\bm{q}$ (see also the discussion in
Sec.~\ref{sec:rpa}).
Since the fermionic spinons carry $S=1/2$ quantum numbers, this susceptibility
can be obtained from dynamical response of the fermionic degrees of freedom.
Below, we first infer key qualitative features of the dynamic spin structure
factor in a mean-field approximation (both in the continuum limit, where an
analytical treatment is feasible and on the square lattice geometry), and then
present magnetic spectra in the CF phase upon including gauge fluctuations
obtained from our DQMC simulations.

\subsection{Continuum field theory analysis}
\label{sec:IVA}
A continuum field theory allows for insights into the dynamic structure factor
near the high-symmetry points $\bm{\Gamma},\bm{M}$ and $\bm{X}$-points.
At these points, in the zero-field limit, the magnetic spectrum is dominated
by particle-hole excitations on top of the single-particle Dirac cones.
Explicitly, these contributions are determined by fermionic bubble diagrams of
the form
\begin{equation} \label{eq:chi}
    \chi_{\bm{Q}}^\pm(\ii \omega, \bm{k}) = \begin{tikzpicture}[baseline=(b)]
    \begin{feynman}[inline=(b)]
    \diagram*{b[empty dot] -- [fermion, half left,
looseness=1.5,out=45,in=135, edge label=\(\uparrow\)] c[empty dot] --
[fermion, half left, looseness=1.5,out=45,in=135, edge label=\(\downarrow\)]
b};
\end{feynman}\end{tikzpicture}
 + \dots,
\end{equation}
and similarly for the longitudinal component. 
Here, $\bm{k}$ refers to small momenta close to the high-symmetry points
$\bm{Q} = \bm{\Gamma},\bm{X}$ and $\bm{M}$.
The choice of lattice momentum $\bm{Q}$ is encoded in the vertices that enter
the bubble diagram -- these are determined by starting from a microscopic
lattice model and making a gradient expansion to read off how microscopic
lattice translation or rotation symmetries act on the sublattice and valley
indices
of the Dirac fermions (we refer the reader to Sec.~I of the SM~\cite{suppl} for a lattice
expression of
this diagram, and Sec.~II of the SM~\cite{suppl} for technical details concerning our
continuum field-theory calculations).
Further, in Eq.~\eqref{eq:chi}, the dots correspond to corrections
arising from interactions with the emergent U(1) gauge field.

We henceforth work on a ``bare'' level and neglect such interactions with the
U(1) gauge field, corresponding to a mean-field approximation. For an
effective RPA-level treatment of the fluctuating gauge field, see
Sec.~\ref{sec:rpa}.

\begin{figure}
\centering
\includegraphics[width=\linewidth]{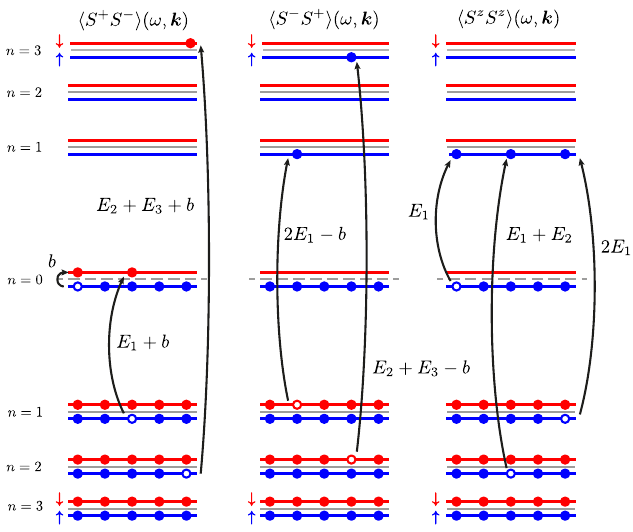}
\caption{\textbf{Illustration of excitations that contribute to the transverse
($\langle S^+ S^- \rangle$ and $\langle S^- S^+ \rangle$) and longitudinal
($S^z S^z$) channels of magnetic spectra.} The finite flux of the gauge field
leads to Landau levels of the Dirac fermions, with energies $\pm E_n - \alpha
b/2 = \pm \sqrt{2n}/\ell - \alpha b/2$, where the finite Zeeman fields lead to
a spin splitting for $\alpha = \uparrow (+1), \downarrow(-1)$ spinons.
For the former, spin-flip excitations can occur either within the spin-split
$n=0$ Landau level, or between distinct sets of spin-split Landau levels,
while the longitudinal channel
only receives contributions from same-spin inter-Landau-level transitions.}
    \label{fig:illu_exc_longi_transv}
\end{figure}

We first give a physical picture for possible excitations that can give rise
to
poles in the structure factor: noting the low-energy expressions of the spin
operators in terms of spinon fields listed in Eqs. (S11a)--(S11c) of Sec.~II of the SM~\cite{suppl}, applying the spin-lowering operator
$S^- \sim \psi^\dagger_\downarrow \psi_\uparrow$ on top of the ground state
with spin-split Landau levels can be seen to create a hole in a filled
spin-$\uparrow$ Landau level and create a particle in an otherwise empty
spin-$\downarrow$ Landau level. This is illustrated in
Fig.~\ref{fig:illu_exc_longi_transv}.
Transitions belonging to the same (spin-split) Landau level are in general not
allowed (as they are fully occupied or empty for $|n|>0$), \emph{except} in the
0-th Landau level, of which only the spin-$\uparrow$ copy is filled (for
$b>0$).
This excitation corresponds to a pole at frequency $\omega = b$ in the
transverse contribution to the structure factor.
Conversely, now consider the action $S^+ \sim \psi^\dagger_\uparrow
\psi_\downarrow$ on top of the ground state with $b \geq 0$.
Creating a spin-$\downarrow$ hole and a spin-$\uparrow$ particle is 
allowed only if they respectively occur in Landau levels $|n| \geq 1$, such that
these particle-hole excitations can occur only with energies $\omega = E_n+E_m
-b$ with $n,m \geq 1$.

\begin{figure}[tb]
    \centering
    \includegraphics[width=0.9\linewidth]{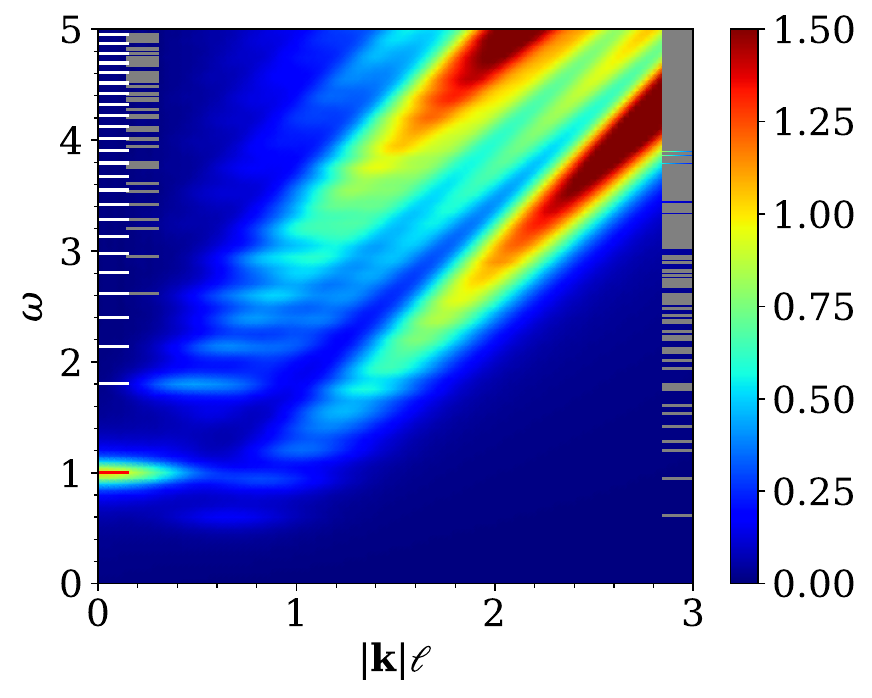}
    \caption{\textbf{Transverse structure factor computed in continuum field
theory.} For $b=1$, with flux $\Phi = b^2/(2\sqrt{2} \times 0.62)^2 \approx
0.33$ (using $b/2 = 0.62 \omega_c $). The white horizontal lines on the left-hand
axis denote poles with frequency $\omega = E_n + b$, the gray horizontal lines
on the left- and right-hand axes denote poles with $\omega = E_n + E_m +b$ and
$\omega = E_n + E_m -b$, respectively. The red line on the left axis indicates
the Larmor pole with frequency $\omega = b$. }
    \label{fig:transverse-field-theory}
\end{figure}

In contrast to the transverse response, the longitudinal susceptibility, and
thus the structure factor, will involve excitations that preserve the
magnetization.
Hence, only inter-Landau-level transition are allowed. These can in general
give rise to poles at frequencies $\omega = E_n + E_m$, where $n,m \geq 0$
(but not $n= m = 0$).

With a qualitative understanding of excitations (in the mean-field limit) that
can give rise to poles in the transverse and longitudinal structure factors,
we now turn to their respective spectral weights at the $\bm{\Gamma},\bm{M}$
and $\bm{X}$ points, where the spinon particle-hole excitations dominate the
low-energy response.
We obtain the spectral weights via an explicit evaluation of
Eq.~\eqref{eq:chi},
with technical details provided in Sec.~II of the SM~\cite{suppl}.

We note the following key predictions from our field-theoretical analysis:
\begin{enumerate}
    \item At the $\bm{\Gamma}$-point, only the pole of frequency $\omega = b$
contributes finite spectral weight to the \emph{transverse} structure factor
(arising from spin-flip particle-hole excitations in the spin-split 0th-Landau
level). This is in accordance with Larmor's theorem: since the ground state (at
zero field $b=0$) is SU(2)-symmetric, the only dynamic response to the
SU(2)-breaking Zeeman field $b$ consists in a transverse spin wave with
frequency $b$, with the spectral weight determined by the system's
magnetization
$m^z$~\cite{balentsCollective2020,agarwalCollective2024,
chenUniversal2024},
$\langle S^+ S^- + S^- S^+ \rangle(\omega,\bm{\Gamma}) = 4\pi m^z
\delta(\omega-b)$. We further note that this mode of frequency $b$ also has
finite spectral weight in the transverse structure factor at $\bm{M}$ (in
addition to many other modes), while it is absent in $S^\pm$ at the
$\bm{X}$-point.
    \item Turning to the longitudinal structure factor $S^{zz}$, poles occur
at energies proportional to the cyclotron frequency (which is \emph{a priori}
independent of the Zeeman-energy). We find that $S^{zz}(\bm{\Gamma},\omega) =
0$, consistent with Larmor's theorem (see above). We further observe that all
low-energy poles at the $\bm{X}$-point (i.e. $\omega = E_1, E_1 +E_2,\dots$)
generally carry finite weight, while poles with finite weight at the
$\bm{M}$-point occur at much higher frequencies, i.e. $\omega = 2 E_1 , 2E_2
\dots$.
    \item One may further find the structure factor at finite (small) momenta
      $\bm{q}$ relative to these high-symmetry points. We have explicitly
      obtained $S^{\pm}_{\bm{\Gamma}}(\bm{q},\omega)$, shown in
      Fig.~\ref{fig:transverse-field-theory}, see also Sec.~II of the SM~\cite{suppl} for
details: Going away from $\bm{\Gamma}$, other poles (in addition to the Larmor
mode) generally acquire finite spectral weight. Here, $\langle S^+ S^-
\rangle$ contains poles with frequencies $\omega \geq b$, while $\langle S^-
S^+ \rangle$ may contain poles with energies below the Larmor mode.
    The maximum spectral weight at a given frequency follows a \emph{linear}
dispersion $\nu_\mathrm{max.}(\bm{k}) = |\bm{k}| + b$ -- this is consistent
with continuity arguments wherein going toward the zero-field limit the
Landau levels become dense, and linearly dispersing cones emerge in the
structure factor (resulting from particle-hole excitations on the spinon Dirac
cones).
\end{enumerate}

Below, we compare these insights based on a low-energy theory (assuming a
static gauge field) with our numerical results.

\subsection{Lattice mean-field calculations}
\label{sec:IVB}

\begin{figure}[tbp]
\includegraphics[width=\columnwidth]{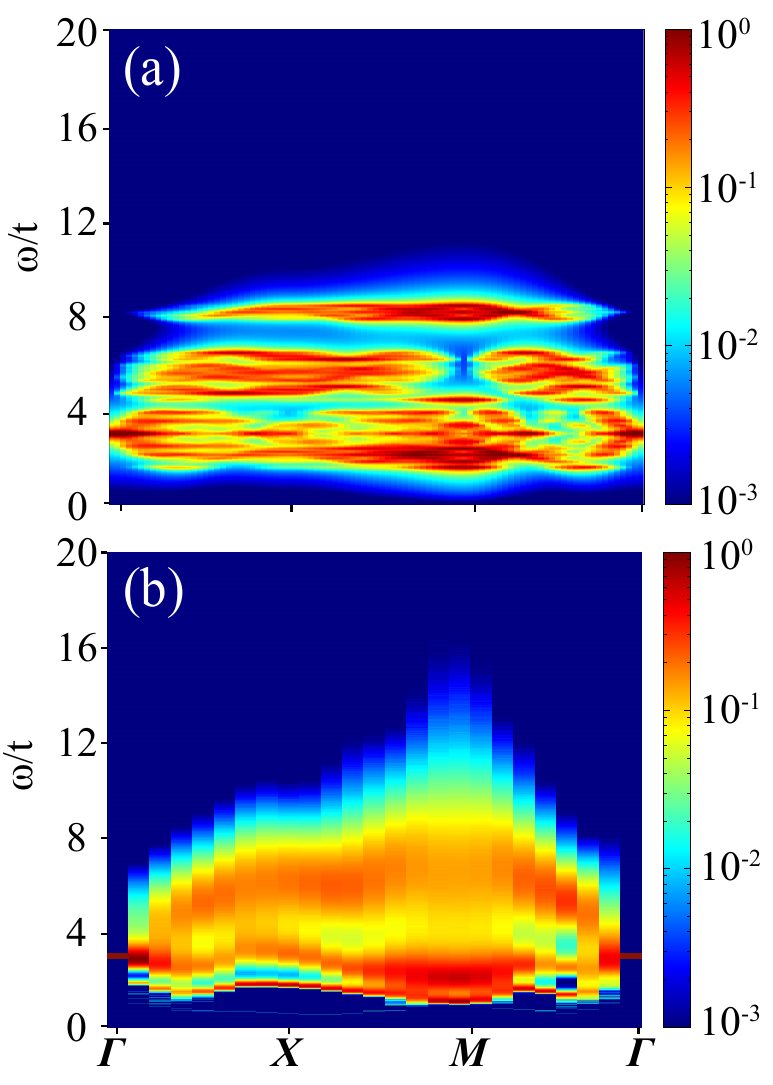}
\caption{\textbf{Transverse magnetic spectra at
          Zeeman field $B/t=3$ from lattice mean field and QMC+SAC
calculations.} (a) $S^\pm(\mathbf{q}, \omega)$
          from mean field with lattice size $L=60$. (b) $S^\pm(\mathbf{q},
\omega)$ from DQMC simulations with $L=16, \beta=16, J/t=0.1$. Mean-field data show three sets of bands, while spectra from QMC simulations show at least two broad spectral features, with high-energy structure difficult to resolve.
          Larmor mode at $\bm{\Gamma}$ is clearly resolved both in mean-field
and QMC data. QMC data also show a reduction of weight at $\bm{M}$ point at frequencies corresponding to the second mean-field band set. }
        \label{fig:fig_spec_spsm}
\end{figure}

\begin{figure}[tbp]
\includegraphics[width=\columnwidth]{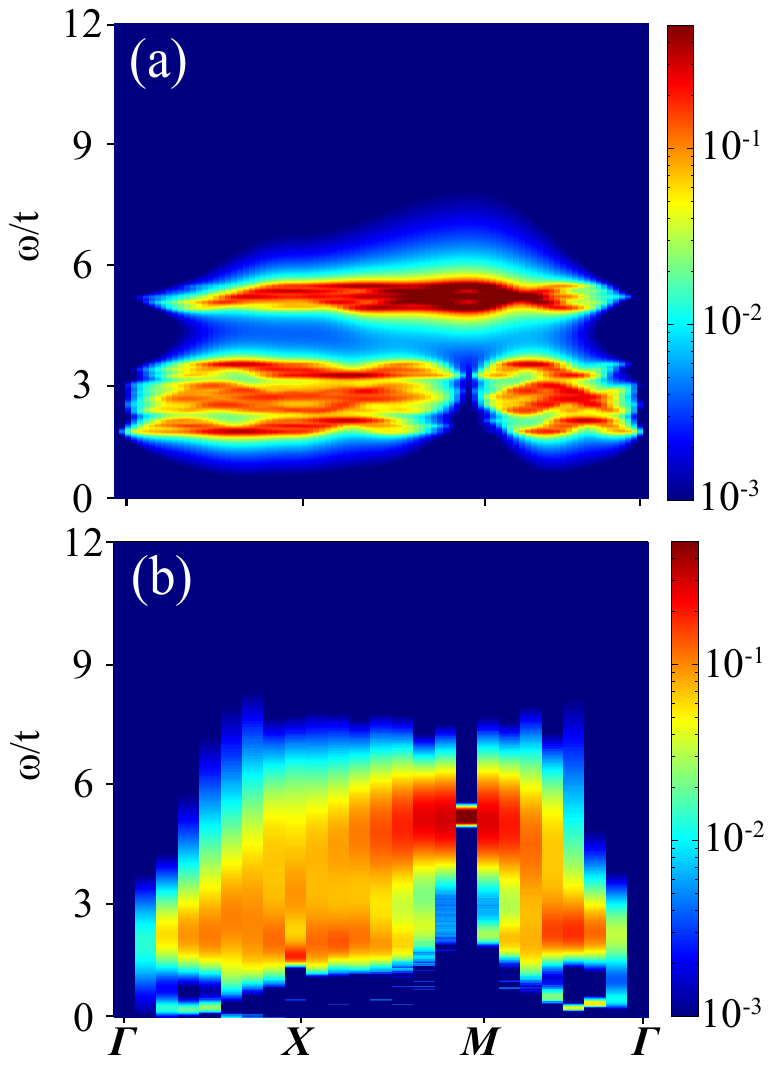}
\caption{\textbf{Longitudinal magnetic spectra at
          Zeeman field $B/t=3$ from lattice mean field and QMC+SAC
calculations.} (a) $S^{zz}(\mathbf{q}, \omega)$
          from mean field with lattice size $L=60$. (b) $S^{zz}(\mathbf{q},
\omega)$ from DQMC simulations with $L=16, \beta=16, J/t=0.1$. QMC and mean-field results are quite
          consistent, both showing lower- and higher-frequency regions of spectral weight and the absence of spectral weight at $\bm{M}$ in the lower-frequency region. Crucially, QMC data show the emergence
of a low-energy mode near $\Gamma$, possibly consistent with the analytical
prediction Eq.~\eqref{eq:Sz=flux} in Sec.~\ref{sec:CF} of the gapless photon.
Such a mode is absent in the mean-field analysis.}
        \label{fig:fig_spec_szsz}
\end{figure}

\begin{figure}[tbp]
        \includegraphics[width=\columnwidth]{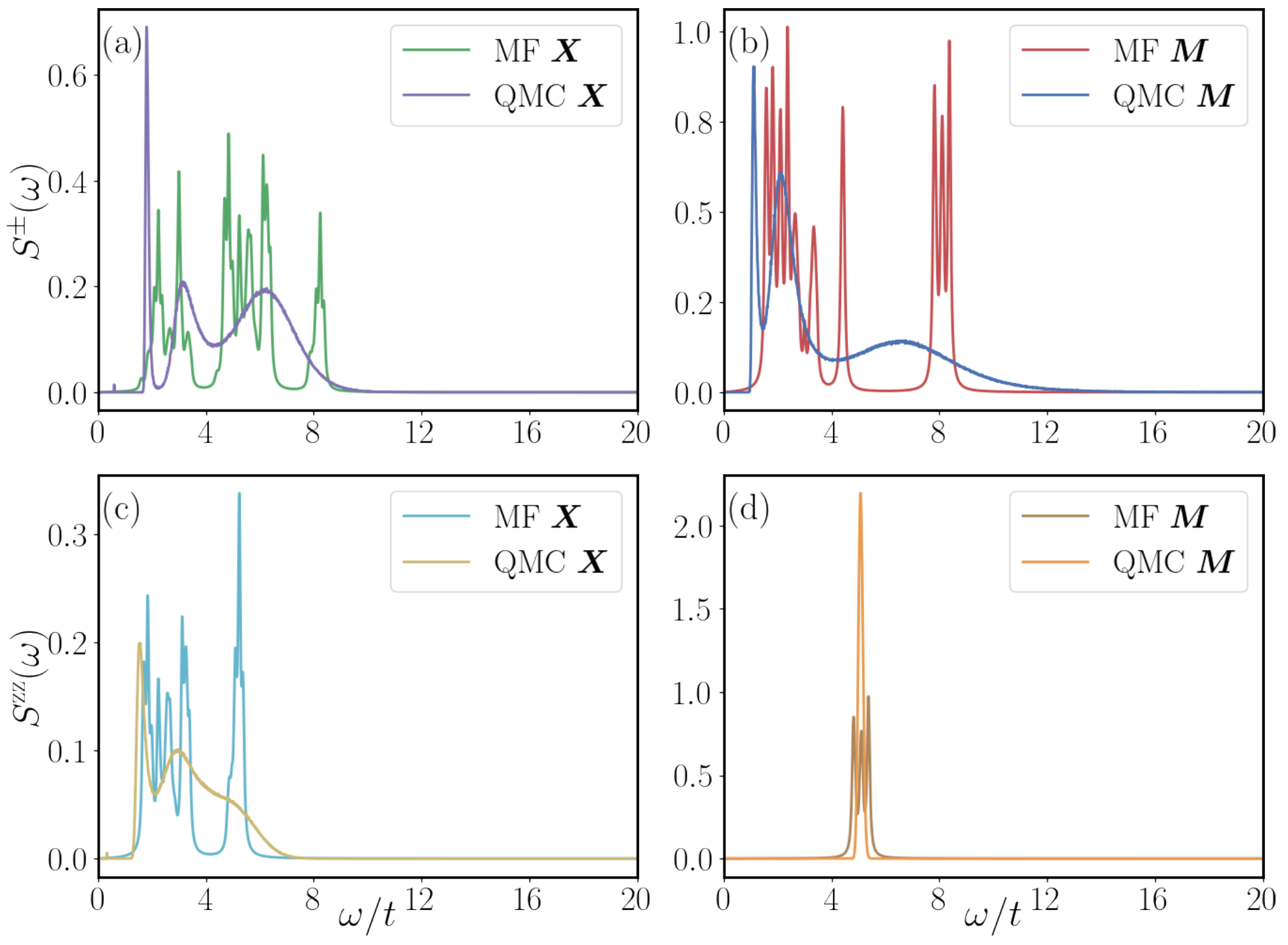}
        \caption{\textbf{Transverse and longitudinal magnetic spectra at high-
symmetry
          points.} Comparison between QMC and lattice mean-field
          results for Zeeman field $B/t=3$ at $\bm{X}(\pi, 0)$ and
$\bm{M}(\pi, \pi)$. (a) and (b) for $S^\pm(\omega)$. (c)
          and (d) for $S^{zz}(\omega)$. QMC and mean-field results are in
better agreement
          for low frequencies, with the QMC results showing smoother spectra (compared
to the mean-field result) at high energies.}
        \label{fig:fig7}
\end{figure}

We now turn to the lattice action in Eq.~\eqref{eq:action}.
In the limit of static gauge field, $J=0$, we perform lattice mean-field
calculations (see Sec.~I of the SM~\cite{suppl} for details)

\begin{figure}[tbp]
        \includegraphics[width=\columnwidth]{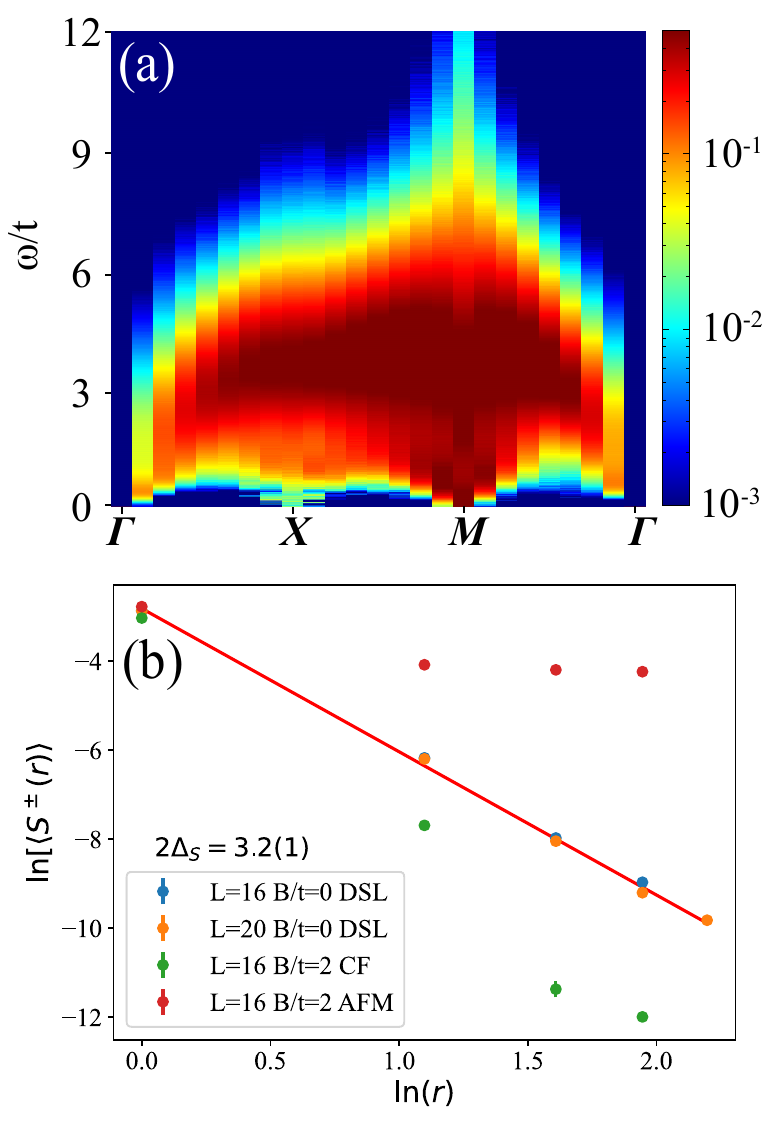}
        \caption{\textbf{Magnetic spectra inside the DSL phase and the scaling
dimension of fermion bilinear operator.} (a) Transverse magnetic spectra
$S^\pm(\mathbf{q}, \omega)$ for $
          J/t=1, B/t=0$ with system size $L=16, \beta=16$, mixed case. The
model is in the U(1) DSL phase. The spectra show well-known
          spinon continuum characteristics. (b) Real-space spin
          correlation $S^\pm(r)$ with $r$ denoting the odd site distance along
          the $x$ direction. Blue and orange data points are with the same
          parameters as (a), inside DSL phase. Data exhibit a power-law decay
          with exponent $2\Delta_S=3.2(1)$, consistent with the expectation of
          the scaling dimension of fermion bilinear operators in the CFT of
          SU(4) \qed, discussed in Sec.~\ref{sec:IIC1}. Green and red data
          points are in either CF or AFM phase with $B/t=2$. Green dots (CF)
show
          fast decay which is consistent with gapped transverse spin spectra,
          while the red dots (AFM) show long-ranged order.  }
        \label{fig:fig_u1d_spectra}
\end{figure}

Our results for the real-frequency transverse and longitudinal spin structure
factors are shown along a high-symmetry path in the Brillouin zone in
Fig.~\ref{fig:fig_spec_spsm}(a) and ~\ref{fig:fig_spec_szsz}(a), and detailed
energy scans at a few representative momenta are shown in Fig.~\ref{fig:fig7}.

We first note that the lattice mean-field spectra (both transverse and
longitudinal) exhibit a fine structure of many separate nearly dispersionless
bands. Equipped with theoretical analysis in Secs.~\ref{sec:CF} and
\ref{sec:IVA}, this discrete spectrum can be directly attributed to
inter- or intra-Landau-level transitions. We further observe that the individual
bands in the transverse (longitudinal) can be grouped into three (four) sets.
We attribute this additional structure to lattice effects that lie beyond our
continuum field-theory analysis.

\paragraph{Transverse structure factor.}

Comparing to our field-theoretic predictions [points (1) and (3) in
Sec.~\ref{sec:IVA}], we observe that exactly at the $\bm{\Gamma}$-point, there
is only a single mode (at frequency $\omega = 3 t=B$) in the transverse
structure factor which carries finite weight -- this is precisely the Larmor
mode.
For small momenta near the $\bm{\Gamma}$-point, additional modes with finite
weight emerge, and we observe that there is a linear scaling for the frequency
of the spectra weight onset as a function of momenta $\bm{q}$ as a remnant of
the linear dispersion of the Dirac fermions in the zero-field case.
We further note the low-frequency response at the $\bm{M}$-point features
rather intense spectral weight, as visible in Fig.~\ref{fig:fig_spec_spsm}(a)
and in Fig.~\ref{fig:fig7}(b).
This low-energy spectral weight could be an indication of the eventual
emergence of a Goldstone magnon at the $\bm{M}$-point in the AFM phase, upon
increasing the interaction strength.

\paragraph{Longitudinal structure factor.} A particularly striking feature of
the longitudinal structure factor in Fig.~\ref{fig:fig_spec_szsz}(a) at the $\bm{M}$ point is the full concentration of spectral weight in the topmost bands, at frequencies
$\omega/t \sim 5$ (and a concomitant absence of weight at low and intermediate
frequencies). This strong concentration of weight at large frequencies becomes
also visible when inspecting the line cuts at the $\bm{M}$-point in
Fig.~\ref{fig:fig7}(d).
We argue this to be consistent with our field-theoretic finding that many
poles in $S^{zz}(\omega)$ have vanishing spectral weight at the
$\bm{M}$-point.

\subsection{QMC plus stochastic analytic continuation (SAC) simulations}
\label{sec:IVC}

For finite $J\neq 0$ we obtain the imaginary-time correlation function from
our DQMC simulations and then perform a stochastic analytic continuation (SAC)
to obtain the real-frequency
results~\cite{sandvikStochastic1998,beachIdentifying2004,
syliuaasenUsing2008,sandvikConstrained2016}.
This QMC+SAC scheme has been applied to a variety of lattice models, including
the Dirac fermions under Hubbard-type interaction and the magnetic
field~\cite{chenUniversal2024}, producing reliable spectral properties ranging
from magnon and amplitude modes in a magnetically ordered state~\cite{shaoNearly2017,zhouAmplitude2021}, fractionalized excitations in
quantum spin liquid and spin ice
models~\cite{sunDynamical2018,huangDynamics2018,
wangFractionalized2021}, as
well as an emergent Dirac spinon spectrum at deconfined quantum critical
points~\cite{maDynamical2018} and single-particle spectrum of the angle-tuned
Gross-Neveu quantum criticality in twisted bilayer
graphene~\cite{huangAngle2025}.

\paragraph{Qualitative effect of small \texorpdfstring{$J>0$}{J>0}.} In comparing the QMC+SAC
spectra (with small nonzero $J/t=0.1$) to the mean-field spectra (with
$J=0$), we observe two main effects beyond the mean-field.  First, the
non-mean-field spectra are broadened relative to the finely discretized nature
of the mean-field spectra associated with inter- or intra-Landau-level
excitations.  This is expected on general grounds: the interaction with the
fluctuating gauge field will yield a broadening (finite lifetimes) of both
single-particle spectra as well as the two-particle response functions.
Second, in some spectral functions (see below) new collective modes emerge
below the support of the mean-field spectra.

However, we note that even in the presence of such broadening, the DQMC
structure
factors in Fig.~\ref{fig:fig_spec_spsm}(b) and Fig.~\ref{fig:fig_spec_szsz}(b)
possess several bandlike high-intensity features that resemble the lattice
mean-field spectra in Fig.~\ref{fig:fig_spec_spsm}(a) and
Fig.~\ref{fig:fig_spec_szsz}(a).
This correspondence is further corroborated by inspecting the line cuts in
Fig.~\ref{fig:fig7}.
We stress that the key qualitative features identified in the continuum field
theory and lattice mean-field calculations above persist also at finite
interaction strengths.

\paragraph{Transverse structure factor in the CF phase.} In this component of
the structure factor, the main difference of the mean-field and $J/t=0.1$
spectra is the broadening.  However, broadening is absent, as is seen from
Fig.~\ref{fig:fig_spec_spsm}(b), at the $\bm{\Gamma}$-point, where there is a
sharp mode at frequency $\omega = 3t$: this is the Larmor mode, with frequency
$\omega = B$ and weight $4 \pi m^z$ protected against interactions.
Going away from the $\bm{\Gamma}$-point, one can further observe a linear
dispersion $\omega(\bm{k}) \sim |\bm{k}|$ of the spectral weight, in line with
our theoretical predictions.

\paragraph{Longitudinal structure factor in the CF phase.} At the
$\bm{M}$-point, there is a strong concentration of spectral
weight in the longitudinal structure factor near $\omega \simeq 5t$, and
vanishing spectral weight at all other frequencies (see
Fig.~\ref{fig:fig_spec_szsz}(b) and Fig.~\ref{fig:fig7}(d)), in qualitative
agreement with our continuum field theory and lattice mean-field calculations.
We further comment that such a concentration of spectral weight at high
frequencies at the $\bm{M}$-point is reminiscent of the high-energy
single-magnon excitation in the high-field regime of the square-lattice
antiferromagnet which emerges upon increasing the interaction strength $J/t$
\cite{syljuUnstable08, luescherED2009}.

In the longitudinal structure factor, the $J/t=0.1$ spectra clearly exhibit a
low-energy collective mode near $\Gamma$ not present in the corresponding
mean-field spectra -- compare  Fig.~\ref{fig:fig_spec_szsz} panels (a) and
(b).
Although the finite-size and finite-temperature effects ($L=\beta=16$) prevent
us from resolving the predicted linear dispersion quantitatively, the observed
low-energy spectral weight near $\Gamma$ is consistent with, but does not by
itself establish, the photon-related response expected from
Eq.~\eqref{eq:Sz=flux} in Sec.~\ref{sec:CF}. Very similar
spectra, inside the CF phase at $B/t=2$, are shown in Sec.~VII of the SM~\cite{suppl}.

\paragraph{Structure factor in the U(1) DSL phase.} We show the magnetic
spectra in $\pi$-flux U(1) deconfined phase (mixed theory) at
zero Zeeman field ($B=0$) in Fig.~\ref{fig:fig_u1d_spectra}(a), which
demonstrates well-known continuum characteristics with gapless Dirac cones at
the $\bm{\Gamma}$, $\bm{X}$ and $\bm{M}$-points.
The spectra can be qualitatively reproduced by a simple RPA calculation of the
noninteracting structure factor of the $\pi$-flux fermion hopping
model \cite{maDynamical2018}.   However, a quantitative analysis shows the
effects of gauge fluctuations.
Specifically, we have also computed the real-space spin correlation function
inside the DSL phase up to $L=20$.
As shown in Fig.~\ref{fig:fig_u1d_spectra}(b), our data exhibit a power-law
decay with exponent $2\Delta_S=3.2(1)$, consistent with the expectation of
$2\Delta_{adj} \in (2.8,3.5)$ for the fermion bilinear operators in the CFT of
SU(4) \qed, discussed in Sec.~\ref{sec:IIC1}.
We note that similar power-law spin and dimer correlations with the exponent
of
$2\Delta \sim 3$ have also been seen in the DSL phase in
Ref.~\cite{xu2019monte}.
For comparison, real-space correlations inside CF and
AFM phase with Zeeman field $B/t=2$ are shown in
Fig.~\ref{fig:fig_u1d_spectra}. Both display non-power-law decay, with fast
decay in CF phase which indicates short-range transverse spin correlations and gapped transverse spectra. In AFM, spin correlations show long-range order.

\begin{figure}[tbp]
    \includegraphics[width=\columnwidth]{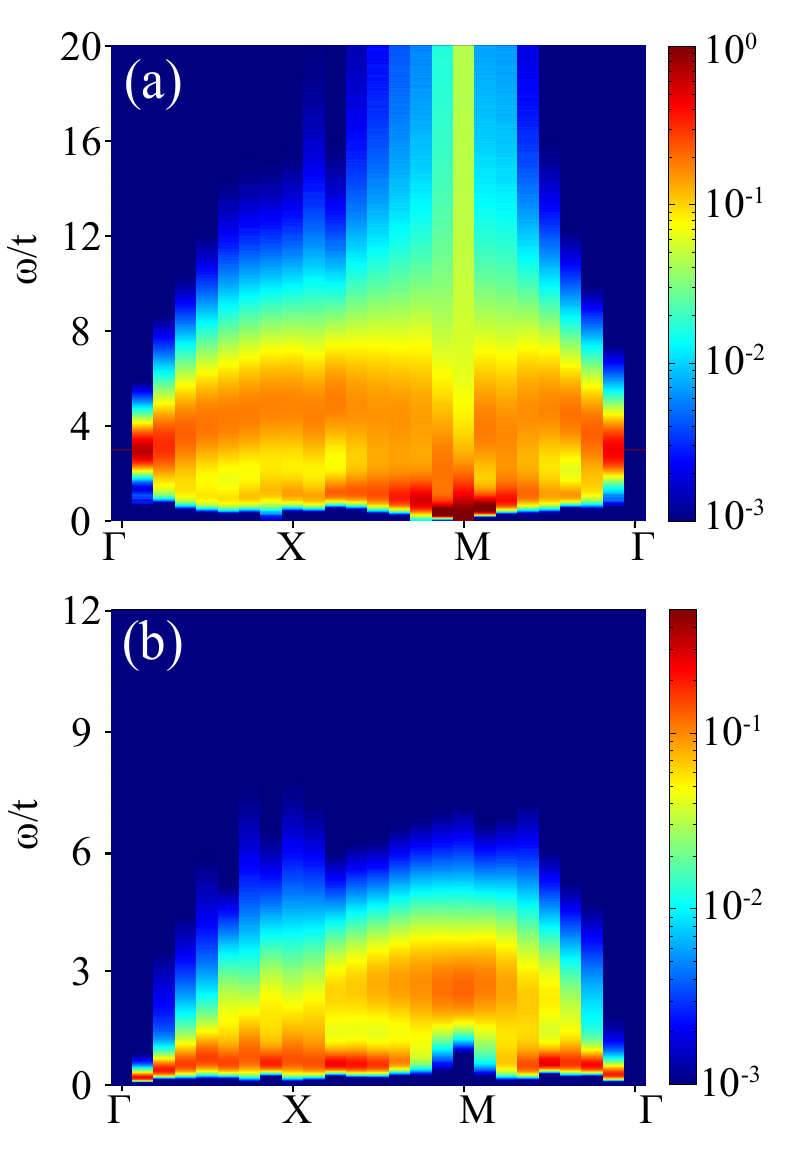}
        \caption{\textbf{Transverse and longitudinal magnetic spectra inside
the AFM phase.} (a)
          $S^\pm(\mathbf{q}, \omega)$ and (b)
          $S^{zz}(\mathbf{q}, \omega)$ for $J/t=3,B/t=3$ with
          system size $L=16, \beta=16$. At point $\bm{M}$ of
$S^\pm(\mathbf{q},
          \omega)$, there is considerable weight at zero frequency, indicating
the Goldstone mode. In parallel, $S^{zz}(\mathbf{q},\omega)$ shows a peak
around $\omega = B$ at $\bm{M}$, in agreement with the spin-wave expectations
for the ordered phase \cite{Mourigal2010}.}
        \label{fig:fig_J3_spec}
\end{figure}

\paragraph{Magnetic spectra in the AFM phase.}
When $J/t$ is larger, there can be substantial deviations from the mean-field
spectrum.
Physically, the probability for fermions to bind or recombine into collective
modes becomes large, and such modes may be more dominant in the spectral
response.
The QMC+SAC results for the transverse and longitudinal spectra in the AFM phase at $J/t=3$ for $B/t=3$ are shown in Fig.~\ref{fig:fig_J3_spec}.
In Fig.~\ref{fig:fig_J3_spec}(a), one observes spin-wave-like features near
$\bm{M}$ point, with the $\bm{M}$-point itself hosting a gapless Goldstone
mode (see also Sec.~\ref{sec:AFM}).
In Fig.~\ref{fig:fig_J3_spec}(b), the longitudinal spectrum is fully gapped,
and in particular there is a high-intensity feature ($\omega/t \approx 3$) at
the ordering wave vector $\bm{M}$.  This feature is similar to that of the
Heisenberg square-lattice antiferromagnet in strong Zeeman fields
\cite{luescherED2009,syljuUnstable08}, which has only localized bosonic
states.  This is in agreement with a picture of substantial recombination of
the fermions into bosonic local excitations.

\section{Discussion}
\label{sec:conclusion}

In this work, we have studied the response of \qed\ to a flavor chemical
potential, which emerges in a condensed matter context as the low-energy
theory for the response of a U(1) Dirac spin liquid to an externally applied
Zeeman field.

Utilizing large-scale quantum Monte Carlo simulations for the case of a
mixed gauge field, we unambiguously establish that for finite flavor chemical
potentials (Zeeman field), there exists a stable ``chiral flux'' phase
which is characterized by the emergent gauge field developing a finite net
average flux, with the system's magnetization being proportional to the
induced flux, $m^z \propto \pi-\Phi_\Box$.

\subsection{Key results and implications}

The field-induced generation of \emph{emergent gauge flux} has several key
implications:
\begin{enumerate}
    \item The fermionic spectrum of the theory is given by spin-split
relativistic Landau-Hofstadter levels and thus becomes gapped. The transverse
and longitudinal magnetic structure factors exhibit signatures of the
Landau-level spectrum, and we match a number of qualitative features obtained
within a continuum field-theory analysis with our numerical results, for
example the presence of a Larmor mode in the transverse magnetic structure
factor~\cite{agarwalCollective2024,chenUniversal2024}.
    \item Owing to the suppression of monopoles in the time-continuum limit
(see \cite{suppl} for an explicit demonstration in an analytically tractable
model) in the model at hand, the U(1) gauge field remains coherent in the CF
phase and features a gapless photon excitation. The effective theory predicts
a photon-related low-energy contribution to the longitudinal magnetic
structure factor. The QMC+SAC spectra show low-energy spectral weight near
$\Gamma$ consistent with this expectation; further work is required to resolve
the predicted linear dispersion quantitatively. Such a response also
distinguishes the strongly correlated nature of the CF phase at finite $J$,
that is beyond the mean-field analysis at $J=0$.
    \item In an effective field-theory framework, the CF state can be
described by a mixed Chern-Simons term for fluctuations of the emergent gauge
field and an external ``spin'' gauge field, which encodes the attachment of
spin $S^z$ to flux of the emergent gauge field. This implies that the CF state
is characterized by a composite order parameter $\langle \mathcal{M} S^+
\rangle \neq 0$, where $\mathcal{M}$ is a $2\pi$-gauge flux insertion
operator. This is consistent with our numerical observation that the
transverse spin spectra show a full gap in the CF phase.
\end{enumerate}

We argue that these results are directly related to the generation of a finite
flux in the \qed\ gauge theory.
Connecting with quantum magnets, we stipulate that observing such behavior of
a quantum spin liquid candidate material in an applied magnetic field would be
hard to reconcile with more conventional descriptions (i.e.~which do not rely
on fractionalization and emergent gauge fields).

\subsection{Effects due to dynamical monopoles}

When \qed\ arises as an emergent gauge field in a microscopic spin model, the
theory is inevitably \emph{fully} periodic in the gauge field.
In general, this allows dynamical \emph{monopole instantons}, which are
suppressed in our study.
It is known that monopoles carry distinct quantum numbers under different
microscopic (lattice-scale) symmetries \cite{song19,WietekQED3}, and that
their presence may have severe consequences for the stability of U(1) Dirac
spin liquids at the lowest energies on different lattices.
The similarity of the results in this paper at zero field for the mixed theory
to zero field simulations of the \emph{compact} theory in
Ref.~\cite{xu2019monte,feng25} suggests, however, that such effects may become
significant only at very low (and vanishing in some cases) energy scales. We
leave a \emph{quantitative analysis of} such very low energy effects to future
(numerical) work. This limitation does not affect the controlled conclusion of
the present work: within the mixed theory, the field-induced chiral-flux phase
and its characteristic dynamical spin-response signatures are established
numerically. The following discussion instead addresses how monopole effects
may modify this physics in a fully compact gauge theory.

While abovementioned effects are crucial for understanding the stability of
U(1) Dirac spin liquids \emph{in general}, we now turn to the central object
of study in this paper, namely the field-induced chiral flux phase with
nonzero gauge flux:
here, the eventually inevitable compactness of the U(1) gauge field, and the
associated monopole excitations will have a particularly prominent effect. We
argue that they render the ``hidden'' order (of joint $2\pi$ flux insertion
and simultaneous change of $\Delta S^z = 1$, see also Sec.~\ref{sec:CF} and in
particular Eqs.~\eqref{eq:m_ord} and \eqref{eq:m_ord_exp} therein) to be
physical, thus allowing it to be resolved in terms of experimentally
accessible local order parameters.
To demonstrate this, we posit that one may relate expectation values of
observables $\langle \mathcal{O} \rangle_\mathrm{comp}$ in the full compact
theory (with dynamical monopoles) to expectation values in the mixed theory
(absent of monopoles) by means of an expansion in the monopole fugacity
$\lambda$, i.e., we deform the action of the mixed theory by $2\pi$ flux
insertion operators, $\mathcal{S}_\mathrm{mix} \to \mathcal{S}_\mathrm{mix} -
\int d^3 x [ \lambda \mathcal{M} + \lambda^\ast \mathcal{M}^\dagger]$ (for
simplicity, we adopt a continuum formulation, but the argument also applies to
discretized space-time).
Now, consider the in-plane spin density $S^+$ as an observable.
As argued in Sec.~\ref{sec:CF}, in the mixed theory $\langle S^+
\rangle_\mathrm{mix} = 0$ vanishes in the chiral flux phase. However, in the
compact theory, we observe
\begin{multline}
    \langle S^+(x) \rangle_\mathrm{comp} = \langle S^+(x) \rangle_\mathrm{mix} \\
{}+ \int d^3 y\, \lambda \langle S^+(x) \mathcal{M}(y) \rangle_\mathrm{mix} +
\dots  \\
    \approx + \lambda \langle S^+(x) \mathcal{M}(x) \rangle_\mathrm{mix} \neq
0.
\end{multline}
The fugacity expansion (to leading order) thus produces the \emph{composite
order parameter} which, within the mixed theory, is finite in the chiral flux
phase: therefore, the presence of dynamical monopoles in the fully compact
theory makes the hidden composite order in the chiral flux phase visible as
\emph{finite in-plane magnetic order}.

We emphasize that the nature of the in-plane magnetic order, for example the
ordering wave vector, will depend on the UV details of the theory which
determine how the microscopic symmetries act on the low-energy degrees of
freedom in the QED$_3$ field theory.
For the challenges elucidated above, a controlled study of this
monopole-induced effect within numerical simulations of the QED$_3$ low-energy
field theory appears challenging at present.
However, from a complementary standpoint, one may construct microscopic
(variational) wave functions associated with a ``chiral flux'' state by
Gutzwiller-projecting Slater determinants of fermionic spinons on a lattice,
thereby accounting for aforementioned UV details. A recent analysis
\cite{WangChirality2026} of such a family of finite-flux wave functions for
triangular-lattice systems has indeed revealed highly pronounced
\emph{antiferromagnetic (120\,$^\circ$) in-plane spin correlations},
consistent with the general field-theoretic argument provided above.

\subsection{Outlook}

We note that recently several candidate materials have been identified that
exhibit signatures consistent with a (proximate) U(1) Dirac spin liquid ground
state, such as the triangular-lattice antiferromagnets
YbZn$_2$GaO$_5$~\cite{BagEvidence24,WuSpinDynamics25} and the $A$-YbSe$_2$
delafossites~\cite{ScheieKYbSe2, ScheieNaYbSe2} as well as the kagom\'e
antiferromagnet
YCu$_3$(OD)$_6$Br$_2$[Br$_{x}$(OD)$_{1-x}$]~\cite{zengSpectral2024,
hanSpin2024,luUnraveling2026,
mengPerspective26},
where inelastic neutron scattering experiments have found spectra that are
consistent with a Dirac cone filled with a continuum of excitations.
In light of our results, it will be highly interesting to scrutinize the
behavior of these systems in applied magnetic fields.
In particular, we foresee the experimental observation of a gapless magnetic
spectral feature related to the photon mode in the longitudinal spin structure
factor, possibly via inelastic neutron scattering, to be interesting evidence
of the magnetized DSL state.
An interesting direction for further study is to investigate what
perturbations might lead to different filling of the spinon Landau levels in
the CF phase (while maintaining gauge invariance or particle-hole symmetry), and
establish possible connections to experimentally observed unusual magnetic
oscillations in the YCu$_3$(OD)$_6$Br$_2$[Br$_{x}$(OD)$_{1-x}$] kagom\'e
antiferromagnet~\cite{zhengUnconventional2025} (see also
Ref.~\cite{SodemannPseudoscalar21}).

Going beyond quantum magnets, we stress that our main result, the generation
of the Chiral Flux phase due to a flavor chemical potential, could also be of
relevance to other systems with phases (or phase transitions) which have been
proposed to be described by ($N \geq 2 $)-\qed\,
\cite{lukasConfinement2020,wangDynamics2019,SongEmergent23,
Paoletti2025,WangEmergent2025}. We are hopeful that our results are a key step
toward establishing experimentally \emph{testable predictions} that will
eventually enable the identification of a fractionalized phase with deconfined
gauge fields in a condensed matter system.

{\em Note added.---}
Recently, Ref.~\cite{DumitrescuCommentsQED325} appeared, which considers symmetry
breaking in QED$_3$ in a mean background gauge flux (without an externally
imposed Zeeman field) and is in agreement with our conclusions in
Secs.~\ref{sec:IIC1} and \ref{sec:CF} where results overlap.
Further, Refs.~\cite{KeselmanTriangular2025,BaderTriangular2025} also appeared recently, which study field-induced phases of
frustrated triangular-lattice quantum antiferromagnets, to which our results
may be applicable.
In a similar vein, Ref.~\onlinecite{WangChirality2026} studies
Gutzwiller-projected finite-flux wave functions for triangular-lattice quantum
antiferromagnets.
We also note recent progress in hybrid Monte Carlo simulations of the lattice
gauge theory (at zero Zeeman field) \cite{feng25} closely related to the
current investigation.

\begin{acknowledgments}
We acknowledge discussions with A. Capelli, J. Knolle, A. Rosch, I. Sodemann,
and A. Vainshtein.
C.~C. and Z.~Y.~M. acknowledge Y.~D. Liao and Y. Qi for collaboration on a previous
project concerning magnetized Dirac cones in the Hubbard
model~\cite{chenUniversal2024}.
U.~F.~P.~S., O.~A.~S., L.~B., and Z.~Y.~M. are grateful to the Pollica Physics Centre for
hospitality during the stimulating workshop ``Exotic quantum matter: from
quantum spin liquids to novel field theories''.
C.~C., K.~F., and Z.~Y.~M. acknowledge the support
from the Research Grants Council (RGC) of Hong Kong (Project Nos. C7037-22GF,
17302223, 17301924, 17301725), the ANR/RGC Joint Research Scheme
sponsored by RGC of Hong Kong and French National Research Agency (Project No.
A\_HKU703/22). We thank HPC2021 system under the Information Technology
Services at the University of Hong Kong, as well as the Beijing
PARATERA Tech CO., Ltd for providing HPC resources that have
contributed to the research results reported within this paper. U.~F.~P.~S.
acknowledges support from the Deutsche Forschungsgemeinschaft (DFG, German
Research Foundation) via SFB 1238, Project ID No. 277146847, the Emmy Noether
Program, Project ID No. 544397233. L.~B. is supported by the NSF CMMT program
under Grant No. DMR-2419871, and by the Simons Collaboration on Ultra-Quantum
Matter, which is a grant from the Simons Foundation (Grant No. 651440).
\end{acknowledgments}

\section*{Data availability}
Selected numerical data used to generate the figures are publicly available
in the authors' GitHub data repository~\cite{mixed_qed3_cf_data}.
The complete bin-resolved DQMC outputs and the original imaginary-time
correlation functions used for SAC are available from the authors upon
reasonable request because of their volume. The DQMC code is not publicly
available.

\nocite{gradshteyn15}

\bibliographystyle{longapsrev4-2}
\bibliography{main}

\clearpage
\onecolumngrid
\makeatletter
\@ifundefined{ltx@footnote@pop}{}{\ltx@footnote@pop}
\@booleanfalse\footinbib@sw
\makeatother
\begin{center}
\textbf{\large Supplemental Material for\\
``Emergent Gauge Flux in Mixed QED$_3$ with Flavor Chemical Potential:\\
Implications for Magnetized U(1) Dirac Spin Liquids''}
\end{center}
\setcounter{equation}{0}
\setcounter{figure}{0}
\setcounter{table}{0}
\setcounter{section}{0}
\setcounter{subsection}{0}
\renewcommand{\theequation}{S\arabic{equation}}
\renewcommand{\thefigure}{S\arabic{figure}}
\renewcommand{\thetable}{S\Roman{table}}
\renewcommand{\theHequation}{SM.\arabic{equation}}
\renewcommand{\theHfigure}{SM.\arabic{figure}}
\renewcommand{\theHtable}{SM.\arabic{table}}
\renewcommand{\theHsection}{SM.\arabic{section}}
\renewcommand{\theHsubsection}{SM.\arabic{section}.\arabic{subsection}}
\renewcommand{\theHsubsubsection}{SM.\arabic{section}.\arabic{subsection}.\arabic{subsubsection}}

The Supplemental Material provides details both in analytic derivations and
quantum Monte Carlo simulations, as well as the benchmark data that are
referred to in the main text.

\section{Lattice mean field magnetic spectra calculation}
\label{appA}
The formalism of the calculation has been applied in
Ref.~\cite{chenUniversal2024} for the $\pi$-flux Hubbard model on the square
lattice under the Zeeman field. We have employed GPU computation to accelerate
the evaluation of the transverse
and longitudinal spin susceptibilities, which enables access to larger lattice
sizes with $L$ upto 60.

Let's first look at the transverse channel. Start from real space
and Matsubara frequency
\begin{align}
\chi^{\pm}(i,j,i\omega_{n}) & =\sum_{i\nu_{n}}G_{\uparrow}\left(j,i,i\nu_{n}
\right)G_{\downarrow}\left(i,j,i\nu_{n}+i\omega_{n} \right)\nonumber\\
 & =\sum_{i\nu_{n}}\left(\frac{1}{(i\nu_{n}+\mu)\mathbb{I}_{N\times N}-H}
\right)_{ji}\cdot\left(\frac{1}{(i\nu_{n}+i\omega_{n}-\mu)\mathbb{I}_{N
\times N}-H}\right)_{ij}\nonumber\\
 & =\sum_{i\nu_{n}}\sum_{m}\sum_{l}U_{jm}
\frac{1}{i\nu_{n}+\mu-D_{m}}U_{mi}^{-1}\cdot U_{il}
\frac{1}{i\nu_{n}+i\omega_{n}-\mu-D_{l}}U_{lj}^{-1}\nonumber\\
 & =\sum_{m}\sum_{l}U_{jm}U_{mi}^{-1}U_{il}U_{lj}^{-1}
\frac{n_{F}(D_{m}-\mu)-n_{F}(D_{l}+\mu)}{i\omega_{n}+D_{m}-D_{l}-2\mu}
\end{align}
where the singular value decomposition of the non-interacting $(J=0)$
Hamiltonian $H$ at $\mu=\frac{1}{2}B$ is performed. After analytical
continuation $i\omega_{n}\rightarrow\omega+i\eta^{+}$, one has
\begin{equation}
  \chi^{\pm}(i,j,\omega) =\sum_{m}\sum_{l}U_{jm}U_{mi}^{-1}U_{il}U_{lj}^{-1}
\frac{n_{F}(D_{m}-\mu)-n_{F}(D_{l}+\mu)}{\omega+i\eta^{+}+D_{m}-D_{l}-2\mu},
\end{equation}
then we perform Fourier transformation to obtain momentum space dynamical
susceptibility
\begin{align}
\chi^{\pm}(q,\omega)=\frac{1}{N}\sum_{ij}\chi^{\pm}(i,j,\omega)\cdot
e^{-iq\cdot r_{ij}},
\end{align}
and the spectra is $-2\text{Im}\chi^{\pm}(q,\omega)$. 

In the actually lattice model computation, we can further analyze
the formula to make it suitable for GPU simulation to access larger system
sizes in shorter time
\begin{align}
\chi^{\pm}(q,\omega) & =\frac{1}{N}\sum_{ij}\chi^{\pm}(i,j,\omega)\cdot
e^{-iq\cdot r_{ij}}\nonumber\\
 & =\frac{1}{N}\sum_{ij}\sum_{m}\sum_{l}U_{jm}U_{mi}^{-1}U_{il}U_{lj}^{-1}
\frac{n_{F}(D_{m}-\mu)-n_{F}(D_{l}+\mu)}{\omega+i\eta^{+}+D_{m}-D_{l}-2\mu}
\cdot e^{-iq\cdot r_{ij}}\nonumber\\
 & =\sum_{m}\sum_{l}\left[\frac{1}{N}\sum_{i}U_{im}^{*}U_{il}e^{-iq
\cdot r_{i}}\sum_{j}U_{jl}^{*}U_{jm}e^{+iq
\cdot r_{j}}(n_{F}(D_{m}-\mu)-n_{F}(D_{l}+\mu))\right]
\frac{1}{\omega+i\eta^{+}+D_{m}-D_{l}-2\mu}\nonumber\\
 & =\sum_{ml}\Gamma(m,l,q)G(m,l,\omega)
\label{eq:chipm}
\end{align}
One notices the matrix size of
$\Gamma(m,l,q)$ is $L^2 \cdot L^2 \cdot \frac{3}{2}L$ and computation of each
element
involves sum over $i,j$, which is perfect for GPU parallelization.
The sum over $m,l$ for each $(\mathbf{q}, \omega)$ is also well-suited for GPU
parallelization. It is in such arrangement that the $60\times60$ mean-field
spectra in the Figs. 7, 8 and 9 of the main text are obtained.

In order to compute $\chi^{\mp}(q,\omega)$, we starts from
$\chi^{\mp}(i,j,i\omega_{n})=
\sum_{i\nu_{n}}G_{\downarrow}(j,i,i\nu_{n})
G_{\uparrow}(i,j,i\nu_{n}+i\omega_{n})$
and in practice
just need to replace $\mu\rightarrow-\mu$ in the expression of
Eq.~\eqref{eq:chipm},
\begin{align}
  \chi^{\mp}(i,j,\omega)&=\sum_{m}\sum_{l}U_{jm}U_{mi}^{-1}U_{il}U_{lj}^{-1}
\frac{n_{F}(D_{m}+\mu)-n_{F}(D_{l}-\mu)}{\omega+i\eta^{+}+D_{m}-D_{l}+2\mu}.
\end{align}

For $\chi^{zz}(q,\omega)$, it is
\begin{align}
  \chi^{zz}(i,j,\omega)=\sum_{m}\sum_{l}U_{jm}U_{mi}^{-1}U_{il}U_{lj}^{-1}
                 \frac{
                     n_{F}(D_{m}-\mu)-n_{F}(D_{l}-\mu)
                     +n_{F}(D_{m}+\mu)-n_{F}(D_{l}+\mu)}
                     {\omega+i\eta^{+}+D_{m}-D_{l}}.
\end{align}
Correspondingly, we follow the practice in Eq.~\eqref{eq:chipm} to modify
$\Gamma(m,l,q)$ and $G(m,l,\omega)$ and GPU compute $\chi^{\mp}(q,\omega)$ and
$\chi^{zz}(q,\omega)$ accordingly.

\section{Dynamical spin spectra from low-energy field theory}\label{sm:field-theory} 
\label{sec:app_LL_spectra}

A continuum field theory approach is applicable at high-symmetry points
$\bm{\Gamma},\bm{X}, \bm{M}$ where the zero-field U(1) Dirac spin liquid
features a gapless dynamical spin susceptibility arising from
intra/inter-valley particle-hole excitations of Dirac fermions.

In the presence of a finite emergent gauge flux, the spinons are confined to
relativistic Landau levels. We find the corresponding spin susceptibility by
first deriving the propagator for Dirac fermions in relativistic Landau
levels, and then subsequently calculate the dynamical response function as a
fermionic bubble diagram.

\subsection{Propagator}

We first derive the propagator for Dirac fermions coupled to the emergent
gauge field $a$ with a non-zero internal gauge flux.
Picking a Landau gauge $a = \phi x \hat{y}$, we take the Hamiltonian for the
$\alpha = \uparrow (+1),\downarrow(-1)$ spinons (suppressing all indices) as
\begin{equation}
        H_\sigma = v \int \di^2 \bm{x} \, \psi^\dagger(x,y,t) \left[\gamma^x
p_x + \gamma^y (p_y - A_y) - b \frac{\alpha}{2}  \right] \psi(x,y,t).
\end{equation}
The solutions to the Dirac equation with energy $\pm E_n = \pm v  \sqrt{2n} /
\ell$ are then of the form $\varphi_{p_y,\pm}^{(n)}(\bm{x},t) = \ei^{\ii p_y
y} \frac{1}{\sqrt{2}} (\varphi_n , \pm \varphi_{n-1}) $ with $\varphi_n =
\frac{1}{\sqrt{2^n n! \sqrt{\pi} \ell}} \ei^{-(x/\ell-p_y \ell)^2/2}
H_n(x/\ell-p_y\ell)$ with Hermite polynomials $H_n$, and $\phi_{n=-1} \equiv
0$. We henceforth take $v=1$ for simplicity.

From this, we obtain the propagator $G_{AB} =
\braket{0|\Psi_A(x)\Psi_B^\dagger(x')|0}$ (in imaginary time) via a
mode-expansion (alternatively, functional methods may be used) as
\begin{multline}
        G^\alpha(x,x') = \frac{\ei^{-\xi/2} \ei^{\ii \Phi(x,x')}}{2 \pi
\ell^2} \int \frac{\di \omega}{2\pi} \ei^{-\ii \omega (t-t')}\Big[
\frac{\mathcal{P}_+}{\ii \omega + \alpha b/ 2}
    + \sum_{n=1}^\infty \Big\{ \frac{(\ii \omega+\alpha b/2)(\mathcal{P}_+
L_n^0(\xi) + \mathcal{P}_- L_{n-1}^0(\xi) )}{(\ii \omega +\alpha b/2)^2
-E_n^2} \\
    + \frac{\ii \bm{\gamma} \cdot (\bm{x}-\bm{x}') L^1_{n-1}(\xi) / \ell^2
}{(\ii \omega +\alpha b/2)^2 -E_n^2} \Big] \Big\},
\end{multline}
where we use that $G_{AB}$ is spin-diagonal to explicitly label the $\alpha =
\uparrow,\downarrow$ components, and make all other indices implicit (via
matrix notation).
Here, $L^{m}_n(x)$ are Laguerre polynomials that arise from using the identity
\begin{equation}
        \int_{-\infty}^\infty \di x \ei^{-x^2} H_m(x+y) H_n(x+z) = 2^n
\sqrt{\pi} m! z^{n-m} L^{n-m}_m(-2 y z)
\end{equation}
and the \emph{Schwinger phase} is given by $\Phi(\bm{x},\bm{x}') =
(x+x')(y-y') / (2\ell^2)$, and $\xi = \xi(\bm{x},\bm{x}') = \left((x-x')^2 +
(y-y')^2\right)/(2 \ell^2)$.

\subsection{Magnetization}

For consistency, we can compute the uniform magnetization from the propagator
(in imaginary time) as $m^z = \frac{1}{2} \langle \psi^\dagger(t+\epsilon,x)
\sigma^z \psi(t,x) \rangle$
with $\epsilon > 0$.
Noting that $\Phi(\bm{x},\bm{x}) = 0$ and $\xi(\bm{x},\bm{x}) = 0$, and using
$\tr \mathcal{P}_\pm = 1$ as well as $L^0_n(0) =1$, we find
\begin{equation}
        m^z_\mathrm{LL} = \frac{1}{2} \tr [G^+(x,t;x,t+\epsilon) -
G^-(x,t;x,t+\epsilon)] =
     \frac{1}{2\pi \ell^2} \left[\left(1 + \sum_{n=1}^\infty 1 \right) -
\left(\sum_{n=1}^\infty 1 \right) \right] = \frac{1}{2\pi \ell^2} =
\frac{\Phi}{2\pi}.
\label{eq:mz-landau-level}
\end{equation}

\subsection{Susceptibility bubble diagram}

By tracing how lattice-scale symmetry operations act on the low-energy spinon
degrees of freedom, we obtain field-theory expressions for microscopic spin
operators at the $\bm{\Gamma}$, $\bm{M}$ and $\bm{X}$-points as
\begin{subequations}
\begin{align} 
    S^\alpha_{\bm{\Gamma}} &\sim \frac{1}{2} \bar{\Psi} \gamma^0 \sigma^\alpha
\Psi\label{eq:s-low-energ-gamma} \\
    S^\alpha_{\bm{M}} &\sim \frac{1}{2} \bar{\psi} \sigma^a \mu^x \psi
\label{eq:s-low-energ-M} \\
    S^\alpha_{\bm{X}} &\sim \frac{1}{2} \bar{\psi} \sigma^a \mu^z \dots \psi
\label{eq:s-low-energ-X}
\end{align}
\end{subequations}
With this identification, we obtain dynamical susceptibilities in imaginary
time as
\begin{equation}
    \chi^{AB}(\ii \nu, \bm{k}) \sim V^{-1}\int \di t \ei^{\ii \nu t} \int
\di^2 \bm{x} \di^2 \bm{x}' \, \ei^{-\ii \vec{k}\cdot (\bm{x}-\bm{x}')} \tr
(\gamma \sigma \mu)^A G(t,\bm{x};0,\bm{x}') (\gamma \sigma \mu )^B
G(0,\bm{x}';t,\bm{x}),
\end{equation}
where $A, B$ are composite indices that are determined by choosing the
respective vertices according to
Eqs.~\eqref{eq:s-low-energ-gamma}-\eqref{eq:s-low-energ-X}.
Using $\xi(\bm{x},\bm{x}') = \xi(\bm{x}',\bm{x})$ and $\Phi(\bm{x},\bm{x}') =
-\Phi(\bm{x}',\bm{x})$ so that the Schwinger phases in the propagators cancel,
one can integrate over the center-of-mass coordinate $\bm{R} = (\bm{x} +
\bm{x}')/2$. The subsequent integrals over the relative spatial coordinate
$\bm{r} = \bm{x} - \bm{x}'$ can be done in polar coordinates, where the
angular integral gives rise to Bessel functions of the first kind.
The remaining radial integrals over products of Bessel functions with Laguerre
polynomials can be reduced to tabulated integrals (see Chapter 7 in
Ref.~\cite{gradshteyn15}).
We then arrive at
\begin{multline}  \label{eq:bubble-almost done}
        \chi^{+-}_{\bm{\Gamma}}(\ii \nu,\bm{k}) = -
2\times\frac{\ei^{-|\bm{k}|^2 \ell^2 /2 }}{2 \pi \ell^2} \bigg[ \frac{1}{\ii
\nu - b} + \sum_{n=1}^\infty \frac{1}{n! } \left(\frac{\ell^2 |\bm{k}|^2}{2}
\right)^n \frac{1}{\ii \nu - b -E_n} \\ +
        \frac{1}{2} \sum_{n=1,m=1}^\infty \frac{(-1)^{n+m} }{(\ii \nu - b)^2 -
(E_m +E_n)^2} \times \big\{(E_m+E_n) \left( \mathcal{L}^{n,m}_{0,0}(\ell^2
|\bm{k}|^2/2) + \mathcal{L}^{n,m}_{1,1}(\ell^2 |\bm{k}|^2/2) \right) \\
        - \frac{E_m+E_n}{E_m E_n} \frac{4 n }{\ell^2}
\mathcal{L}^{n,m}_{0,1}(\ell^2 \bm{k}^2) \big\}
	\bigg],
\end{multline}
where we use the notation $\mathcal{L}^{n,m}_{s,r}(x) = L^{m-n}_{n-s}(x)
L^{n-m}_{m-r}(x)$ for products of Laguerre polynomials.

The structure factor $S^{+-}_{\bm{\Gamma}}(\omega,q)$ can now be obtained from
\eqref{eq:bubble-almost done} via the fluctuation-dissipation theorem,
$S^{+-}_{\bm{\Gamma}}(\omega,q) = -2 \Theta(\omega) \chi^{\prime
\prime}_{+-}(\omega,q)$ where $\chi^{\prime \prime}_{+-}(\omega,q)=
\Im[\chi^{+-}_{\bm{\Gamma}}(\ii \omega \to \omega + \ii 0^+,q)]$.

To compute $S^{-+}(\omega,q)$, we use the KMS condition to write
$S^{-+}(\omega,q) =  \ei^{\beta \omega} S^{+-}(-\omega,q)$.
Then, using the fluctuation-dissipation theorem and letting $\beta \to
\infty$, we have $ \ei^{\beta \omega }S^{+-}(-\omega,q) = \ei^{\beta \omega}
2(1+n(-\omega)) \chi^{''}_{+-}(-\omega,q) \equiv -2 \theta(\omega)
\chi^{\prime \prime}_{+-}(-\omega,q)$ (note that Eq.~\eqref{eq:bubble-almost
done} is even under $q \to -q$).
Here, we see that finite contributions to $S^{-+}(\omega,q)$ only arise from
negative-frequency poles in the analytically continued
Eq.~\eqref{eq:bubble-almost done}. These occur only in the last term of
Eq.~\eqref{eq:bubble-almost done}, and thus at frequencies $\omega = -( E_m +
E_n) + b$, where $m,n \geq 1$.

\subsection{Dispersion of poles}

To find the dispersion relation of poles for small $|\bm{k}|$ away from the
$\Gamma$-point, we focus for simplicity only on the contribution to the
analytically continued expression of Eq.~\eqref{eq:bubble-almost done} and
find the maximum spectral weight as a function of $k = | \bm{k}|$:
\begin{equation}
        0 \overset{!}{=} \frac{\partial}{\partial k} \left[
\ei^{-\frac{k^2\ell^2}{2}} \frac{1}{n!} \left( \frac{\ell^2 k^2}{2} \right)^n
\right],
\end{equation}
which yields $\ell^2 k^2_\mathrm{max} = 2n$. We plug this into the pole $\nu =
E_n + b$ by using that $E_n = \sqrt{2n}/\ell$, yielding the dispersion
relation for the maximum spectral weight located at the poles as
\begin{equation}
	\nu_\mathrm{max}(|\bm{k}|) = |\bm{k}| + b.
\end{equation}
Such behavior is shown in Fig. 6 in the main text.

\section{Larmor's theorem and sum rules}
\label{app:Larmor}
Considering the correlation function $S^{+-}(t) = \langle S^+(t) S^-(0)
\rangle$ and $S^{-+}(t) = \langle S^-(t) S^+(0) \rangle$, we note that
$[S^+,S^-] = 2 S^z$ implies for their Fourier transforms $\int \frac{\di
\omega}{2\pi}  \left( S^{+-}(\omega) - S^{-+}(\omega) \right) =2 \langle S^z
\rangle$.
Using the fluctuation-dissipation theorem, the integrand can be expressed in
terms of the imaginary part of the corresponding susceptibility,
$S^{+-}(\omega) - S^{-+}(\omega) = 2 \Theta(\omega) ( \chi{''}_{+-}(\omega) -
\chi{''}_{+-}(-\omega)) $. We thus obtain the sum rule
\begin{equation}
    \int_{-\infty}^\infty \frac{\di \omega}{2 \pi} \chi^{\prime
\prime}_{+-}(\omega) = \langle S^z\rangle.
\end{equation}
Larmor's theorem \cite{balentsCollective2020, chenUniversal2024} then implies,
upon identifying $\vec{S}$ with the system's total spin, i.e. considering
correlation functions at the $\bm{\Gamma}$-point, the dynamical susceptibility
must thus be of the form
\begin{equation} \label{eq:chi-larmor}
    \chi_{+-}^{\prime \prime}(\omega) = 2 \pi m^z \delta(\omega - b)
\end{equation}
This also implies (at zero temperature, for $b>0$) that the structure factor
$S^{+-}_{\bm{\Gamma}}(\omega) = 2 (2 \pi m^z) \delta(\omega - b)$ while
$S^{-+}_{\bm{\Gamma}}(\omega) \equiv 0$.

As a cross-check, we now show that $\chi_{\bm{\Gamma}}^{+-}(\ii \nu, \bm{k})$
as given in Eq.~\eqref{eq:bubble-almost done} satisfies Larmor's theorem.
To this end, note that the second term in Eq.~\eqref{eq:bubble-almost done}
vanishes as $\bm{k} \to 0$. For the last term, use that
\begin{equation}
        L^{m-n}_{n-1}(0) L^{n-m}_{m-1}(0) = \tbinom{n-1+ m - n}{n-1}
\tbinom{m-1 + n-m}{m-1} \equiv \delta_{n,m},
\end{equation}
and further also
\begin{equation}
        L^{m-n}_n(0) L^{n-m}_{m-1}(0) = \tbinom{n+ m - n}{n} \tbinom{m - 1 +
n-m}{m-1}  \equiv \delta_{n,m}.
\end{equation}
So the Laguerre-polynomials in Eq.~\eqref{eq:bubble-almost done} at $\bm{k}=0$
just give $\delta_{n,m}$, and we can cancel the $m$-summation and get a single
sum over $n$.
The brace then becomes (observe that $2 n \ell^{-2} \equiv E_n^2$):
\begin{equation}
	\sum_{n,m=1}^\infty  (\dots)\delta_{n,m}\big\{2 (E_m+E_n)
	- \frac{E_m+E_n}{E_m E_n} \frac{4 n}{\ell^2}  \big\} 
        = \sum_{n=1}^\infty \left\{ 4 E_n - \frac{2 E_n}{E_n^2} E_n^2\times 2
\right\}  \equiv 0.
\end{equation}
So at $\bm{k} = 0$, the dynamical response is given by
\begin{equation}
        \chi^{+-}(\ii \nu,{k}) \sim - 2\times\frac{1}{2 \pi \ell^2}
\frac{1}{\ii \nu - b}
\end{equation}
Analytic continuation then yields $\chi^{\prime \prime}_{+-}(\omega,0) =
\frac{1}{\ell^2} \delta(\omega - b)$, which is in agreement with
Eq.~\eqref{eq:chi-larmor} upon identifying $m^z = 1/(2 \pi \ell^2)$, which is
precisely the magnetization of the state under consideration (see
Eq.~\eqref{eq:mz-landau-level}).

In Figs. 6 and 7(a) and (c), the Larmor mode at $\Gamma$ in the transverse
magnetic spectra is clearly seen in both field-theoretical calculations and
lattice-model (mean-field and QMC) simulations.
It has also been previously observed in QMC simulations of magnetized Dirac
fermions in a Hubbard-model setting~\cite{chenUniversal2024}.

\section{Analysis of flux-flux correlation functions}
\label{app:correlation_gauge}

\subsection{Free photon theory (non-compact 2+1-dim. electrodynamics without
fermions)}

\begin{figure}[tbp]
\includegraphics[width=0.6\columnwidth]{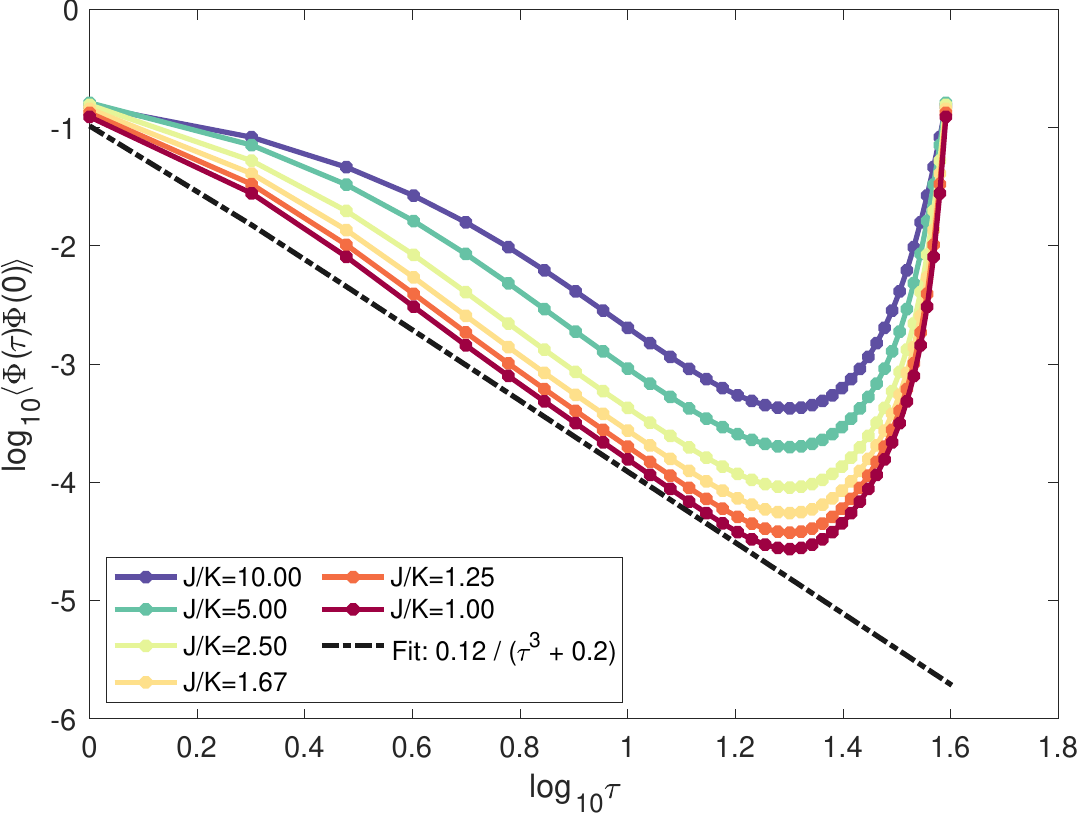}
        \caption{The log-log plot of imaginary-time flux correlation function
$\left\langle \Phi(x, y, \tau)\, \Phi(x, y, 0) \right\rangle$ for the
non-compact QED without fermion coupling. The result is averaged over all
real-space sites. The fit line is $y=0.12/(\tau^3 + 0.2)$ which agrees with
$1/\tau^3$ scaling behavior in the low-frequency limit (large $\tau$) obtained
in the analytical result \refeq{eq:anal_Gtau}, as a consequence of gapless
spectrum of the gauge field.}
        \label{fig:non_compact_qed}
\end{figure}

\begin{figure}[tbp]
\includegraphics[width=0.6\columnwidth]{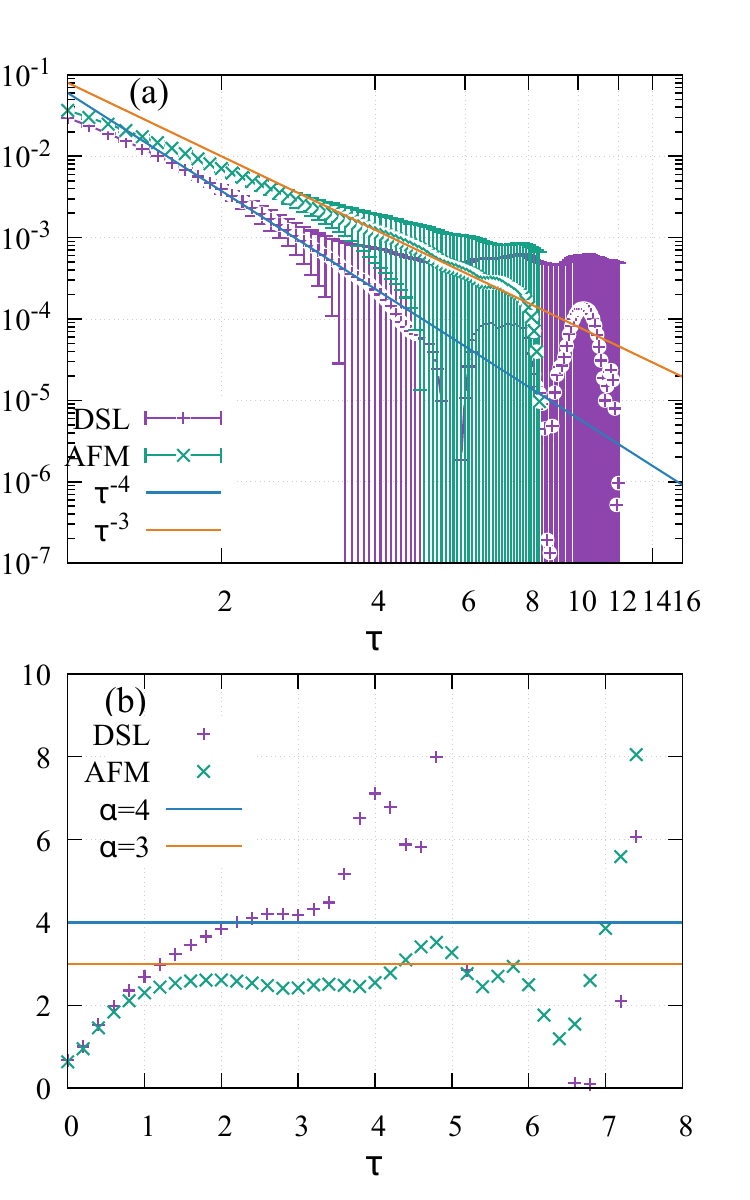}
        \caption{(a) The log-log plot of imaginary-time flux correlation
function $C_\chi(\tau)$
          for the non-compact QED$_3$ with fermion coupling. The purple line
          corresponds to parameters $B/t=0, J/t=3, K/t=1$ and lies in the DSL
          phase, while the green line has parameters $B/t=3, J/t=3, K/t=0$ and
          is in the deconfined AFM phase. The system size and inverse
          temperature are $L=12, \beta=36$. Two solid lines represent
          baselines of $\tau^{-3}$ and $\tau^{-4}$. (b) Power law fitting of
          $C(\tau) \sim A\tau^{-\alpha}$ with sliding window, the window size
is
        $\Delta \tau = 1.0$. $\alpha$ vs $\tau$, which is the imaginary
      time of the start of the fitting window. Two solid lines represent
      baselines of $\alpha = 3$ and $\alpha = 4$.}
        \label{fig:flux_dyn_corr_scaling}
\end{figure}

\begin{figure}[tbp]
\includegraphics[width=0.6\columnwidth]{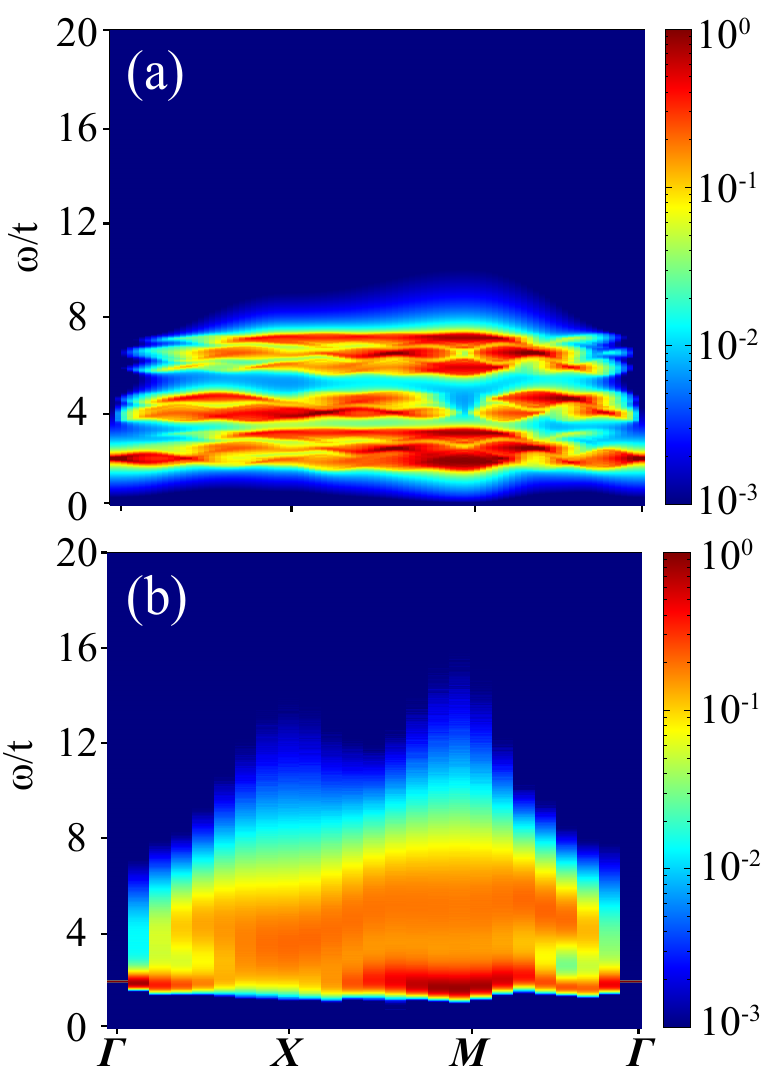}
\caption{\textbf{Transverse magnetic spectra at
          Zeeman field $B/t=2$ from QMC+SAC and
          mean field calculations.} (a) $S^\pm(\mathbf{q}, \omega)$
          from mean field with lattice size $L=60$. (b) $S^\pm(\mathbf{q},
\omega)$ from DQMC
          with $L=16, \beta=16, J/t=0.1$. Mean field data shows three bands,
while spectra
          from QMC shows at least two bands, with high energy properties hard
to resolve.
          QMC data also shows a reduction of weight at $\bm{M}$ point for the
second
          band. }
        \label{fig:fig_spec_spsm_B2}
\end{figure}

\begin{figure}[tbp]
\includegraphics[width=0.6\columnwidth]{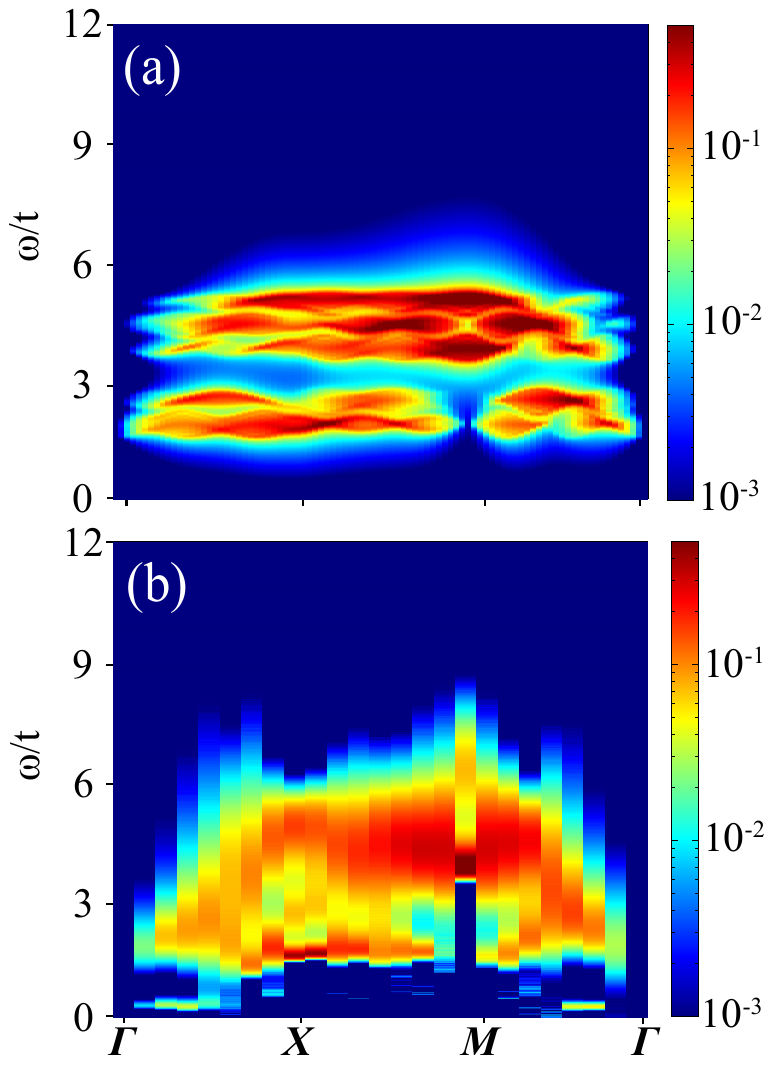}
\caption{\textbf{Longitudinal magnetic spectra at
          Zeeman field $B/t=2$ from QMC+SAC and
          mean field calculations.} (a) $S^{zz}(\mathbf{q}, \omega)$
          from mean field with lattice size $L=60$. (b) $S^{zz}(\mathbf{q},
\omega)$ from DQMC
          with $L=16, \beta=16, J/t=0.1$. QMC and mean field results are more
          consistent, both with two bands and a disappearing of weight at
$\bm{M}$
          for the first band.}
        \label{fig:fig_spec_szsz_B2}
\end{figure}

In this section, we consider a free photon theory and analyze the fluctuation
of the magnetic flux, which arises from the Goldstone mode of a spontaneously
broken U(1)$_m$ symmetry. The Lagrangian of the free photon theory is a
non-compact \qed:
\begin{equation}
    L=\sum_r \frac{1}{2 J} \sum_{a=x, y}\left[\frac{a_\alpha
\left(x, y, \tau+e_\tau\right)-a_\alpha(x, y, \tau)}{\left|e_\tau\right|}
\right]^2 +\sum_r \frac{K}{2}\left[a_x(r)+a_y\left(r+e_x\right)-a_x\left(r+e_y
\right)-a_y(r)\right]^2,
\end{equation}
where $a_\alpha(r)$ is the gauge field which lives on a bond with direction
$\alpha\in\{x, y\}$, whose left or bottom end point is on the site $r=(x, y,
\tau)$. The $K$ term consists of magnetic flux defined as $\Phi(r) = \nabla
\times a_\alpha(r) =
a_x(r)+a_y\left(r+e_x\right)-a_x\left(r+e_y\right)-a_y(r)$, which goes through
a spatial plaquette whose bottom left corner is at site $r$.
We have also chosen the $\tau$ unit cell  $|e_\tau| = 1/J$. Transforming the
Lagrangian to the $k$-space, we can obtain the flux correlation function
\begin{equation}
\left\langle\Phi(r)\Phi\left(0\right)\right\rangle=
\frac{1}{V} \sum_{k_x, k_y, \omega_n} \!\! e^{i k_x x+i k_y y +i \omega_n \tau
} \frac{2\left(\cos \left[k_x\right]+\cos \left[k_y\right]-2\right)}{\cos
\left[\omega_n\right]-1+K/J \cdot \left(\cos \left[k_x\right]+\cos
\left[k_y\right]-2\right)}. \label{eq:anal_Gtau}
\end{equation}
The imaginary-time flux correlation function $\left\langle \Phi(x, y, \tau)\,
\Phi(x, y, 0) \right\rangle$ is plotted in \reffg{fig:non_compact_qed}.
The fit line is $y=0.12/(\tau^3 + 0.2)$ which agrees with $1/\tau^3$ scaling
behavior in the low-frequency limit (large $\tau$) obtained from the
analytical expression above, which is a consequence of gapless spectrum of the
gauge field.

\subsection{Mixed \texorpdfstring{QED$_3$}{QED3} (with fermionic matter)}

We also investigate the scaling of dynamical flux correlation for mixed
QED$_3$ with fermion coupling. The results are shown in
Fig.~\ref{fig:flux_dyn_corr_scaling}. DSL phase with Dirac-type gapless
spinon dispersion shows $C_\chi(\tau) \sim \tau^{-4}$ asymptotic scaling
behavior at large $\tau$, while at
$B=3$, the deconfined AFM with gapped fermion shows better agreement with a
$C_\chi(\tau) \sim \tau^{-3}$ asymptotic behavior at large $\tau$.
An unambiguous power-law exponent is difficult to extract from these results,
due to the challenge of avoiding non-universal short-time features at small
$\tau$ and finite-size cutoffs at large $\tau$ of order $\beta$.
Nevertheless, we see that the theoretically predicted scalings (Sec.~IIC3) are
at least consistent with the data in a range of intermediate $\tau$. In order
to
study the variation of the fitted power $\alpha$ in $C(\tau) \sim
A\tau^{-\alpha}$ versus the fitting window, we perform the fitting starting
from
$\tau=0$ to $\tau=\beta/2$ with fitting window $\Delta \tau = 1$ and show the
results in Fig.~\ref{fig:flux_dyn_corr_scaling}(b).
We observe that the fitted
power $\alpha$ at small $\tau$ and $\tau \sim \beta/2$ deviate from $\alpha
\sim 4$ for DSL and $\alpha \sim 3$ for AFM. In order to study the asymptotic
behavior, we need to use larger $\tau$ but eventually stopped by large
numerical
error before reaching $\tau \sim \beta/2$.

\section{\texorpdfstring{$z$}{z}-direction flux insertion and global update}
\label{app:flux_insert}

In DQMC simulations, one can introduce $z$-direction flux into the model. The
consequence is shifted momentum points in Brillouin zone, thus offering extra
momentum points otherwise unavailable for that system size. Finite size
effects
can be effectively reduced with this method~\cite{alf2.0,jiangMonte2022}. The
flux is
introduced via Peierls phase factors with $a_{o}$ the orbital phase factor,
$\vec{A}_o(\vec{r})$ the vector potential and $\Phi_0$ the flux quanta.
In Landau gauge, we choose
$\vec{A}_o(\vec{r})=-B_o(y,0,0)$ with $B_o$ the introduced orbital magnetic
field.
In order to respect periodic boundary condition, we use a gauge
transformation for the vector potential on the boundary 
\begin{align}
  \vec{A}_o(\vec{r}+L\vec{e}_x) &= \vec{A}_o(\vec{r}) +
\vec{\nabla}\chi_x(\vec{r}) \nonumber \\
  \vec{A}_o(\vec{r}+L\vec{e}_y) &= \vec{A}_o(\vec{r}) +
\vec{\nabla}\chi_y(\vec{r})
\end{align}
and the corresponding boundary condition for fermion
\begin{align}
  \bar{\psi}_{\vec{i}+L\vec{e}_x} &= e^{\frac{2\pi i}{\Phi_0}\chi_x(\vec{i})}
\bar{\psi}_{\vec{i}} \nonumber \\
  \psi_{\vec{i}+L\vec{e}_y} &= e^{\frac{2\pi i}{\Phi_0}\chi_y(\vec{i})}
\psi_{\vec{i}}
\end{align}
We choose $\chi_x({\vec{i}})=0$ and $\chi_y({\vec{i}})=-B_oLx$ with orbital
magnetic
field satisfying
\begin{equation}
  \frac{B_o \cdot L^2}{\Phi_0} = N_\Phi
\end{equation}
with $N_\Phi$ an integer to ensure uniqueness of wave function. In conclusion,
$a_{o,x}=-\frac{2\pi N_\Phi
y}{L^2}$, $a_{o,y} = 0$ away from boundary and $a_{o,y} = \frac{2\pi N_\Phi
x}{L}$ for boundary bonds.

Based on this method, we can propose global space-time gauge field update by
inserting random $N_\Phi$
$z$-direction flux uniformly for all imaginary time slices to the current
gauge field configuration. It helps to quickly evolve to
the desired flux sector and traditional local update will explore the whole
phase space
ergodically. As mentioned in Sec. IIB in the main text, we combine both the
global and the local updates in the QMC sampling process.

\section{\texorpdfstring{U(1)$_m$}{U(1)m} symmetry and duality map}\label{sm:duality}

The action in Eq.~(1) of the main text is of a ``mixed'' compact/non-compact
character: on the one hand, the coupling between gauge field and fermions
suggests $a_{ij}(\tau_n) \sim a_{ij}(\tau_n) + 2\pi$, as common in compact
U(1) lattice gauge theories.
On the other hand, the kinetic term of the gauge field, which couples the
gauge field between different time slices, spoils such a putative compactness
in the time-continuum limit. This unusual behaviour is distinct from fully
compact or non-compact theories, and is the motivation behind the nomenclature
of ``mixed'' QED$_3$ in the main text.

In order to gain insight into the properties of a theory which is periodic in
the gauge-field along certain space-time directions only, we consider a
simplified effective pure gauge theory model. Such a theory will capture the
low-energy dynamics in phases where the fermions have a finite gap and can be
integrated out.

\subsection{Effective action for gapped states}

Integrating-out the fermionic degrees of freedom will give rise to terms
periodic in the field strength on spatial plaquettes (i.e. plaquettes at
constant $\tau$), while the temporal plaquettes feature a non-compact kinetic
term for the gauge field,
\begin{equation} \label{eq:s-eff-gauge-mixed}
        S = - \frac{1}{g_\tau^2} \sum_r \cos (\epsilon_{\tau ij} \Delta_{i}
A_{r,j}) + \frac{1}{2g^2} \sum_{r,i}  (\Delta_\tau A_{r,i} - \Delta_i
A_{r,\tau})^2,
\end{equation}
with the associated partition function $Z= \int_\mathbb{R} D[A] e^{-S}$.
Here, $r$ are coordinates for the sites of the 2+1-dim. space-time lattice
with unit vectors $\hat{x},\hat{y},\hat{\tau}$ (latin indices are used for
spatial components $i=x,y$ while $\tau$ denotes the temporal component), and
$\Delta_\mu f_r = f_{r+\hat{\mu}} - f_r$ denotes the forward difference
operator (with $\mu = x,y,z$), i.e. a lattice discretization of the partial
derivative $\partial_\mu$.
For later convenience, we have also explicitly kept the temporal component of
the gauge field $A_{r,\tau}$, while in Eq.~(1) of the main text we are using
the temporal gauge $A_{r,\tau} \equiv 0$.

Making a Villain approximation of the cosine, $e^{t \cos (x) } \approx
\sum_{k\in \mathbb{Z}} e^{-t/2 (x-2\pi k)^2}$, one obtains
\begin{multline} \label{eq:villain-approx}
        S= \frac{1}{2 g_\tau^2} \sum_r \left( \Delta_x A_{r,y} - \Delta_y
A_{r,x} - 2\pi k_{r,\tau} \right)^2 + \frac{1}{2g^2} \sum_{r,i}  (\Delta_\tau
A_{r,i} - \Delta_i A_{r,\tau})^2 \\
    - \sum_r h_{r,\tau} \left( \Delta_x A_{r,y} - \Delta_y A_{r,x} - 2\pi
k_{r,\tau} \right) - \sum_{r,i,j} h_{r,i} (\Delta_j A_{r,\tau} - \Delta_\tau
A_{r,j}) \epsilon^{ij \tau} .
\end{multline}
Here, for convenience, we have also introduced a three-component background
field $h_{r,\mu}$ that can be associated with a plaquette anchored at $r$ and
normal vector $\mu$.
The field $h$ may be used to compute expectation values on spatial and
temporal plaquettes as
\begin{equation} \label{eq:h-derivs}
    \langle \sin \left( \Delta_x A_{r,y} - \Delta_y A_{r,x} \right) \rangle =
\left.\frac{\delta}{\delta h_{r,\tau}} \log Z[h] \right|_{h=0} \quad
\text{and} \quad \langle \Delta_j A_{r,\tau} - \Delta_\tau A_{r,j} \rangle
\epsilon^{ij \tau} = \left.\frac{\delta}{\delta h_{r,i}} \log Z[h]
\right|_{h=0}.
\end{equation}
The first equality is obtained from Eq.~\eqref{eq:s-eff-gauge-mixed}, after
consistently including the field $h$ therein.
We emphasize that these (local) expectation values are to be taken with
respect to the full action, and in particular include summing over all
possible sectors of the integer variables $k_{r,\tau}$.

In a fixed sector defined by the Villain variables $k_{r,\tau} \in
\mathbb{Z}$, the total flux through a time slice (spanned by sites with $\tau
= \mathrm{const}.$) is given by the sum $F(\tau_n) = \sum_{(r_x,r_y)}
F_{(r_x,r_y,\tau_n),\tau}$, where $F_{r,\tau}$ is obtained as a variation
$F_{r,\tau} = \left.\delta / \delta h_{r,\tau} (-S)\right|_{h_{r,\tau} =0} =
\Delta_x A_{r,y} - \Delta_y A_{r,x} - 2\pi k_{r,\tau}$.
With periodic boundary conditions, the smooth contributions cancel, and we
find a \emph{quantized} total flux through a given time slice, determined by
the Villain variables as
\begin{equation} \label{eq:total-flux}
    F(\tau_n) = \sum_{(r_x,r_y)} (-2 \pi k_{(r_x,r_y,\tau_n),\tau}) \in  2 \pi
\mathbb{Z}.
\end{equation}

\subsection{Duality mapping}

Resumming the first term via $
        \sum_{k \in \mathbb{Z}} e^{-\frac{1}{2t}(x-2\pi k)^2} =
\sqrt{\frac{t}{2\pi}} \sum_{n\in \mathbb{Z}} e^{-\frac{t}{2} n^2 + i x n}$,
dropping a term quadratic in the background field $h$, and using a
Hubbard-Stratonovich decoupling for the second set of terms yields the action
(we neglect terms quadratic in $h$)
\begin{multline}
        S = \sum_r  \left[ \frac{g_\tau^2}{2} (n_{r,\tau})^2 +
\frac{g^2}{2}\left( (b_{r,y})^2 + (b_{r,x})^2\right) \right] - i \sum_r \Big[
n_{r,\tau} ((\Delta_x A_{r,y} - \Delta_y A_{r,x}) - g^2_\tau h_{r,\tau}) \\
    -  \sum_i  \epsilon^{ij \tau} ((\Delta_j A_{r,\tau} - \Delta_\tau A_{r,j})
- g^2 h_{r,i}) b_{r,i} \Big],
\end{multline}
and the partition function becomes $Z = \sum_{\{n \in \mathbb{Z} \}}
\int_\mathbb{R} D[ A] \int_\mathbb{R} D[ b] e^{-S}$.
Here, the spatial components $b_{r,x}$ and $b_{r,y}$ are $\mathbb{R}$-valued,
while the integer-valued nature of the temporal component $n_{r,\tau} \in
\mathbb{Z}$ directly follows from the fact that
Eq.~\eqref{eq:s-eff-gauge-mixed} is periodic in the flux on spatial
plaquettes.

For a unified treatment, we use the Poisson summation formula\footnote{For
some function $f(n)$ of integers, we may write $\sum_{n\in \mathbb{Z}} f(n) =
\int_{-\infty}^\infty {\mathrm{d} \phi} \sum_{n \in \mathbb{Z}} \delta( n-
\phi) f(\phi) \equiv \int_{-\infty}^\infty {\mathrm{d}  \phi} \sum_{m \in
\mathbb{Z}} e^{i 2\pi m \phi} f(\phi)$.} rewrite $n_{r,\tau}$ in terms of a
$\mathbb{R}$-valued field (which we can identify as the temporal component
$b_{r,\tau}$ of a vector field $b_{r,\mu}$) along with a quantization
constraint. Further performing a partial resummation (i.e. the discrete
version of partial integration), one then arrives at
\begin{multline}
        S = \sum_{r} \left [\frac{g_\tau^2}{2}  (b_{r,\tau})^2 + \frac{g^2}{2}
\left( (b_{r,x})^2 + (b_{r,y})^2 \right) \right] \\
{}+ i \sum_r \left[ A_{r,\tau}
(\bar{\Delta}_y b_{r,x} - \bar{\Delta}_x b_{r,y}) + A_{r,x} (\bar{\Delta}_\tau
b_{r,y} - \bar{\Delta}_y b_{r,\tau}) + A_{r,y} (\bar{\Delta}_x b_{r,\tau} -
\bar{\Delta}_\tau b_{r,x})  \right] \\
        - 2\pi i \sum_r m_r b_{r,\tau} + i g^2_\tau \sum_r h_{r,\tau}
b_{r,\tau} + i g^2 \sum_{r,j} h_{r,j} b_{r,j},
\end{multline}
where $\bar{\Delta}_\mu f_r = f_{r} - f_{r-\hat{\mu}}$ denotes the backward
difference operator, and $m_r \in \mathbb{Z}$ is integer-valued, so that the
partition function now reads $ Z = \int_\mathbb{R} D[b] \int_\mathbb{R} D[A]
\sum_{\{m_r \in \mathbb{Z} \}} e^{-S}$.
We may now integrate over the gauge-field variables $A$, which act as Lagrange
multipliers to enforce that $b_{r,\rho}$ is flat, i.e. $\epsilon^{\mu \nu
\rho} \bar{\Delta}_\nu b_{r,\rho} = 0$. We solve this constraint by writing
$b_{r,\rho} = \bar{\Delta}_\rho \phi_{r}$ (note that we may write $b_{r,\rho}$
\emph{globally} as a gradient since any putative winding, i.e. topological
defects, are now encoded in the explicit quantization of $b_{r,\tau}$ via the
last term $\propto m_r b_{r,\tau}$).

\begin{figure}
    \centering
    \includegraphics[width=0.5\linewidth]{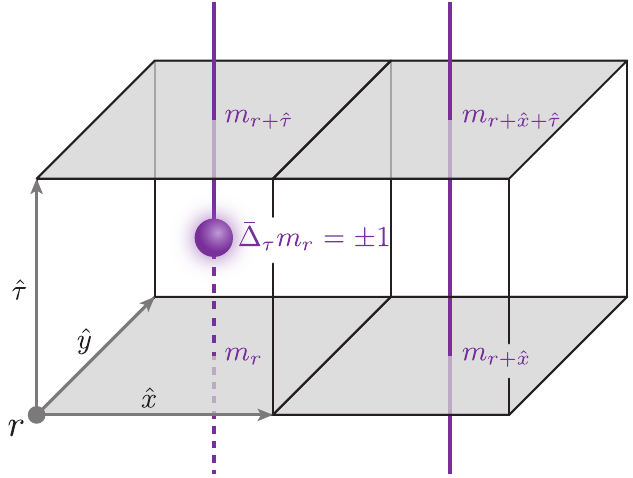}
    \caption{Illustration of dual theory on the space-time lattice. The action
is periodic in the flux through spatial plaquettes (i.e. within a time slice
defined by $\tau = \mathrm{const}.$), shaded in grey. As a result, the gauge
field may wind by $2
    \pi$ between two plaquettes. Integer multiples of $2
    \pi$ flux passing through the plaquettes are encoded by the variables
$m_{r}$ (purple). A monopole insertion $\bar{\Delta}_\tau m_r \neq 0$
corresponds to a defect with an integer multiple of $2\pi$ emanating a closed
surface (e.g. an elementary space-time cube). The flux through temporal
plaquettes (i.e. within a spatial slice define by $x = \mathrm{const}.$ or $y
= \mathrm{const}.$) is $\mathbb{R}$-valued, and accordingly no flux quanta
penetrate these surfaces.}
    \label{fig:monopole-illu-cut}
\end{figure}

Upon partially integrating the last term, we then arrive at the action
\begin{equation} \label{eq:s-mixed}
        S = \sum_r \Big[\frac{g_\tau^2}{2} (\bar{\Delta}_{\tau} \phi_r)^2 +
\frac{g^2}{2} \left( (\bar{\Delta}_{x} \phi_r)^2 +  (\bar{\Delta}_{y}
\phi_r)^2  \right) \Big] + 2\pi i \sum_r (\bar{\Delta}_{\tau} m_r) \phi_r + i
\sum_r \Big[ g_\tau^2 h_{r,\tau} \bar{\Delta}_\tau \phi_r + g^2 \sum_{j=x,y}
h_{r,j} \bar{\Delta}_j \phi_r \Big].
\end{equation}
with the associated partition function $Z = \sum_{\{m\in \mathbb{Z} \} }\int
D[\phi] e^{-S}$ (recall that $h_{r,\mu}$ is a static background/source term).
For a given configuration of $m_r \in \mathbb{Z}$, the theory
\eqref{eq:s-mixed} is a (discretized) action of a scalar field in 2+1-dims.
which couples to integer variables $\bar{\Delta}_\tau m_r \in \mathbb{Z}$.
While here we have index $\phi_r$ with the sites of the space-time lattice, it
is natural to associate the field $\phi_r$ with the sites of the dual lattice
(i.e. the centers of an elementary cube with a lower left front corner
anchored at $r$), and the field $m_r$ with the horizontal (top/bottom) faces
of such cube, such that $\bar{\Delta}_\tau m_r$ lives on the cube centers, see
also Fig.~\ref{fig:monopole-illu-cut}.

We now consider a given sector of winding variables $k_{r}$ in
Eq.~\eqref{eq:villain-approx}. By varying with respect to $h_{r,\tau}$ and
$h_{r,i}$ we obtain the flux on each plaquette as
\begin{subequations}
    \begin{align}
        F_{r,\tau} \equiv \Delta_x A_{r,y} - \Delta_y A_{r,x} - 2\pi k_r &= -
i g^2_\tau \bar{\Delta}_\tau \phi_r \\
        F_{r,y}\equiv \Delta_\tau A_{r,x} - \Delta_x A_{r,\tau} &= - i g^2
\bar{\Delta}_y \phi_r \\
        F_{r,x} \equiv \Delta_y A_{r,\tau} - \Delta_\tau A_{r,y} &= - i g^2
\bar{\Delta}_x \phi_r.
    \end{align}
\end{subequations}The left-hand side of the above expressions corresponds to
the flux through the horizontal and vertical plaquettes of a cube anchored at
$r$. We may evaluate the total flux through the cube $C$ as a sum over the six
(oriented) plaquettes as
\begin{equation}
    \sum_{p \in C} F_p (-1)^p = 2\pi \bar{\Delta}_\tau m_r \in 2\pi
\mathbb{Z}, \label{eq:flux-cube}
\end{equation}
where we have used a discrete version of the divergence theorem and the
saddle-point equations of motion for $\phi_r$, i.e. $g_\tau^2 \Delta_\tau
\bar{\Delta}_\tau \phi_r + \sum_i g_i^2 \Delta_i \bar{\Delta}_i \phi_r =
\bar{\Delta}_\tau m_r$.

\subsection{Discussion}

From the above treatment, we draw three main conclusions:
\begin{enumerate}
    \item The non-compact theory where the $\cos$ in
Eq.~\eqref{eq:s-eff-gauge-mixed} is replaced by a harmonic term, i.e. the
sector with $k_{r,\tau} \equiv 0$ in Eq.~\eqref{eq:villain-approx}, features $
\sum_{p \in C} F_p (-1)^p \equiv 0$: magnetic flux is conserved. There is an
associated $\mathbb{R}$-symmetry which acts as $\phi_r \to \phi_r + c$, $c\in
\mathbb{R}$ in the dual theory which becomes a free scalar field. The ground
state breaks this symmetry spontaneously, and the linearly dispersing free
scalar (Goldstone mode) yielding a dual description to the free photon.
    \item In the theory \eqref{eq:s-eff-gauge-mixed}, flux \emph{only} through
spatial plaquettes (i.e. within a given time-slice) becomes compactified mod
$2\pi$, as also evidenced by Eq.~\eqref{eq:h-derivs}. Is there an associated
U(1)$_m$ flux conservation symmetry? Our explicit calculation,
Eq.~\eqref{eq:flux-cube} shows that there exist ``monopole'' configurations
where integer multiples of $2\pi$ flux emanate a space-time cube: one may
define a corresponding monopole charge $q_r = \bar{\Delta}_r m_r$, and
interpret the $m_r$ variables as encoding the corresponding ``Dirac string''.
In contrast to usual (2+1)-dim. compact U(1) lattice gauge theory, these
strings can only extend along the temporal direction as the $2\pi$ windings on
vertical plaquettes is heavily penalized by the harmonic terms in
Eq.~\eqref{eq:s-eff-gauge-mixed}. These monopole configurations explicitly
spoil a putative U(1)$_m$ symmetry, as can also be seen from
Eq.~\eqref{eq:s-mixed}: the dual scalar field gets pinned and loses the
``shift'' symmetry $\phi_{\bf r} \to \phi_{\bf r} + c$.
    \item We now consider the continuum limit $\delta \tau \to 0$ along the
imaginary-time ($\tau$) direction which amounts to scaling $g_\tau^2 \to
g_\tau^2 / \delta \tau \to \infty$. In this limit, the kinetic term of the
dual theory \eqref{eq:s-mixed} enjoys an (approximate) symmetry $\phi_r \to
\phi_r + g(x,y)$, i.e. the field $\phi_r = \phi(x,y)$ is approximately
time-independent. Then, the coupling $\sum_r \bar{\Delta}_\tau m_r \phi_r$
reduces to a boundary term in the temporal direction. Hence, the ``monopole''
term in Eq.~\eqref{eq:s-mixed} vanishes (in the bulk). Thus, no confinement
occurs: the shift symmetry of the scalar field $\phi_r$ is restored, and it
gives way to the (dual of the) gapless photon.
\end{enumerate}

\section{\texorpdfstring{$B=2$}{B=2} DQMC and mean field spectra}
\label{app:B2spectra}

We also investigate the spin structure factor inside the CF phase at $B/t=2$
as shown in
Fig.~\ref{fig:fig_spec_spsm_B2} and \ref{fig:fig_spec_szsz_B2}, comparing the
results obtained with DQMC+SAC and mean-field methods. The $B/t=2$ spin
structure
exhibits similar characteristics, featuring separate nearly dispersionless
bands, akin to the findings for $B/t=3$, as seen in Figs. 7
and 8. The Larmor mode in the transverse structure factor
is located at $\omega = 2t = B$. Notably, there is a significant reduction in
spectral weight below the Larmor mode compared to the $B/t=3$ case. The lower
orbital magnetic field induced by $B/t=2$ more closely resembles the scenario
discussed in continuum field theory, where such reduced spectral weight below
the Larmor mode is observed at small momenta, as shown in
Fig. 6. Regarding the longitudinal structure
factor, the DQMC spectra for $J/t = 0.1$ reveals a similar low-energy
collective
mode near $\mathbf{\Gamma}$, also seen in $B/t=3$ case, which is absent in
mean-field results. These
deviations from mean-field analysis might be attributed to gauge fluctuations
discussed in Sec. IIC2.

\section{Finite size analysis of order parameters}\label{sm:finite-size}
\begin{figure}[tbp]
        \includegraphics[width=0.6\columnwidth]{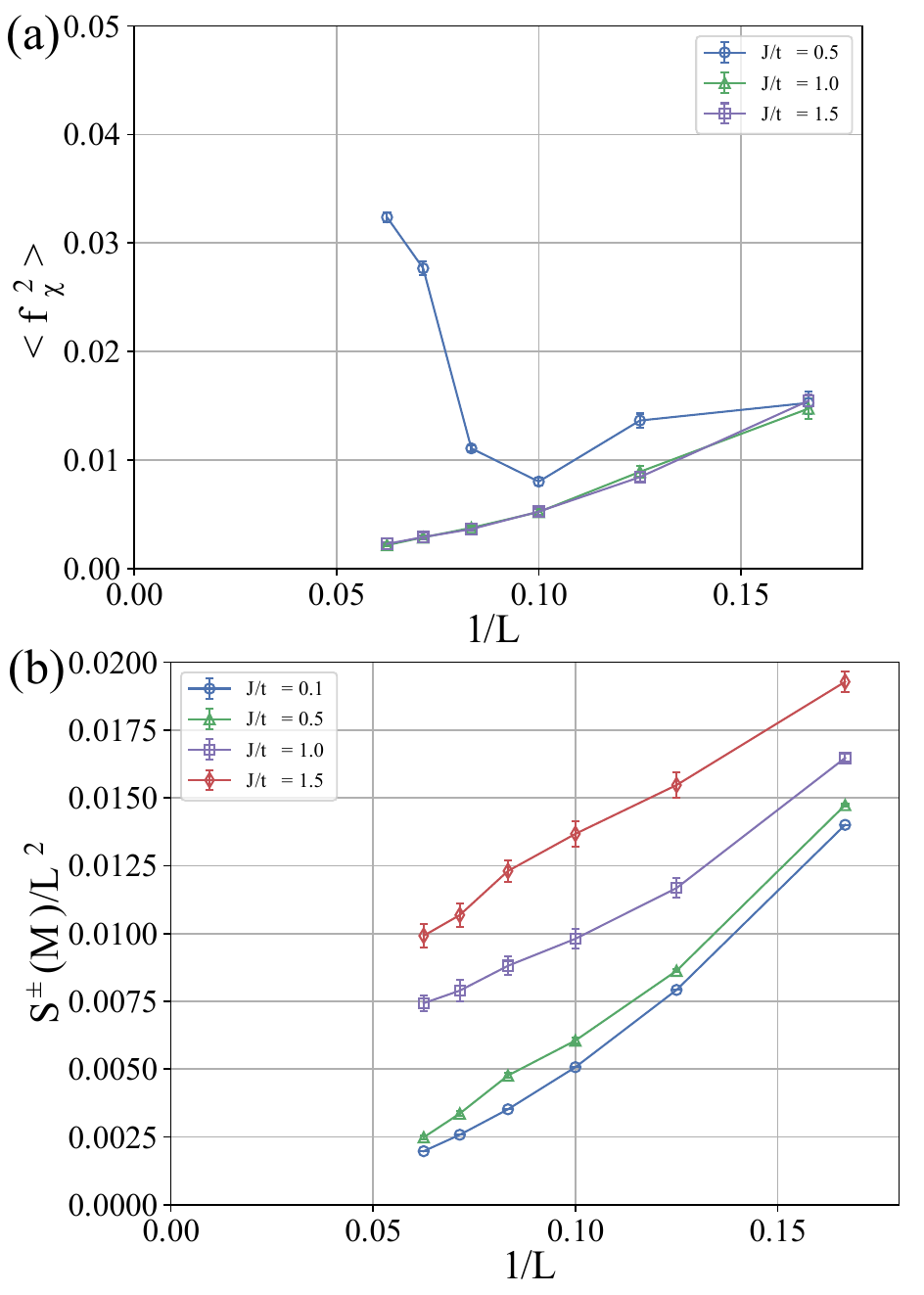}
        \caption{Finite size analysis of CF (a) and AFM (b) respectively. The
          parameter is $B/t=2$ and $\beta=2L$ with $L=6, 8, 10, 12, 14, 16$.
(a)
        For $J/t=0.5$ inside CF phase, CF order parameter $<f^2_\chi>$ goes to
        finite value as $L$ increases. While for $J/t=1.0, 1.5$, $f^2_\chi$
goes
      to zero as $L$ increases. (b) AFM order parameter
$S^\pm(\mathbf{M})/L^2$
      verse $1/L$. $J/t=0.1, 0.5$ extrapolates to zero and $J/t=1.0, 1.5$
      extrapolates to finite value indicating long range order. The data is
      consistent with the phase diagram as shown in Fig.1(b) of the main
text.}
        \label{fig:order_parameters_finitesize}
\end{figure}

Apart from the crossing point analysis in main text that manages to determine
the transition points of CF and AFM phases, it is also worthwhile to do finite
size
analysis of CF and AFM order parameters. Namely study the scaling behavior of
$<f^2_\chi>$ and $S^\pm(\mathbf{M})/L^2$ versus $1/L$. Note that in the DQMC
simulations, we already scale inverse temperature $\beta$ with $2L$ to ensure
that we are studying zero temperature physics. The finite size analysis
results
are shown in Fig.~\ref{fig:order_parameters_finitesize}. The order parameters
extrapolate to finite values inside long range ordered phase, whose boundaries
are shown in the phase diagram, which is determined by crossing point
analysis, in Fig.1(b) of the main text.

\end{document}